\documentclass[aps,floats,superscriptaddress,twocolumn]{revtex4-2}
\usepackage{graphicx,epsfig}
\usepackage{graphics,dcolumn,bm,epic, eepic,float}
\usepackage{amssymb,amsmath,amsfonts,multirow,rotate,color}
\usepackage{soul,xcolor}

\newcommand{\avg}[1]{{\langle #1 \rangle}}

\begin{document} 

\title{A link model approach to identify congestion hotspots}

\author{Aleix Bassolas}
\affiliation{Departament d'Enginyeria Inform\`atica i Matem\`atiques, Universitat Rovira i Virgili, 43007 Tarragona, Spain}
\affiliation{Instituto de F\'{\i}sica Interdisciplinar y Sistemas Complejos IFISC (CSIC-UIB), Campus UIB, 07122 Palma de Mallorca, Spain}
\author{Sergio G\'omez}
\affiliation{Departament d'Enginyeria Inform\`atica i Matem\`atiques, Universitat Rovira i Virgili, 43007 Tarragona, Spain}
\author{Alex Arenas}
\affiliation{Departament d'Enginyeria Inform\`atica i Matem\`atiques, Universitat Rovira i Virgili, 43007 Tarragona, Spain}
\date{\today}

\begin{abstract}
Congestion emerges when high demand peaks put transportation systems under stress. Understanding the interplay between the spatial organization of demand, the route choices of citizens, and the underlying infrastructures is thus crucial to locate congestion hotspots and mitigate the delay. Here we develop a model where links are responsible for the processing of vehicles, that can be solved analytically before and after the onset of congestion, and providing insights into the global and local congestion. We apply our method to synthetic and real transportation networks, observing a strong agreement between the analytical solutions and the Monte Carlo simulations, and a reasonable agreement with the travel times observed in 12~cities under congested phase. Our framework can incorporate any type of routing extracted from real trajectory data to provide a more detailed description of congestion phenomena, and could be used to dynamically adapt the capacity of road segments according to the flow of vehicles, or reduce congestion through hotspot pricing.
\end{abstract}

\maketitle

\section{Introduction}

Almost $25\%$ of greenhouse gas (GHG) emissions in the United States and Europe are a consequence of road transportation, which are only worsened by the congestion produced by the stress of the infrastructures \cite{barth2009traffic,berechman2010evaluation}. Pollution is, however, only one of the many adverse outcomes of congestion as it also affects the safety of pedestrians and drivers \cite{albalate2019congestion} and strongly impacts local and global economies \cite{weisbrod2003measuring,goodwin2004economic}. All together, it makes the mitigation of congestion an imperative. 

The interest in understanding the emergence of congestion and improving urban mobility led to the development of a wide variety of models \cite{gipps1981behavioural,helbing1998generalized,olak2016,chodrow2016demand,w2016multi,wang2018dynamic}: going from the more microscopic car-following approaches aiming to reproduce the interaction between vehicles to the more aggregated based on the distribution of flows. In between, we find models based on critical phenomena that have been useful to assess the role of topology in the emergence of congestion and its optimization \cite{guimera2002optimal,barthelemy2006optimal,danila2006optimal,youn2008price,ramasco2010optimization,arenas2010optimal,li2010towards,scellato2010traffic}.

Although a majority of traffic models are link-oriented \cite{olak2016,chodrow2016demand,w2016multi,Ganin_2017,wang2018dynamic}, most of the approaches based on critical phenomena and complex networks were done at the node level, as they were originally intended to understand the movement of packets in communication networks; later, they were extended to transportation systems \cite{echenique2005dynamics,de2009minimal,kim2009jamming,Manfredi_2018,sole2016model,echague2018effective,bassolas2020scaling}. Nodes, or junctions, are responsible for processing the vehicles and are the basic units that can become congested. However, in the case of road transportation, it only provides a partial picture, since not all the segments that arrive at a junction might get necessarily congested at the same time. In fact, most of the sophisticated models aiming to fit on-road congestion are done at the link level \cite{Wang2012,Li2014,olak2016,Ganin_2017,Zeng2018,Hamedmoghadam_2021} and empirical evidence of a jamming transition at the level of link has been recently found \cite{taillanter2021empirical}. Moreover, the congestion data provided by common-day applications such as Google Maps \cite{gmaps} or Uber data \cite{uber} usually appears at the link level. Besides traffic dynamics, a link approach has allowed the implementation of more efficient policies of epidemic containment \cite{Matamalas2018}.

In this work, we develop an extension of the microscopic congestion model (MCM) developed in \cite{sole2016model} where the links are now responsible for processing the vehicles and are the main entities that suffer congestion. Together with the model, we derive a set of transport balance equations that provide analytical predictions for the local and global levels of congestion. We first validate our analytical approach in spatial synthetic graphs built through two models that resemble inter and intra-city transportation infrastructures, showing how the link model gives rise to some unique features compared to the node one. In the second part, we investigate how our model can provide us with useful insights into the dynamics of urban congestion.

\begin{figure}[!tb]
  \begin{center}
  \includegraphics[width=3in]{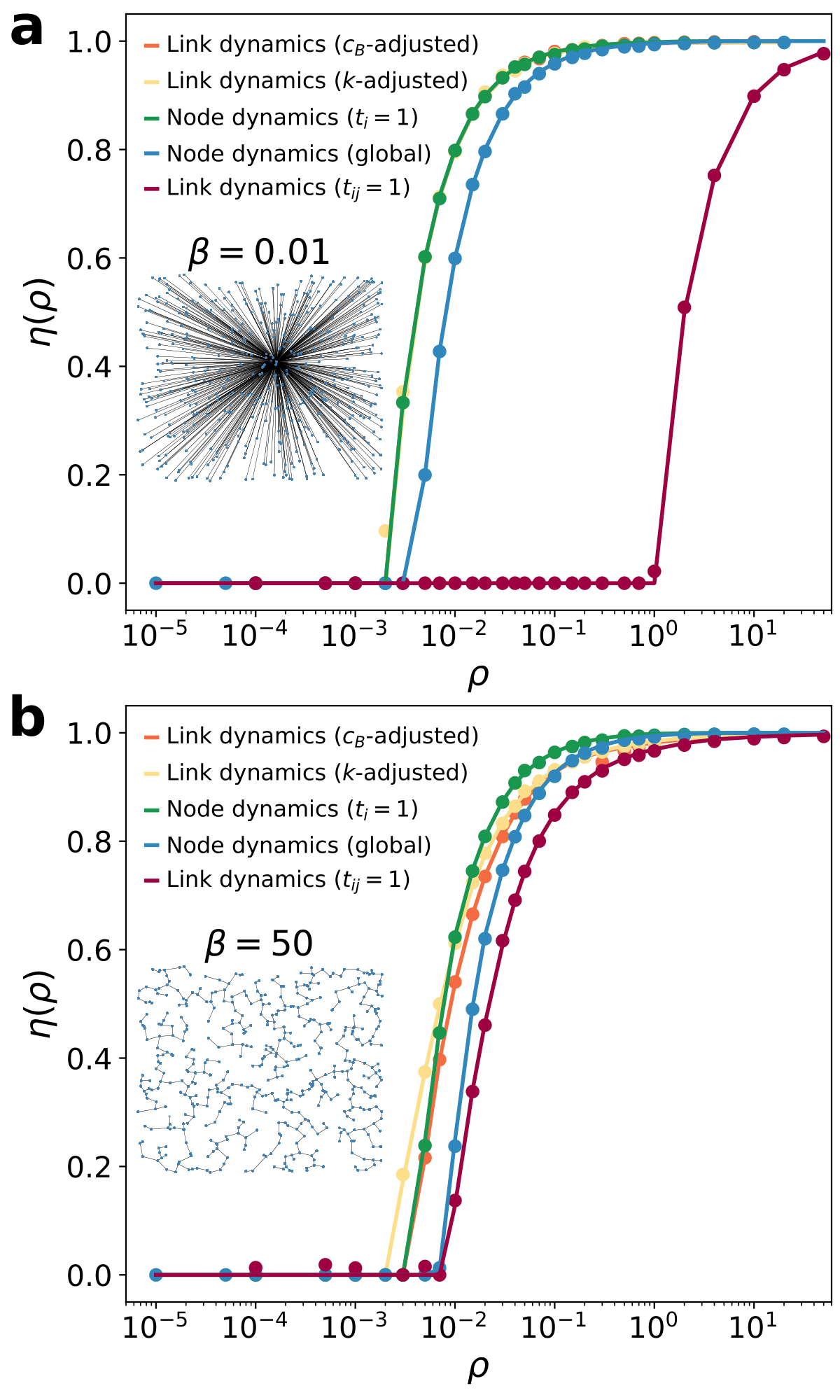}
  \end{center}
  \caption{\textbf{Congestion transition in synthetic graphs for the link and node models.} Evolution of the order parameter $\eta$ as a function of the injection rate $\rho$ for the node model with $\tau_i=1$ (green) and global adjusted capacity (blue), the link model with $\tau_{ij}=1$ (red), the link model with $k$-adjusted capacity $\widetilde{\tau}^k_{ij}$ (yellow) and the link model with $c_B$-adjusted capacity $\widetilde{\tau}^{c_B}_{ij}$ (orange) in two networks with (\textbf{a}) $\beta=0.01$ and (\textbf{b}) $\beta=50$. As an inset we depict the corresponding graphs. Markers correspond to the simulations and lines indicate the analytical solution of the balance equations. With the exception of the link model without normalization, the rest of them are equivalent in terms of total capacity.} \label{fig1}
\end{figure}

\section{Results}

\subsection{Link microscopic congestion model}

We develop here a model to mimic the on-road urban mobility based on \cite{sole2016model} and the critical phenomena on complex networks, but where the links are now responsible for delivering the vehicles instead of the nodes. The framework allows for a more realistic depiction of urban mobility since congestion usually occurs in the road segments ---or links--- rather than in the intersections. The node model assumes that all the road segments that arrive at an intersection get congested at the same time once the flow of vehicles is higher than its capacity, a property that might not necessarily be observed in real-world scenarios. 

In our model, vehicles are generated in a node or junction $i$ at a rate of $\rho_i$ per time step, with a destination that can be drawn from any type of probability distribution, and they move using the road segments or links following a routing algorithm. In this section, we assume that origins and destinations are homogeneously distributed among the nodes, with an equal injection rate $\rho$ ($\rho_i=\rho,\forall i$), and routes follow the shortest path to facilitate the calculations, yet it could be extended to any other type of routing and origin-destination matrices. Each of the links in the system connecting a node $i$ with a node $j$ has a fixed capacity  $\tau_{ij}$ ---the number of vehicles they can process per time step---, and their congestion level at time $t$ is determined by the balance equation
\begin{equation}
\Delta q_{ij}(t)=g_{ij}(t)+\sigma_{ij}(t)-d_{ij}(t),
\end{equation}
where $g_{ij}(t)$ is the number of vehicles generated at $i$ entering the road segment $ij$, $\sigma_{ij}(t)$ is the number of vehicles entering link $ij$ from the adjacent links, and $d_{ij}$ corresponds to the vehicles processed by the road segment. Given that the maximum value $d_{ij}$ can attain is bounded by the road capacity $\tau_{ij}$, as long as $\Delta q_{ij}(t)$ is equal to zero ($g_{ij}(t)+\sigma_{ij}(t)<\tau_{ij}$) the road segment is not congested, while congestion will grow if $g_{ij}(t)+\sigma_{ij}(t)>\tau_{ij}$.

In detail, $g_{ij}(t)$ is calculated by multiplying the injection rate at node $i$, $\rho_{i}(t)$, and the probability that a vehicle generated in $i$ goes through the segment from $i$ to $j$, $p^{\rm origin}_{ij}$,
\begin{equation}
g_{ij}(t)=\rho_{i}(t)p^{\rm origin}_{ij},
\end{equation}
where $p^{\rm origin}_{ij}$ is given by the total number of paths that start at $i$ and go through $ij$ divided by the total number of paths starting at node $i$.
The quantity $\sigma_{ij}(t)$ can be obtained by solving the set of coupled link flow equations
\begin{equation}
\sigma_{ij}(t)=\sum_{k=1}^N P_{kij}(t)p_{ki}(t)d_{ki}(t),
\end{equation}
where $N$ is the number of nodes, $P_{kij}$ is the probability that a vehicle traversing the link from $k$ to $i$ traverses then the link from $i$ to $j$, $p_{ki}$ is the probability of traversing the link from $k$ to $i$ but not finishing in $i$, and $d_{ki}$ accounts for the total number of vehicles traversing the link going from $k$ to $i$. The quantities $P_{kij}$ and $p_{ki}$ are general and can be adapted to any type of routing algorithm yet we focus here on the shortest path that minimizes the sum of weights. Although the shortest path approach might seem rather simplistic, the betweenness centrality has been repeatedly related to the patterns of road usage and congestion \cite{Wang2012,de2014navigability,Kirkley2018,barthelemy2018morphogenesis,lampo2021multiple,lampo2021emergence}. Further details on how to solve the coupled system of equations can be found in the Methods section.

\subsection{The link model in synthetic networks}

To determine the global level of congestion we use the long-established order parameter  \cite{echenique2005dynamics,de2009congestion,sole2016model}
\begin{equation}
\eta(\rho)=\lim\limits_{t \rightarrow \infty}\frac{\avg{\Delta Q}}{N\rho},
\end{equation}
where

\begin{equation}
\avg{\Delta Q}_t=\sum_{i,j=1}^N a_{ij}\Delta q_{ij}
\end{equation}
is the temporal average of the increment of vehicles trapped in the system, $a_{ij}$ are the elements of the adjacency matrix of the network (1 if road segment $ij$ exists, 0 otherwise), and $\rho$ is the global injection rate. When no vehicles are trapped in the system, $\avg{\Delta Q}\sim 0$, there is no congestion and thus $\eta(\rho)=0$. In the extreme case, when most of the vehicles entering the system get trapped, $\avg{\Delta Q}\sim N\rho$ and, thus, $\eta(\rho)\sim 1$. For the sake of simplicity, and without loss of generality, we have assumed an homogeneous generation of vehicles at each node ($\rho_i=\rho$) and distribution of destinations.

\begin{figure}[!tb]
  \begin{center}
  \includegraphics[width=\columnwidth]{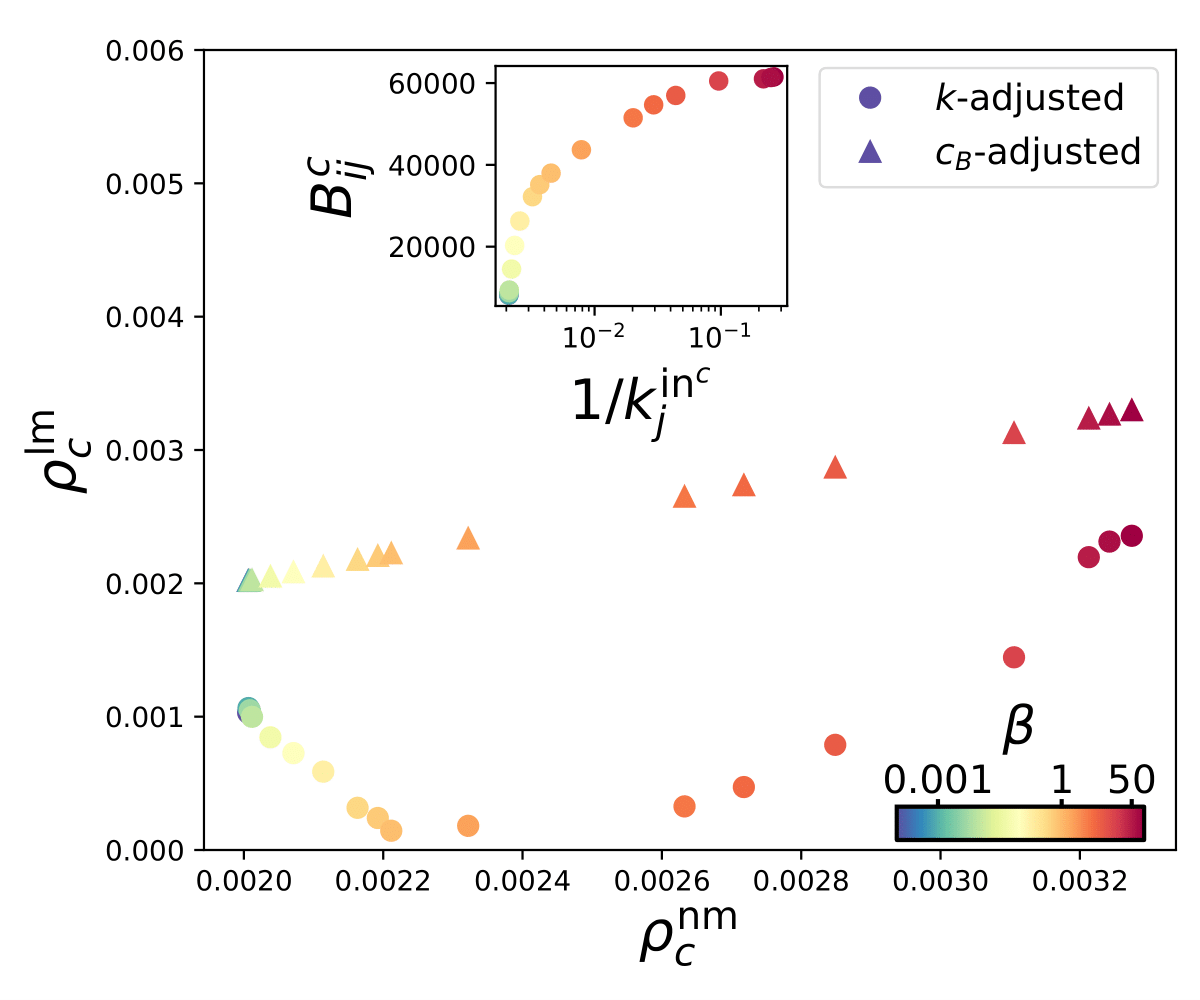}
  \end{center}
  \caption{\textbf{Critical generation rates for the node model with $\tau_{i}=1$ and the link model with degree and betweenness centrality normalization.} Critical generation rate for the link model $\rho^{lm}_c$ with degree (dots) and betweenness centrality (triangles) rescaling as a function of the critical generation rate in the node model with $\tau_{i}=1$. The inset depicts the algorithmic betweenness centrality of the first link to become congested $B^c_{ij}$ as a function of the inverse of its degree $1/k^{\mathrm{in}^c}_j$.} \label{fig2}
\end{figure}

\begin{figure}[!tb]
  \begin{center}
  \includegraphics[width=\columnwidth]{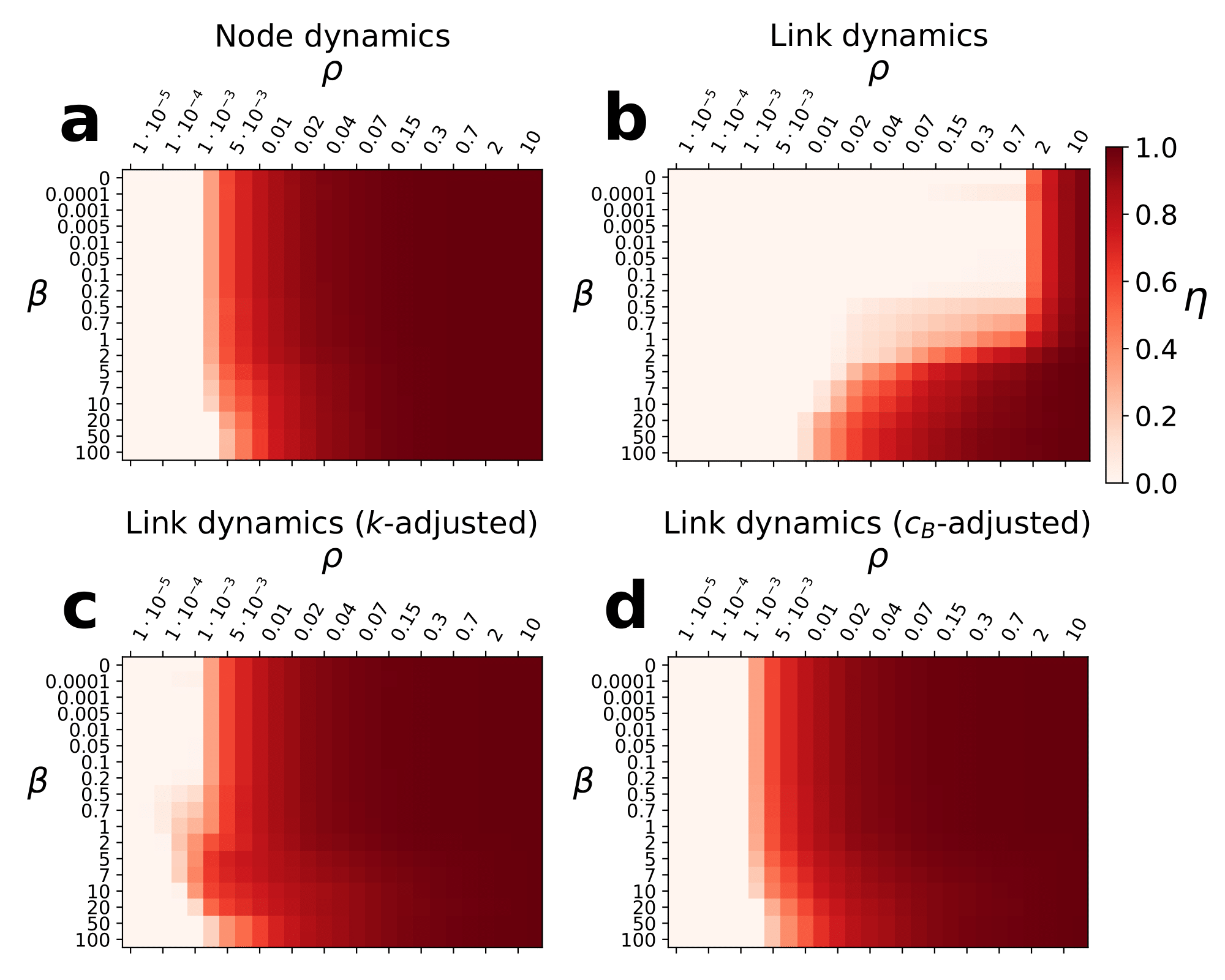}
  \end{center}
  \caption{\textbf{Evolution of the order parameter $\eta$ as a function of the injection rate $\rho$ for each of the models and several values of $\beta$.} Evolution of $\eta(\rho)$ for (\textbf{a}) the node model, (\textbf{b}) the link model without capacity normalization ($\tau_{ij}=\tau_{i}$), (\textbf{c}) the link model with $k$-adjusted normalization ($\widetilde{\tau}_{ij}^k$), and (\textbf{d}) the link model with $c_B$-adjusted normalization ($\widetilde{\tau}_{ij}^{c_B}$). The parameter $\beta$ determines the underlying structure of the spatial graph, low values produce a highly hierarchical layout in front of high values that produce a layout closer to Random Geometric Graphs (RGG) (see inset of Fig.~\ref{fig1}). Results are averaged over 50 network realizations.} \label{fig3}
\end{figure}

\begin{figure}[!tb]
  \begin{center}
  \includegraphics[width=\columnwidth]{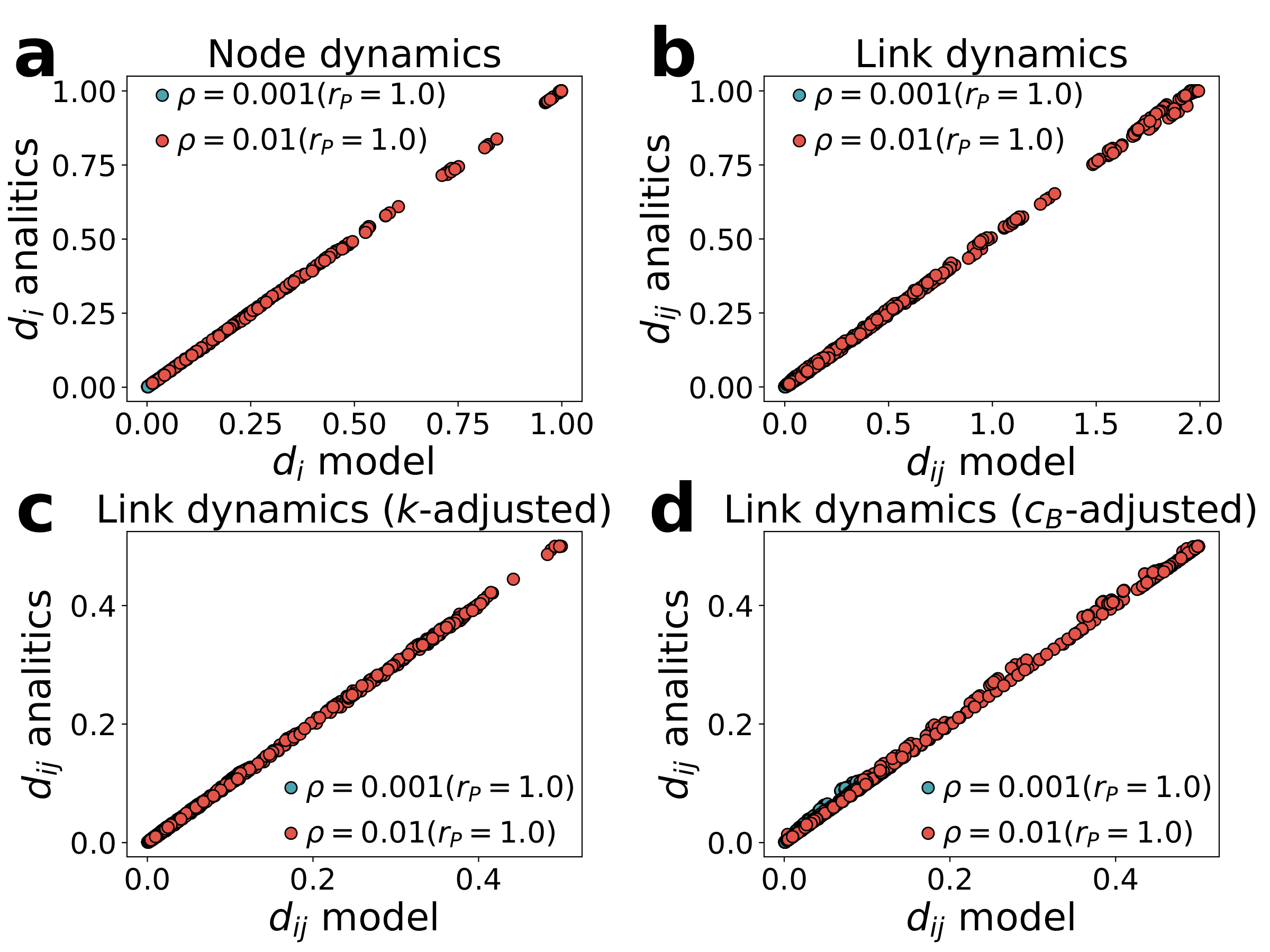}
  \end{center}
    \caption{\textbf{Comparison between number the vehicles traversing each road in the model and the analytical prediction for each model flavour for a network with $\beta=50$}. (\textbf{a}) Node model with $\tau=1$, (\textbf{b}) link model with $\tau=1$, (\textbf{c}) link model with degree adjusted capacities, and (\textbf{d}) link model with betweenness-adjusted capacity.}
 \label{fig4}
\end{figure}

\begin{figure*}[!tb]
  \begin{center}
  \includegraphics[width=\textwidth]{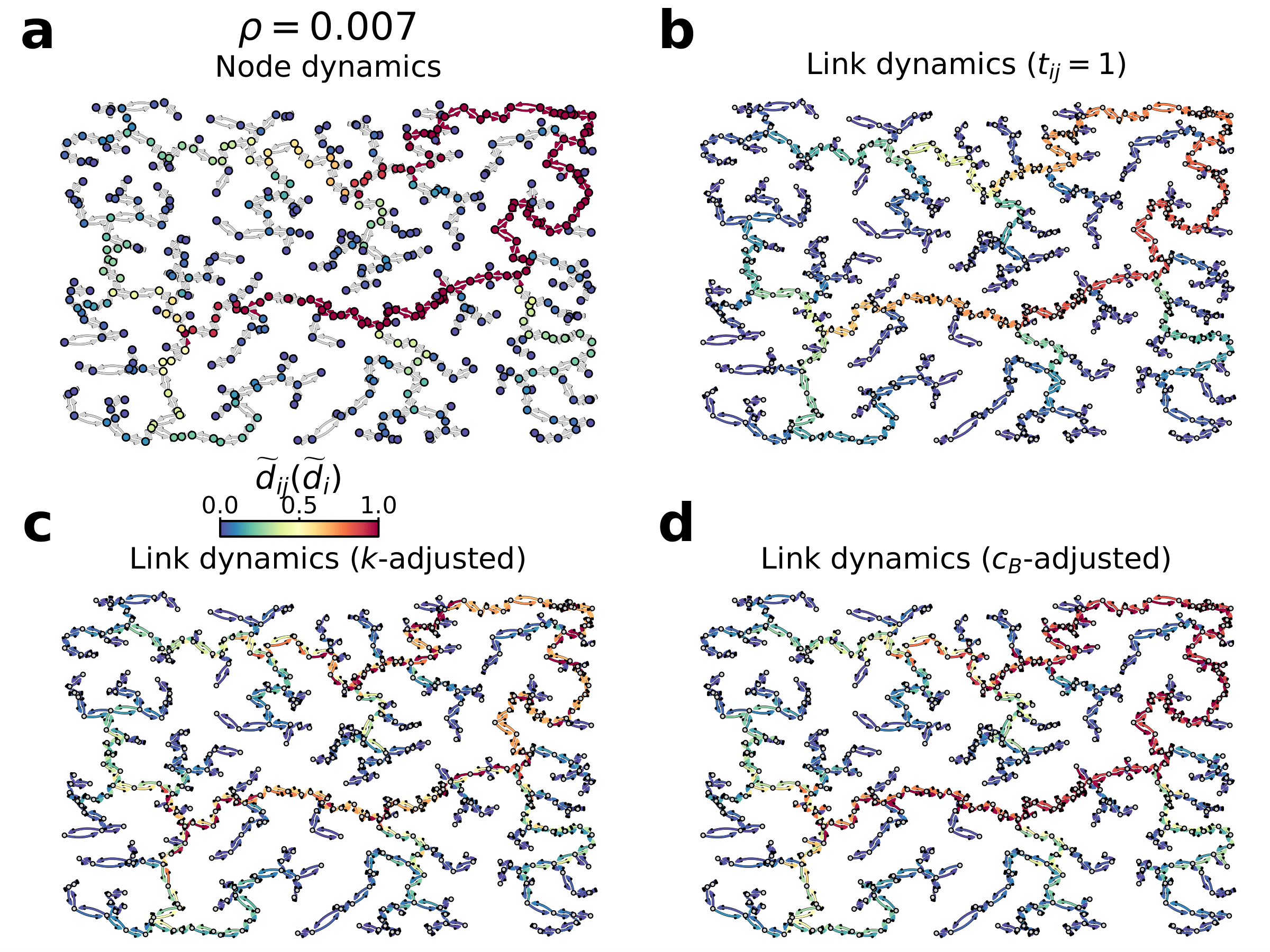}
  \end{center}
  \caption{\textbf{Congestion maps produced by the link and node models in cost-driven spatial networks.} Average number of vehicles traversing each segment (junction)  $d_{ij}$ ($d_{i}$) obtained from the node and link models for $\rho=0.007$ in a spatial graph generated with $\beta=50$. (\textbf{a}) Node model with $\tau_i=1$, (\textbf{b}) link model with $\tau_{ij}=1$, (\textbf{c}) link model with degree adjusted capacities ($\widetilde{\tau}_{ij}^k=1/k_j$) and (\textbf{d}) link model with betweenness-adjusted capacity ($\widetilde{\tau}_{ij}^{c_B}=\frac{\sigma_{ij}}{\sum_{j'}\sigma_{ij'}}$). In the node model we have highlighted in red the links arriving to a congested node.}
 \label{fig5}
\end{figure*}

We start by validating our framework in synthetic spatial graphs obtained from a cost-driven growth model designed to generate road networks depending on the trade-off between cost and efficiency \cite{louf2013emergence}. The model aims to reproduce the interurban transportation infrastructures and has a parameter $\beta$ that tunes the relevance of the cost: when $\beta$ is low, the cost term is negligible and the links connect peripheral nodes to the most influential hubs. Conversely, as $\beta$ increases the cost term becomes important producing shorter links and patterns compatible with random geometric graphs \cite{penrose2003random}.

\begin{figure}[!tb]
  \begin{center}
  \includegraphics[width=\columnwidth]{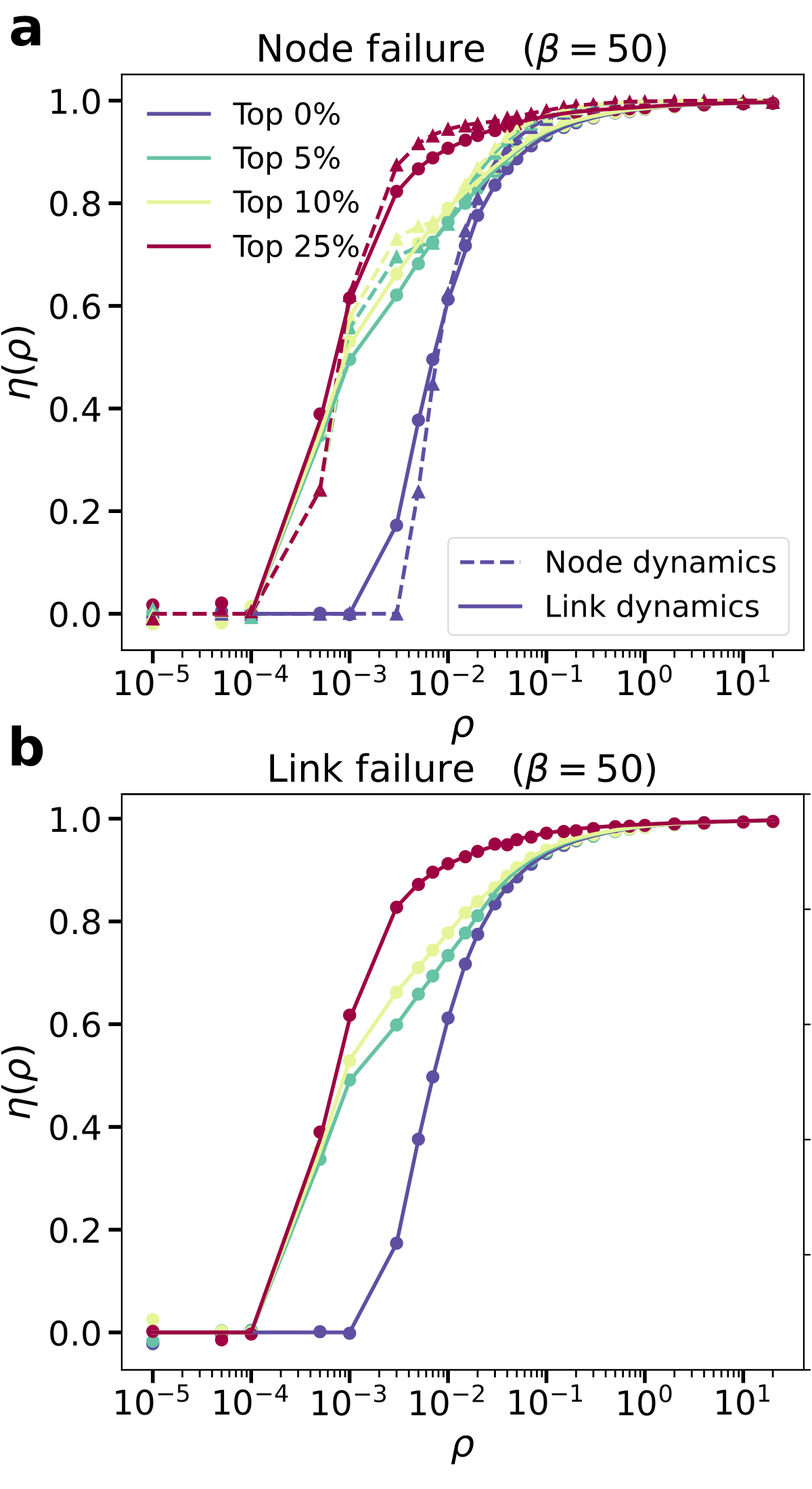}
  \end{center}
  \caption{\textbf{Robustness of synthetic graphs under failures.} Effect of reducing the capacity of (\textbf{a}) junctions and (\textbf{b}) segments. In \textbf{a} we display the updated congestion level $\eta$ after the capacity of the  junctions with top $10\%$ and top $25\%$ is reduced to one tenth of the original value. Dashed lines and triangles correspond to the node model and regular lines and squares to the link model with $k$-adjusted capacity. To simulate the failure of a junction in the link model, all the links arriving to a junction see their capacity reduced to one tenth. In (\textbf{b}) we display the updated congestion level $\eta$ after the capacity of the  junctions with top $10\%$ and top $25\%$ is reduced to one tenth of the original value} \label{fig6}
\end{figure}

To properly compare the node model developed in \cite{sole2016model} with our link model, either link or node capacities need to be rescaled; otherwise, by setting $\tau_{ij}=\tau_{j}$ for all links, each node would have a capacity equivalent to its in-degree. Although multiple capacity normalizations could be implemented, we focus here on either a global rescaling of the node model equivalent to the link model with $\tau_{ij}=\tau$, or a local rescaling of the link capacity equivalent to the node model with $\tau_{i}=\tau$. 

To achieve a node model equivalent to the link model with $\tau_{ij}=\tau$, we set the capacity of each node equal to $\tau_i=\tau E/N$ where $E$ is the total number of edges and $N$ the total number of nodes in the graph. 

To rescale the link model so that the capacity is equivalent to the node model with $\tau_{i}=\tau$, we have implemented both a degree and a betweenness centrality normalization. In the degree normalization, each road segment~$ij$ has a capacity $1/k^{\mathrm{in}}_j$, where $k^{\mathrm{in}}_j$ is the in-degree of node~$j$, while in the betweenness centrality one, the capacity of each junction is distributed through the incoming segments proportionally to their betweenness centrality while preserving  $\sum_i \tau_{ij}=\tau_{j}$ (see Methods). The degree rescaling can be understood as the implementation of a traffic light that allows an intermittent flow of vehicles between the incoming links. In the case of the betweenness centrality, the behavior would be more similar to that of a roundabout in which the capacity across the links is adjusted to the number of vehicles passing through them.

In Fig.~\ref{fig1}, we compare the standard node and link models ---$\tau_{i}=1$ and $\tau_{ij}=1$---, together with the node model with global rescaling and the two versions of the link model with adjusted capacities, in extreme network topologies with $\beta=0.01$ and $\beta=50$ (see inset). We observe an overall good agreement between the analytical solution and the Monte Carlo simulations. Comparing first the link model with $\tau_{ij}=1$ (red) and the node model with global rescaling (blue), the first features lower congestion despite they are comparable in terms of total capacity, likely a consequence of their different capacities at the local level. Whereas all junctions have the same capacity in the node model with global rescaling, in the link model the effective capacity of junctions ---obtained by summing the capacity of incoming links--- is equal to their in-degree. Such difference, combined with the fact that the betweenness centrality of nodes with high degree tends to be higher, makes the link model more efficient in managing traffic. The differences are particularly high for $\beta=0.01$, where the networks resemble a star graph and the node of highest degree accumulates most of the paths. Our results evince that when capacities are equivalent at the system level but not at the local one, the link model features a more efficient processing of vehicles.

If, instead, we consider the normalized versions (orange and yellow) of the link model that have an equivalent total capacity to the node model with $\tau_{ij}=\tau=1$ (green), we get a more nuanced message. The $k$-adjusted model suffers from higher congestion than the node model for low $\rho$, but as the injection rate increases, a crossover appears and the node model appears more congested. In the $k$-adjusted model, the homogeneous distribution of capacities across the links finishing in a junction leads to an early appearance of congestion, compared with the node model where the full junction capacity can be allocated to the segment with the highest vehicle flow. Conversely, large values of $\rho$ induce the congestion of most of the junctions in the node model affecting almost all road segments, while in the link model those links with lower flows can still operate regardless of the congestion of other segments.

The capacity adjustment by betweenness centrality compensates for the initial disadvantage of the $k$-adjusted model for low injection rates, producing similar values of $\eta$ to the node model at low injection rates but outperforming it as $\rho$ increases. The transition to the congested phase happens at the same value of the injection rate $\rho_c$ for the node and betweenness centrality adjusted models. 
By comparing the transitions for both networks, we observe little to no difference between the link model with degree normalization and the node model for $\beta=50$ but a more significant difference for $\beta=0.01$.

We can gain further insights by comparing the critical generation rates for the node model \cite{sole2016model}
\begin{equation}
\rho^{\mathrm{nm}}_c=\min_i\frac{\tau(N-1)}{B_i+2(N-1)},
\end{equation}
with the one for the link model
\begin{equation}
\rho_c^{\mathrm{lm}}=\min_{i,j}\frac{\tau_{ij}(N-1)}{B_{ij}},
\end{equation}
which in the degree normalization becomes
\begin{equation}
\rho^{\mathrm{lm-k}}_c=\min_{i,j}\frac{\tau(N-1)}{k^{\mathrm{in}}_jB_{ij}}.
\end{equation}

We compare both critical generation rates in Fig.~\ref{fig2}, where $\rho^{\mathrm{nm}}_c$ progressively increases with $\beta$ and $\rho^{\mathrm{lm-k}}_c$ features a U-shape. It decreases with $\beta$ until it reaches a minimum around $1$, and increases thereafter. To explain such behavior, the inset displays the algorithmic betweenness centrality of the first link that gets congested as a function of the inverse of its degree. For low values of $\beta$ the first link to become congested has a low capacity but also a low betweenness centrality. As $\beta$ increases, the spatial graphs display a combination of long and short-range links producing a steeper increase of the betweenness centrality compared to $1/k^{\mathrm{in}^c}_j$. As long-range links disappear and the networks resemble random geometric graphs, the degree of nodes decreases inducing an increase of $\rho^{\mathrm{lm-k}}_c$. For completeness, we provide the critical generation rate in the link model with betweenness centrality normalization, which as expected, is equal to that of the node model with $\tau_i=\tau=1$.

To investigate in detail the interplay between the network topology and each of the models, we display in Fig.~\ref{fig3} their whole congestion transition $\eta(\rho)$ as a function of $\beta$. The link model without normalized capacities shows a qualitative and quantitative different behavior, performing better for networks with low $\beta$, as opposed to the rest of the models that display less congestion for high $\beta$. In hierarchical graphs, most of the flows traverse the high degree junctions, whose capacity is heavily affected by the renormalization. In fact, in the non-normalized link model, the total junction capacity is equivalent to its incoming degree, improving the performance of networks where most of the flows traverse high-degree nodes.
The comparison between the link models with adjusted capacity and the node model for other values of $\beta$ are similar to the results for $\beta=50$ (Fig.~\ref{fig1}), with the $k$-adjusted model displaying a transition before the node model but with a crossover for higher values of $\rho$. In a similar line, we have analyzed in Supplementary Note~1 the DT+MST model developed in \cite{lampo2021multiple,lampo2021emergence} that mimics the intraurban road structure, observing a similar global behavior.

\begin{figure*}[!tb]
  \begin{center}
  \includegraphics[width=6in]{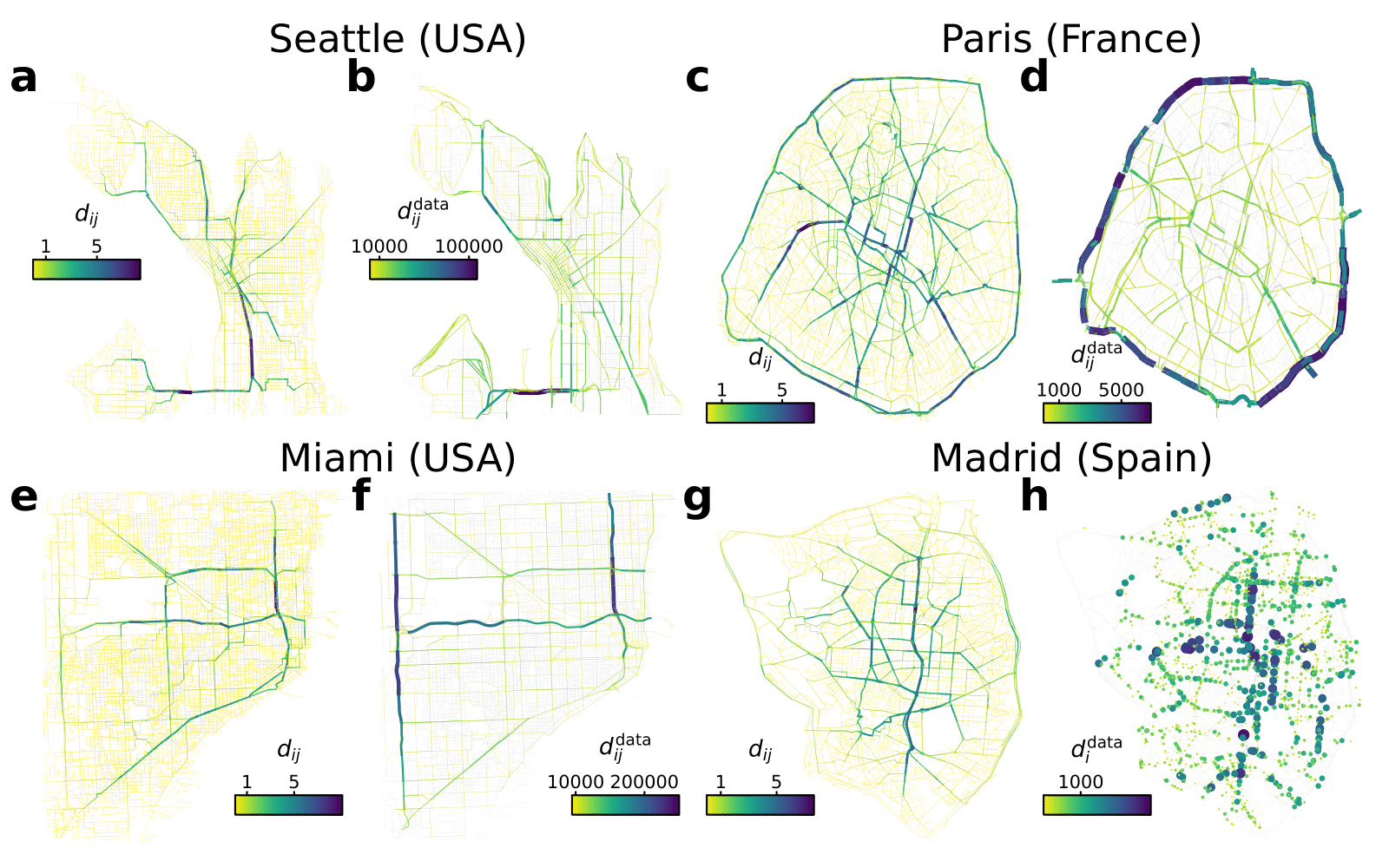}
  \end{center}
  \caption{\textbf{Analysis of congestion hotspots in real cities.} Comparison between the congestion obtained for our model and real traffic counts in (\textbf{a}--\textbf{b}) Seattle (USA) with $\rho=0.31$, (\textbf{c}--\textbf{d}) Paris (France) with $\rho=0.39$, (\textbf{e}--\textbf{f}) Miami (USA) with $\rho=0.31$ and (\textbf{g}--\textbf{h}) Madrid (Spain) with $\rho=0.23$. In (\textbf{a},\textbf{c},\textbf{e},\textbf{g}) we display the total number of vehicles traversing link $d_{ij}$ and in (\textbf{b},\textbf{d},\textbf{f},\textbf{h}) the observed traffic counts $d^{\rm data}_{ij}$ ($d^{\rm data}_i$) either at the level of link as in Seattle, Paris and Miami or at a concrete counter as in the case of Madrid. See Methods section for details on how the traffic counts were computed and Supplementary Figures S2-S13 for the maps in the rest of the cities.} \label{fig7}
\end{figure*}

Our analytical approach provides us with a prediction for the number of vehicles traversing each of the road segments. In Fig.~\ref{fig4} we display the comparison between $d_{ij}$ in the simulations and the analytical solution for each of the link models (\textbf{b}--\textbf{d}) with varying capacities as well as the results for the node model (\textbf{a}). Two injection rates have been tested, one in the free-flow regime ($\rho=0.001$) and another in the congested phase ($\rho=0.01$). As the Pearson correlation coefficient indicates, the agreement between them is high for several values of the injection rate. It is worth noting that the maximum value $d_{ij}$ can attain in the adjusted versions is not one due to the renormalization of capacities.

We further inspect the differences between the node model and the three flavors of the link model proposed in Fig.~\ref{fig5}, where we show the level of congestion in a $\beta=50$ network for $\rho=0.007$. To ease the comparison between them, we show the normalized flow of vehicles $\widetilde{d}_{ij}$ ($\widetilde{d}_{i}$) given by the ratio $d_{ij}/\tau_{ij}$ ($d_{i}/\tau_{i}$). In (\textbf{a}) and (\textbf{b}) we show the standard versions where the capacity of nodes and links is equal, $\tau_{ij}=\tau_{i}=1$. Given the higher total capacity of the latter, for $\rho=0.007$ we only observe two links close to the congested state, in contrast to the node model where several congested junctions appear. Instead, the link model with degree-adjusted capacity (\textbf{c}) has a similar congestion on account of their equivalent capacity, with congested links in many cases connected to the corresponding most congested junctions. Still, there are many low-congested links connected to nodes that appear heavily congested in the node model. The fact that the congestion of a node affects all the incoming links allows the link model with degree normalization to outperform it for high congestion levels. Finally, in the model normalized by betweenness (\textbf{d}), we observe that the congestion level of the links sharing destination junction is much more similar than in the other approaches since capacity has been adjusted to the flow of vehicles. However, that only happens before the congestion onset since the block of links affects the effective flow of vehicles.

\subsection{Robustness of synthetic networks}

Accidents, storms, and other types of rare events can affect the normal functioning of transportation infrastructures, reducing the effective capacity of road segments or junctions, depending on the case. We test next if our analytical approach can predict the aftermath of such failures by using the node model and the link model where capacities are adjusted by the degree. In Fig.~\ref{fig6}(a) we evaluate the increase of congestion that appears as a consequence of a $90\%$ capacity reduction of up to $25\%$ of the nodes with the highest betweenness centrality. Therein we confirm the validity of our analytical framework despite the more complex behavior and provide further proof that the dynamics of both models are significantly different. As before, the link model adjusted by degree exhibits more congestion for low injection rates but outperforms the node model as $\rho$ increases. From an infrastructure planning perspective, authorities should be aware that a reduction in only $5\%$ of links significantly increases the congestion of the system.

Unlike to the node model, our framework now allows us to assess not only the failure of junctions but also that of links which are also common in real-world systems. In Fig.~\ref{fig6}(b) we report how the failure of the links with higher betweenness centrality impacts the congestion, with a strong increase even for a failure of a scarce $5\%$ of the links. Interestingly, as the reduction of capacity is equivalent in both cases, there are no significant differences between the failure of a given link or a junction.

\subsection{Application to real-world scenarios}

\begin{figure}[!tb]
  \begin{center}
  \includegraphics[width=3in]{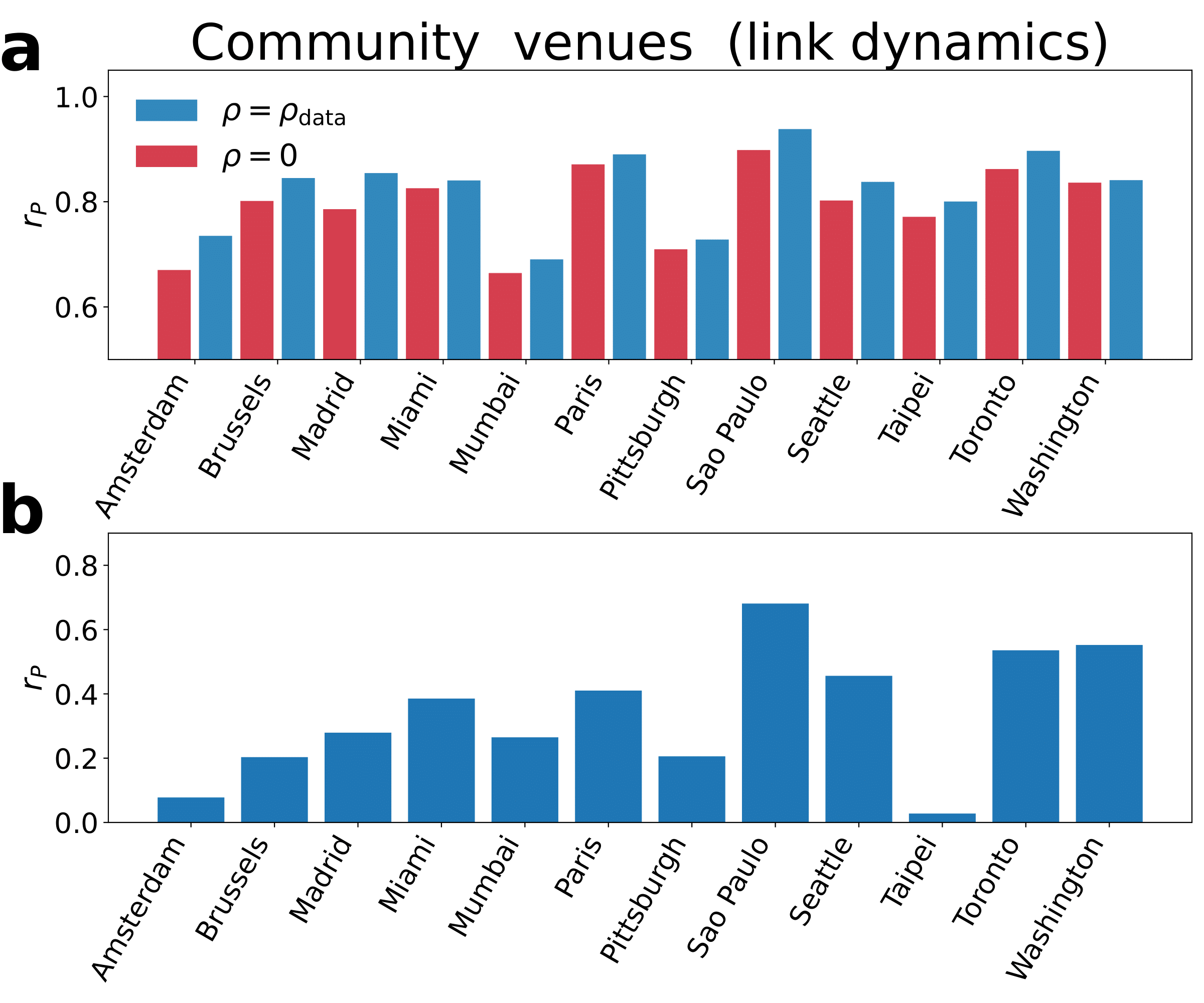}
  \end{center}
  \caption{\textbf{Correlation between the real and modelled delays in the link dynamics when destinations are distributed according to the community POIs.} (\textbf{a}) Comparison between the Pearson correlation coefficient obtained between the travel times from Uber Data \cite{uber} during the morning peak ($8-10$am) in a set of cities and the travel times obtained for $\rho=0$ (red) and $\rho=\rho_{\rm data}$ (blue) as detailed in Eq.~\ref{traveltimes}. (\textbf{b}) Pearson correlation coefficient between the delay observed in the data and the model. All the cities except Taipei display a high level of significance (p-value$<0.001$). The injection rate for each city $\rho_{\rm data}$ is set to match $\eta$ with the percentage of delay observed in the Tom Tom traffic index data \cite{tomtom}. Further comparisons and congestion maps can be found in Supplementary Notes 2 and 3. It is important to note that in Taipei and Mumbai most of the roads do not have precise information on maximum speeds.} \label{fig8}
\end{figure}

We assess next if our framework can effectively provide insights into the congestion level of 12 cities, namely Amsterdam, Brussels, Madrid, Miami, Mumbai, Paris, Pittsburgh, Sao Paulo, Seattle, Taipei, Toronto, and Washington. To better mimic the real congestion dynamics, we have distributed the destinations according to the points of interest (POI) extracted from the location-based social network Gowalla \cite{gowalla}, and set an injection rate $\rho_{\rm data}$ that induces a congestion equivalent to the percentage of delay provided by the TomTom index by 2019 \cite{tomtom}. In other words, if the congestion level in Madrid was of $23\%$ by 2019, we set the $\rho_{\rm data}$ that satisfies $\eta(\rho_{\rm data})=0.23$. To do so, we increase progressively $\rho$ until we match the corresponding value of $\eta$. Since our model does not include any parameter besides the choice of the routing algorithm, this is the only calibration in place. Our approach allows for more realistic OD patterns compared to the random case, althought more sophisticated methods exist using mobile phone data or location-based social networks \cite{yang2015origin,dong2018empirical,cui2018forecasting,huang2018modeling,bachir2019inferring}.
Further details on the modeling of congestion in cities can be found in the Methods section. In Fig.~\ref{fig7} we show the average value of vehicles traversing each segment $d_{ij}$ obtained from our model in Seattle, Miami, Paris and Madrid, together with the observed traffic counts $d^{\rm data}_{ij}$. Depending on the analyzed city, traffic counts are given either on an average yearly volume or an hourly based.
As can be observed, the flows obtained from our model are compatible with the real traffic counts in all four cities, with most of the vehicles going through the main arterial roads. From a policy-oriented perspective, either the congestion hotspots highlighted by our model would need capacity reinforcement or the spatial distribution of destinations could be modified.
In Madrid and Paris, the flow of vehicles through the main ring roads is significantly lower than in the real data since a vast majority of that traffic comes from trips originating outside the city. We provide in Table~S1 of the Supplementary Material the Pearson correlation between the real and the observed traffic counts in Madrid for each type of venues.

Beyond the correct evaluation of the spatial patterns of congestion, we also probe if the link model provides useful insights into the delay it produces. More concretely, we aim to match the travel times and delays observed in our model with those provided by the Uber movement data \cite{uber}. The two main quantities we analyze are thus the travel time $T_{ij}$ between cells $i$ and $j$ in a city, and the delay $\lambda_{ij}$ measured as the ratio between the travel times under the congested phase and free-flow conditions. Each of the quantities has been computed in the Uber data during the morning peak ($T^{\rm data}_{ij}$ and $\lambda^{\rm data}_{ij}$) and in our model ($T^{\rm model}_{ij}$ and $\lambda^{\rm model}_{ij}$) as detailed in the Methods section.

In Fig.~\ref{fig8}(a) we display in blue, for the set of 12~cities, the Pearson correlation coefficient $r_P$ between the travel times in the link model under the congested phase, and those observed in the data during the morning peak when destinations are distributed according to the community POIs. For comparison, we display in red the correlation with the shortest paths without congestion. Comparing both, the correlation increases when we take into account the information provided by the congestion as compared to the travel times in free-flow. When including the congestion, the correlation is greater than $0.6$ for all cities and, in most of cases, greater than $0.8$. If instead of the travel times we focus on the delay $\lambda_{ij}$, there is also a significant positive correlation in all the cities studied except for Taipei, Fig.~\ref{fig8}(b). The Pearson correlation coefficient is especially high for Sao Paulo, Seattle, Toronto, and Washington. We show further results for the link model in the case of Entertainment, Food, and Shopping venues in Supplementary Figs.~S15, S17, and~S19. As shown in Supplementary Figs.~S14, S16, S18, and~S20, a similar analysis conducted with the node model provides a lower increase in the correlations with the travel times, and suffers from a decrease in the correlations regarding the delay itself. We have conducted additional analysis of the residuals for the regression analysis in Supplementary Figs.~S21--S32. It is important to note that, unlike other models, our framework does not have any parameters and focuses only on the shortest paths. By having more detailed information on the real routes that vehicles follow or the road capacities we might be able to provide a better fit with the real delays \cite{de2015personalized}. The amount of detailed information regarding the maximum speeds and the number of lanes is still scarce, limiting the predictive power of our framework.

\section{Discussion}

While there is a majority of works that model the congestion phenomena in complex networks as pure node dynamics, it appears to be a rough approximation as it assumes that the links arriving at a junction get congested at the same time. In certain contexts such as transportation networks, a road segment can still operate regardless of the congestion status of the links sharing destination junction. Thus, to provide a more nuanced picture of congestion phenomena, we have extended the MCM model developed in \cite{sole2016model} at the level of links. Our framework, that can be solved analytically, naturally brings together the two-folded dynamics with links being responsible for processing the vehicles but a capacity limited by the junctions, allowing non-congested road segments to operate regardless of the congestion status of the other links sharing a terminal. The different distribution of junction capacities across the incoming links gives rise to new and interesting rich behaviors in graphs, with certain topologies displaying better management of congestion than others. By incorporating a limited buffer to the links \cite{Manfredi_2018}, our model could provide further insights on the spreading of congestion \cite{saberi2020simple}, allowing an analytical derivation of the most sensible links leading to the global congestion of cities. Moreover, it could provide a better approximation to simulate the delay of vehicles, as the time needed to traverse each link could be adjusted by the density of vehicles traversing each link, $d_{ij}/\tau_{ij}$.

In real scenarios, our methodology yields spatial patterns compatible with the real flow of vehicles and it provides a better assessment of the delay product of congestion. From the perspective of potential applications, and in contrast to the node approach, the link model allows for the incorporation of harmful events of different kinds that reduce the capacity not only of junctions but also of road segments. Our model allows for a better management of transportation infrastructures, either by optimizing the available capacity, allowing a preferential pass of vehicles when needed, or by modifying the patterns of origins and destinations so that the strength of congestion hotspots is mitigated \cite{newman2006environmental,glaeser2010greenness,verbavatz2019critical}. Notwithstanding more sophisticated approaches such as the implementation of efficient pricing schemes or the incorporation of bus lanes that facilitate public transportation \cite{sole2018decongestion,bassolas2019mobile}.

\section{Methods}

\subsection{Analytical derivation of congestion}

The set of balance equations presented in this work can be solved analytically in an iterative form. We let node $i$ denote a junction, $a_{ij}$ the adjacency matrix component for the connection between nodes $i$ and $j$, $w_{ij}$ the corresponding travel time, $L$ the total number of links in a network, $N$ the total number of nodes, $N^{\rm in}_{ji}$ the number of edges arriving at junction $i$ and $N^{\rm out}_{ji}$ the ones departing from it.
Term by term, we can decompose the flux of vehicles arriving to the segment $ij$ 
\begin{equation}\label{basis}
\sigma_{ij}=\sum_{k}P_{kij}p_{ki}d_{ki},
\end{equation}

We can decompose the probability that a vehicle traversing link $ki$ goes through link $ij$ as
\begin{equation}\label{pkij}
P_{kij}=p^{\rm rgen}_{ki}P^{\rm loc}_{kij}+(1-p^{\rm rgen}_{ki})P^{\rm ext}_{kij}, 
\end{equation}
where the first term corresponds to the vehicles generated in $k$ that go through the links $kij$ and the second term corresponds to the vehicles not generated in $k$ that go through the links $kij$. In detail, the fraction $p^{\rm rgen}_{ki}$ is given by the ratio between the vehicles generated in $k$ that go through link $ki$ and the total number of vehicles going through $ki$
\begin{equation}\label{rgen}
p^{\rm rgen}_{ki}=\frac{g_{ki}}{g_{ki}+\sigma_{ki}p^{\rm ext}_{ki}}.
\end{equation}
$g_{ki}$ is again given by $\rho_{i}p^{\rm origin}_{ij}$ and $p^{\rm ext}_{ki}$ is the fraction of vehicles going through $ki$ that do not finish in $i$ divided by the total number of vehicles that traverse the link
\begin{equation}\label{pext}
p^{\rm ext}_{ki}=\frac{\widetilde{B}_{ki}}{\widetilde{B}_{ki}+\widetilde{e}_{ki}}.
\end{equation}
where $\widetilde{B}_{ki}$ and $\widetilde{e}_{ki}$ are, respectively, the expected number of vehicles traversing link $ki$ but not finishing in $i$ and the expected number of vehicles traversing $ki$ and finishing in $i$. 

In the second term of Eq.~\ref{pkij}, the probability $P^{\rm ext}_{kij}$ is obtained by normalizing the total number of paths not starting in $k$ that go through the combination of junctions $kij$ $\widetilde{E}^{\rm ext}_{kij}$ by the total number of vehicles that traverse junction $ki$ 
\begin{equation}
P^{\rm ext}_{kij}=\frac{\widetilde{E}^{\rm ext}_{kij}}{\sum_{j}\widetilde{E}^{\rm ext}_{kij}}.
\end{equation}

The second element in the multiplication of Eq.~\ref{basis} $p_{ki}$ corresponds to the vehicles traversing edge $a_{ki}$ not finishing in $i$ and can be broke down as
\begin{equation}
p_{ki}=p^{\rm gen}_{ki}p^{\rm loc}_{ki}+(1-p^{\rm gen}_{ki})p^{\rm ext}_{ki},
\end{equation}
where the first term accounts for the vehicles generated in $k$ whose destination is not $i$ and the second term accounts for the vehicles not generated in junction $k$ whose destination is not $i$. More in detail, $p^{\rm gen}_{ki}$ is the fraction of vehicles generated in $k$ traversing link $ki$, $p^{\rm loc}_{ki}$ is the probability that a vehicle generated in $k$ do not end in $i$ and $p^{\rm ext}_{ki}$ is the probability that a vehicle not generated in $k$ traverses $ki$ but do not finish in $i$ (Eq.~\ref{pext}). The probability $p^{\rm gen}_{ki}$ is just the fraction of vehicles generated in $k$ that go through $i$ divided by the total number of vehicles entering link $ki$ and can be written as
\begin{equation}
p^{\rm gen}_{ki}=\frac{g_{ki}}{g_{ki}+\sigma_{ki}},
\end{equation}
where $p^{\rm origin}_{ki}$ is again the probability that a vehicle generated in $k$ goes through link $ki$.   The probability $p^{\rm loc}_{ki}$ depends on the concrete distribution of origins and destinations and is equal to $(N-1)/N$ when vehicle destinations are homogeneously distributed.

Similarly to $P^{\rm ext}_{kij}$, the probability $P^{\rm loc}_{kij}$ is obtained by normalizing the total number of paths starting in $k$ that go through the combination of junctions $kij$ $\widetilde{E}^{\rm loc}_{kij}$ by the total number of vehicles that traverse junction $ki$
\begin{equation}
P^{\rm loc}_{kij}=\frac{\widetilde{E}^{\rm loc}_{kij}}{\sum_{j}\widetilde{E}^{\rm loc}_{kij}}.
\end{equation}
To properly describe the system after the congested phase, if a link $ki$ gets congested, the contributions to the quantities $\widetilde{E}^{\rm loc}$, $\widetilde{E}^{\rm ext}$, $\widetilde{B}$ and $\widetilde{e}$ of the paths from a source/destination pair $(s,t)$ traversing that link need to be rescaled by the quantity $\frac{\tau_{ki}}{g_{ki}+\sigma_{ki}}$.

\subsection{Capacity renormalization}

To allow for a fair comparison between the node and link models, the networks should have the same total capacity. In addition, we also require that the capacity is equivalent at the junction level by satisfying the condition
\begin{equation}
\sum_{i}a_{ij}\tau_{ij}=\tau_{j},
\end{equation}
where $\tau_{ij}$ is the capacity of the road segment connecting junctions $i$ and $j$ in the link model and $\tau_{j}$ is the capacity of junction $j$ in the node model. The need for a renormalization is in agreement with the limited capacity of junctions that is distributed across links with, for example, the use of traffic lights. 

We propose here two different capacity renormalizations based either on the degree of a junction or on the betweenness centrality of road segments. In the former, the capacity of a road segment $a_{ij}$ is normalized by the degree of $j$ so that
\begin{equation}
\widetilde{\tau}^k_{ij}=\tau_{ij}\frac{1}{k^{\mathrm{in}}_j},
\end{equation}
which now yields $\sum_i \widetilde{\tau}_{ij}=\tau_{j}$ when $\tau_{ij}=\tau_j$ and implies that the capacity of a link arriving at a junction $j$ is lower if there are more links arriving at it and capacity is homogeneously distributed across them. However, the capacities could be further optimized using the number of paths that traverse the segment from $i$ to $j$ or the edge betweenness centrality ${c_B}_{ij}$, allowing for a normalization 
\begin{equation}
\widetilde{\tau}^{c_B}_{ij}=\tau_j\frac{c_B(ij)}{\sum_{j'}c_B(ij')},
\end{equation}
which again satisfies the relation $\sum_i a_{ij}\widetilde{\tau}_{ij}=\tau_j$ when $\tau_{ij}=\tau_j$ but now distributes the capacity in agreement with the flows traversing each link in the non-congested phase. This frame would resemble the situation in which the road segments with a higher flow of vehicles are favored by an increased capacity through traffic lights.

\subsection{Construction of real transportation networks}

We have extracted the information on the junctions and road segments from OpenStreetMap \cite{openstreetmap} using the Python package OSMnx \cite{boeing2017osmnx}. The data includes the geographical location of junctions and the edges connecting them, together with some metadata such as the maximum speed. Given the limited availability of data regarding the capacity of each road segment $\tau_{ij}$, we have set it according to its maximum speed in meters per second. For those segments that do not include information on the maximum speed, we have set it equal to $40$km/h.

To simulate real scenarios we have focused only on the node model and the link model with degree adjusted capacity. In the former the capacity of a junction is given by 
\begin{equation}
\tau_{j}=\frac{1}{k^{\mathrm{in}}_j}\sum_{i}a_{ij}\tau_{ij},
\end{equation}

while in the latest it is given by 
\begin{equation}
\widetilde{\tau}^k_{ij}=\tau_{ij}\frac{1}{k^{\mathrm{in}}_j}. 
\end{equation}

The average capacity per road junction $\avg{\tau_{j}}$ observed in the cities studied is close to $15$ as used in \cite{sole2016model}. The details on the networks analyzed and the congestion level $\eta$ \cite{tomtom} used in each of the cities are reported in Table~\ref{citydata}.

\begin{table}[ht!]
\begin{tabular}{lrrrr}
\hline
City & Nodes & Links & $\eta_{\rm data}$ & $\langle \tau_i \rangle$\\  \hline\hline
Amsterdam & 13415 & 30802 & 0.26 & 10.7 \\  \hline
Brussels & 15465 & 33729 & 0.38 & 9.0\\  \hline
Madrid & 7954 & 14929 & 0.23 & 11.6\\  \hline
Miami & 17961 & 48734 & 0.31 & 10.2\\  \hline
Mumbai & 17082 & 41790 & 0.65 & 11.0\\  \hline
Paris & 9630 & 18744 & 0.39 & 8.4\\  \hline
Pittsburgh & 11717 & 30208 & 0.21 & 10.4\\  \hline
Sao Paulo & 11346 & 23440 & 0.45 & 11.3\\  \hline
Seattle & 6935 & 18630 & 0.31 & 6.6\\  \hline
Taipei & 6336 & 14676 & 0.35 & 11.1\\  \hline
Toronto & 14270 & 37332 & 0.33 & 11.2\\  \hline
Washington & 9818 & 26547 & 0.29 & 10.7\\  \hline
\end{tabular}
\caption{Main statistics of the real transportation network analyzed including the number of nodes, the number of links, the injection rate observed in the Tom Tom data and the average junction capacity.} \label{citydata}
\end{table}

\subsection{Distribution of destinations according to Gowalla POIs}

To simulate real scenarios we have implemented a distribution of destinations that is not homogeneous but obeys to the spatial distribution of points of interest (POIs) in the Location-based social network Gowalla \cite{gowalla,liu2014exploiting}. Each of the venues can be classified into six main categories which are travel, food, nightlife, outdoors, shopping, entertainment, and community yet in our case we have focused just on community and shopping venues. To obtain the distribution of destinations as a function of the spatial distribution of venues, we first divide each city using a grid of $500\times500m^2$ and assign each of the venues and road junctions to its corresponding grid cell. The probability of a vehicle to have a given road junction as a destination is calculated then as the normalized number of venues within that grid cell -- the number of venues in a cell divided by the total number of venues in a city-- divided by the total number of junctions within that cell. Such operation yields a probability distribution normalized to $1$ when all the junctions are considered.

\subsection{Traffic counts}

For Seattle and Miami, the traffic counts correspond to the Annual Average Daily Traffic (AADT) for the years 2018 and 2020 respectively, which stands for the total volume of traffic on a highway segment for one year, divided by the number of days in the year. For Paris and Madrid, we provide the average hourly flow of vehicles during 2019 at the AM peak going from $7$am to $10$am, at the link-level in Paris and the counter-level in Madrid.

\subsection{Measuring the delay in Uber movement data}

The project Uber movement data \cite{uber} provides the average travel times between regions of heterogenous shape in a city at each hour of the day. First of all, and to fairly compare the cities, we divide each city using a grid of cells of $2\times2$ $km^2$ and calculate the hourly travel times between them according to their spatial overlap with the original shapes. Thus the time needed $T^{h}_{ij}$ to reach cell $j$ when departing from cell $i$ at hour $h$ is an average over the travel times between the regions intersecting $i$ and $j$ weighted by the corresponding area of overlap. Since we want to focus on the congestion produced in the morning peak, the travel times $T^{\rm data}_{ij}$ will be given by the average for $h\in[7am,8am,9am]$.
Besides the quantification of the hourly travel times to evaluate the delay we start by defining the travel time between each pair of cells $i$ and $j$ under free-flow conditions as an average of the $4$ lowest travel times throughout a day $T^{ff}_{ij}$. From that quantity, the delay during the morning peak will be given by $\lambda^{\rm data}_{ij}=T^{\rm data}_{ij}/T^{\rm ff}_{ij}$.

\subsection{Measuring the delay in the MCM model}

In the MCM model, we calculate the travel times between a pair of junctions $i$ and $j$ under the congested phase as
\begin{equation}\label{traveltimes}
W_{ij}(\rho)=\sum_{k,l \in S}\left(1+\frac{\Delta q_{kl}(\rho)}{\tau_{kl}}\right)w_{kl},
\end{equation}
where $k$ and $l$ are the set of nodes that belong to the path shortest path $S$ going from $i$ to $j$ in the non-congested weighted graph, $\Delta q_{kl}$ is the increase of the queue in the link $kl$, $w_{kl}$ is its travel time and $\tau_{kl}$ its capacity. For each city $\rho=\rho_{\rm data}$ in order to meet the congestion observed in the TomTom data \cite{tomtom}. To obtain the travel times $T^{\rm model}_{ij}$ between the same $2\times2$ $km^2$ cells constructed for the Uber data, we average $W_{ij}$ over all the road segments that belong to each of the cells. To measure the delay of trajectories starting and ending within the same cells we follow the same procedure averaging over junctions $i$ and $j$ that belong to the same cell, ensuring that $i\neq j$.

\section*{Data availability statement}

The data on traffic counts is open and accessible in \cite{miami} for Miami,  in \cite{seattle} for Seattle, 
in \cite{paris} for Paris and in \cite{madrid} for Madrid. The coordinates and types of Gowalla POIs is open and accessible at \cite{gowalla}. The Uber movement data was downloaded from \cite{uber} and the road networks were extracted using the python package OSMnx \cite{boeing2017osmnx}. The code used to perform the simulations of the model and extract the road networks using Python3 is available at https://doi.org/10.5281/zenodo.6837557.

\section*{Acknowledgements}

AB acknowledges financial support from the Ministerio de Ciencia e Innovación under the Juan de la Cierva program (FJC2019-038958-I) and the Spanish Ministry of Universities, the European Union - Next Generation EU, the Recovery, Transformation and Resilience Plan and the University of the Balearic Islands. We acknowledge support by Ministerio de Economía y Competitividad (PGC2018-094754-BC21, FIS2017-90782-REDT and RED2018-102518-T), Generalitat de Catalunya (2017SGR-896 and 2020PANDE00098), and Universitat Rovira i Virgili (2021PFR-URV-118). AA acknowledges also ICREA Academia and the James S. McDonnell Foundation (220020325).

\clearpage

\renewcommand\theequation{{S\arabic{equation}}}
\renewcommand\thetable{{Supplementary S\Roman{table}}}
\renewcommand{\figurename}{Supplementary Figure}
\renewcommand\thefigure{{S\arabic{figure}}}
\renewcommand\thesection{{Section S\arabic{section}}}

\setcounter{section}{0}
\setcounter{table}{0}
\setcounter{figure}{0}
\setcounter{equation}{0}

\onecolumngrid

\section*{\large{Supplementary Material}}

\section{Congestion phenomena in the DT+MST model}

In this section we assess the congestion in the DT+MST model developed in \cite{lampo2021multiple,lampo2021emergence} that mimics the structure of real cities with a more densely connected center and a sparse periphery. A set of points is homogeneously distributed in a 2D space of size $LxL$ and are connected according to the the Delaunay triangulation (DT) \cite{lee1980two}. Within a distance smaller than $R_{\rm DT}$ from the center of the domain the network is fully preserved while for the region with radius greater than $R_{\rm DT}$ most of the links are removed in order to keep the maximum spanning tree (MST) that maximizes the betweenness centrality. The value of $R_{\rm DT}$ determines the underlying structure of the spatial graph, low values leading to network dominated by the MST and vice-versa for high values.

\begin{figure*}[!htbp]
  \begin{center}
  \includegraphics[width=6in]{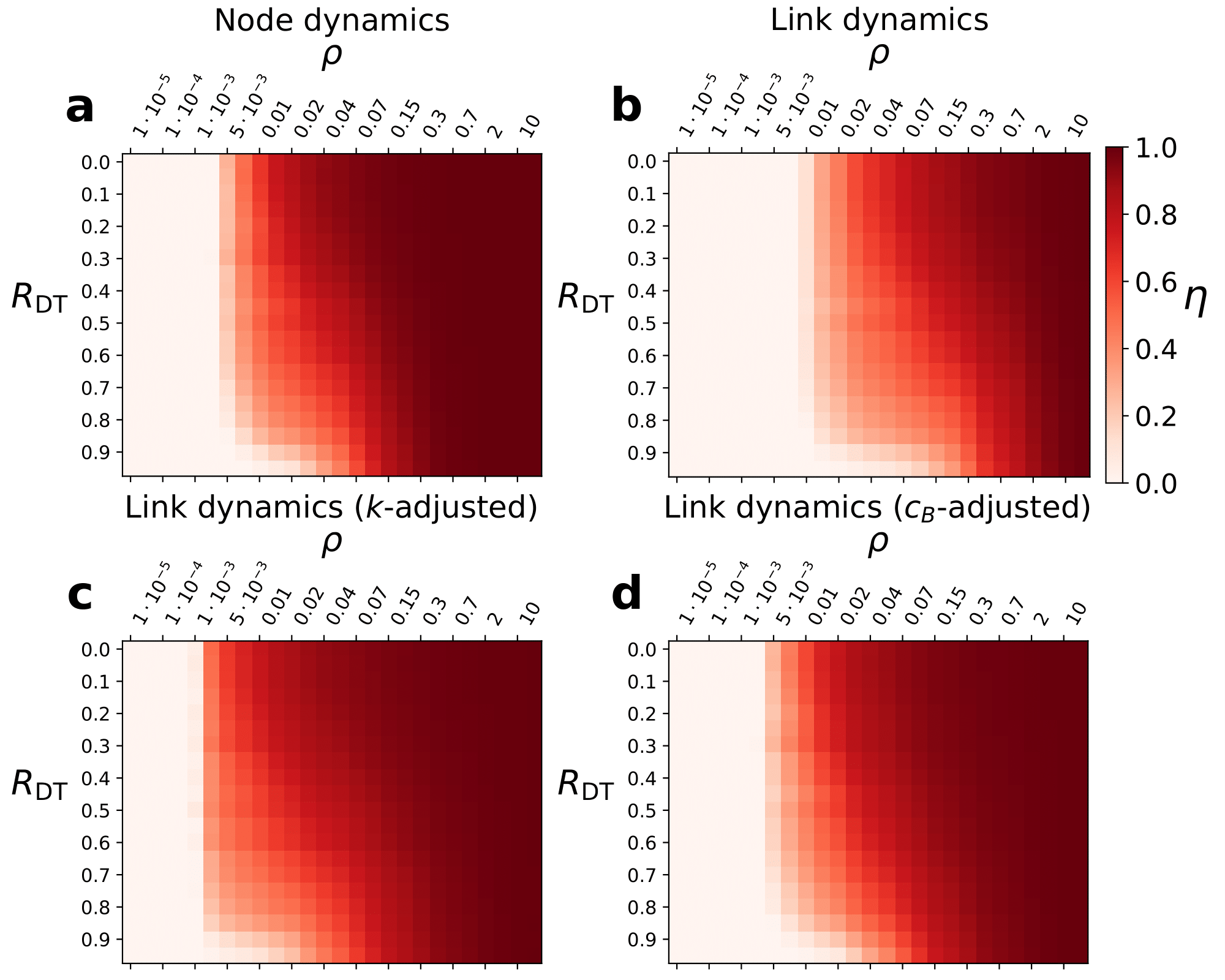}
  \end{center}
  \caption[\textbf{Evolution of the order parameter $\eta$ in DT+MST graphs as a function of the injection rate $\rho$ for each of the models and several values of $R_{\rm DT}$.}]{\textbf{Evolution of the order parameter $\eta$  in DT+MST graphs as a function of the injection rate $\rho$ for each of the models and several values of $R_{\rm DT}$.} Evolution of $\eta(\rho)$ for \textbf{a} the node model, \textbf{b} the link model without capacity normalization ($\tau_{ij}=\tau_{i}$), \textbf{c} the link model with $k$-adjusted normalization ($\widetilde{\tau}_{ij}^k$) and  \textbf{d} the link model with $c_B$-adjusted normalization ($\widetilde{\tau}_{ij}^{c_B}$). The parameter $R_{\rm DT}$ is normalized by $l/2$ where $l$ is the side size of the squared domain $l \times l$  . } \label{dtmst}
\end{figure*}

In Fig. \ref{dtmst} we report $\eta$ as a function of $\rho$ for different values of $R_{\rm DT}$ in networks of 500 nodes distributed in a space of $80\times80$. To better compare the networks we plot them as a function of $\widetilde{R}_{\rm DT}$ which is calculted as $\frac{R_{\rm DT}}{L/2}$. The overall trend seems to indicate that congestion decreases the DT region increases, likely because that there are wider paths alternatives as compared to the MST. In the context of the results for the cost-driven networks shown in Fig. 4 of the main text, congestion seems to increase in a more progressive way, likely as a consequence of the multiple regimes of betweenness centrality already found in \cite{lampo2021multiple,lampo2021emergence}.

\clearpage

\section{Spatial distribution of congestion hotspots}

We report here the congestion hotspots for the rest of cities analyzed in the case of the node model and the link model with a capacity adjusted for node degree. More concretely we have respectively in Figs. S2-S13 the results for Amsterdam, Brussels, Madrid, Miami, Mumbai, Paris, Pittsburgh, Sao Paulo, Seattle, Taipei, Toronto and Washington.

\begin{figure*}[!htbp]
  \begin{center}
  \includegraphics[width=6in]{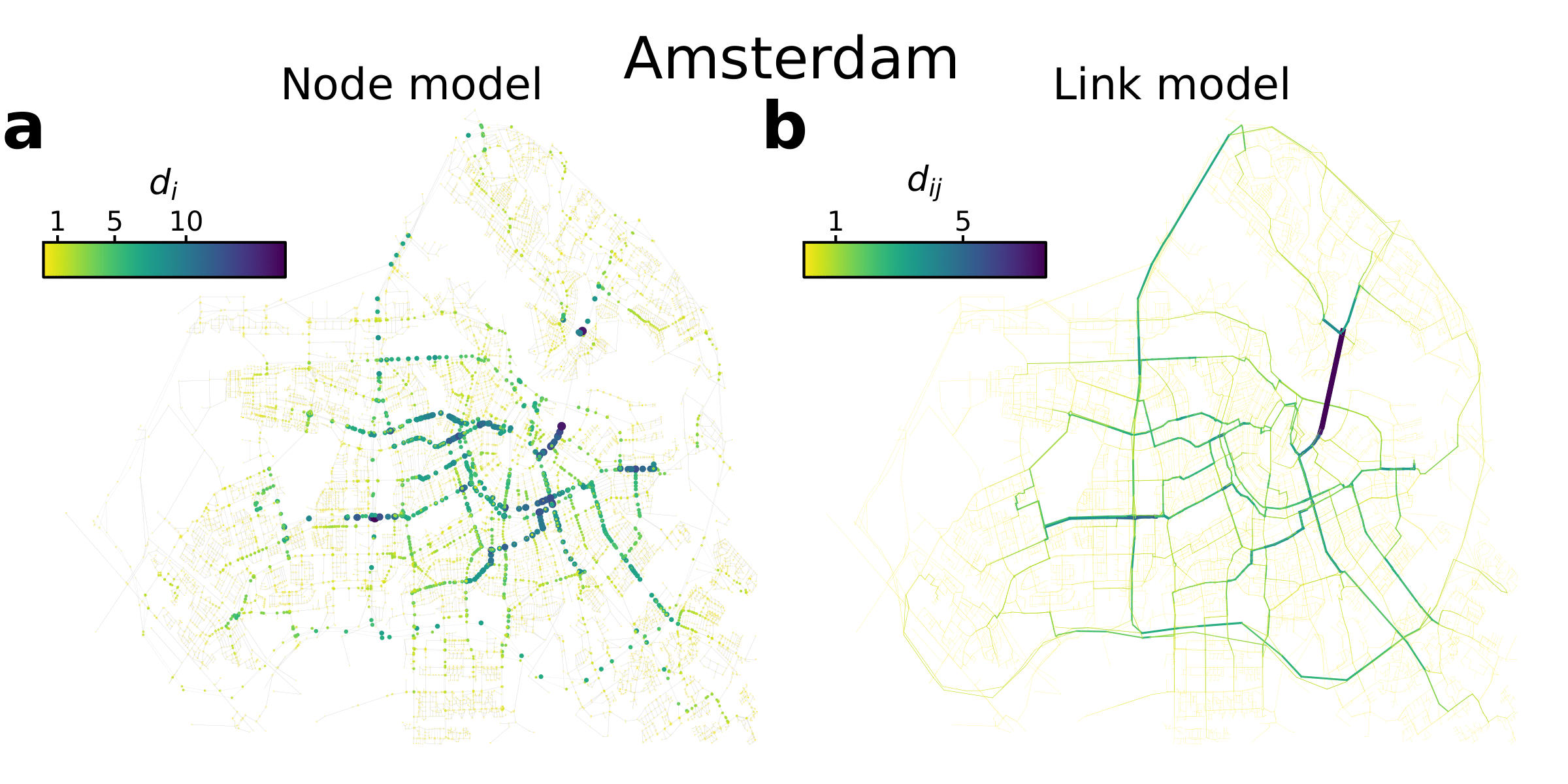}
  \end{center}
  \caption[\textbf{Analysis of congestion hotspots in Amsterdam.}]{\textbf{Analysis of congestion hotspots in Amsterdam.} Congestion hotspots observed in Amsterdam for \textbf{a} the node model and \textbf{b} the link model with destinations distributed according to community venues. Both maps where generated with $\eta_{\rm data}=0.26$.} \label{amsterdam}
\end{figure*}

\begin{figure*}[!htbp]
  \begin{center}
  \includegraphics[width=6in]{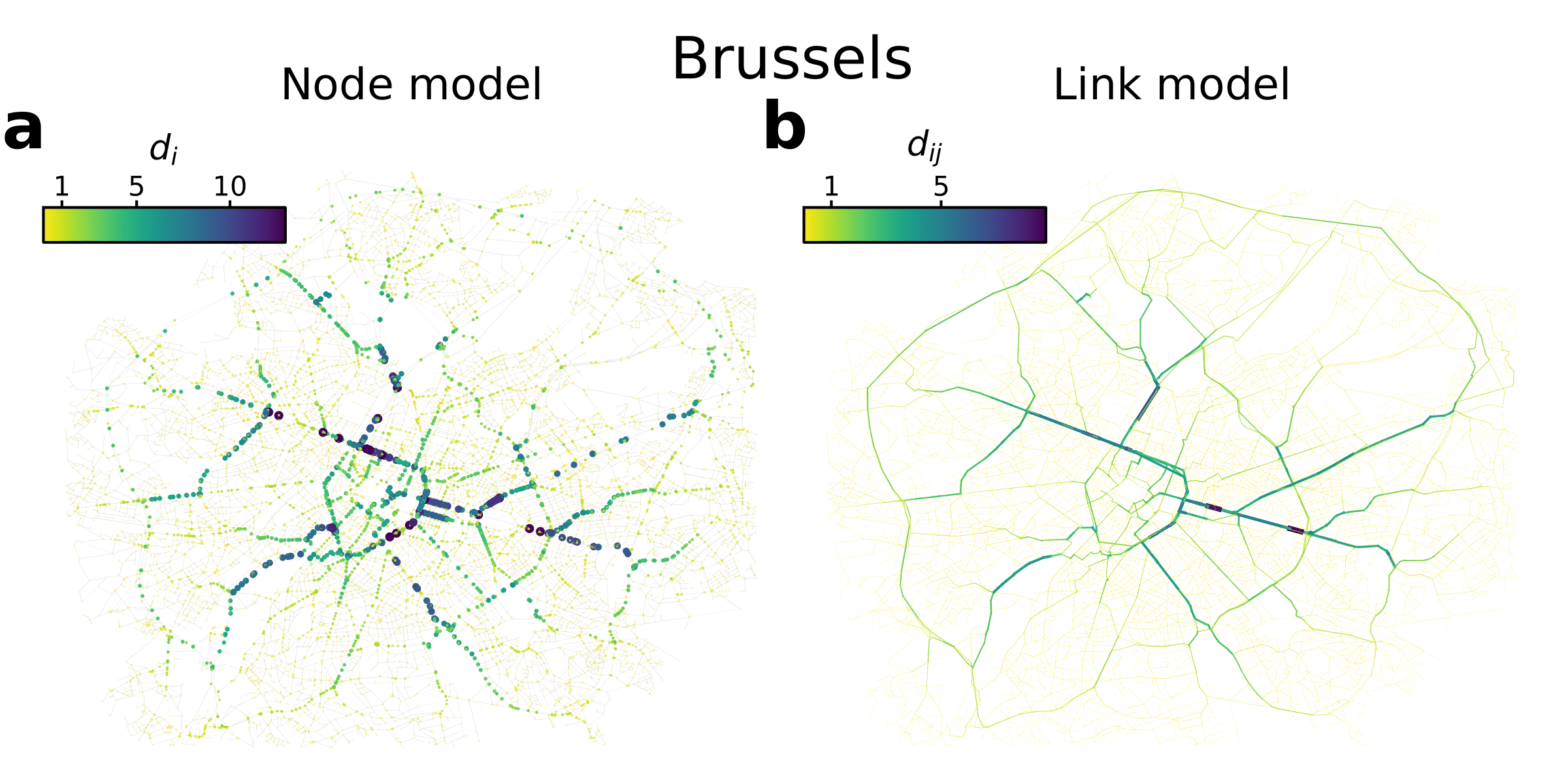}
  \end{center}
  \caption[\textbf{Analysis of congestion hotspots in Brussels.}]{\textbf{Analysis of congestion hotspots in Brussels.} Congestion hotspots observed in Brussels for \textbf{a} the node model and \textbf{b} the link model with destinations distributed according to community venues. Both maps where generated with $\eta_{\rm data}=0.38$.} \label{brussels}
\end{figure*}

\begin{figure*}[!htbp]
  \begin{center}
  \includegraphics[width=6in]{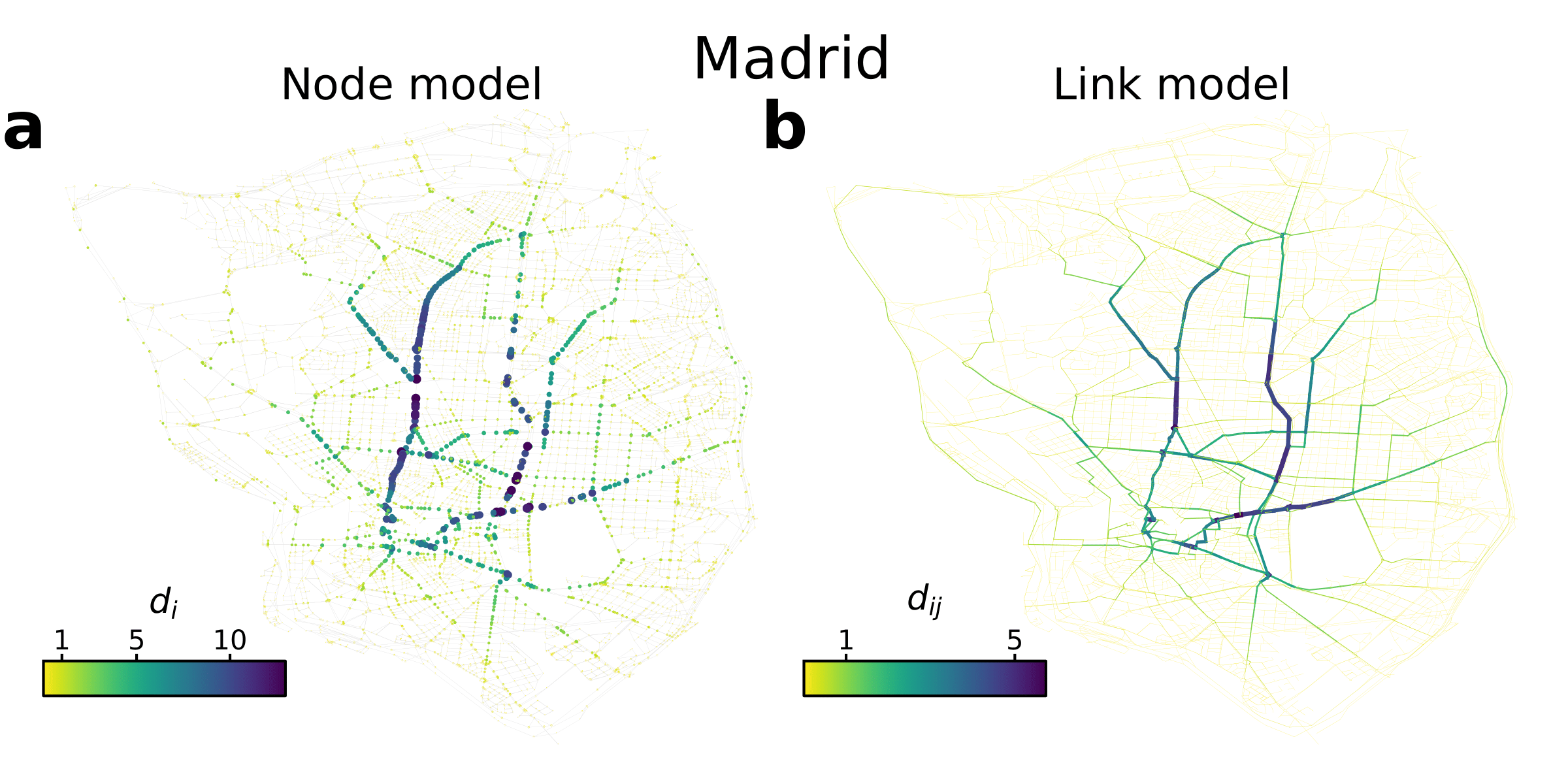}
  \end{center}
  \caption[\textbf{Analysis of congestion hotspots in Madrid.}]{\textbf{Analysis of congestion hotspots in Madrid.} Congestion hotspots observed in Madrid for \textbf{a} the node model and \textbf{b} the link model with destinations distributed according to community venues. Both maps where generated with $\eta_{\rm data}=0.23$.} \label{madrid}
\end{figure*}

\begin{figure*}[!htbp]
  \begin{center}
  \includegraphics[width=6in]{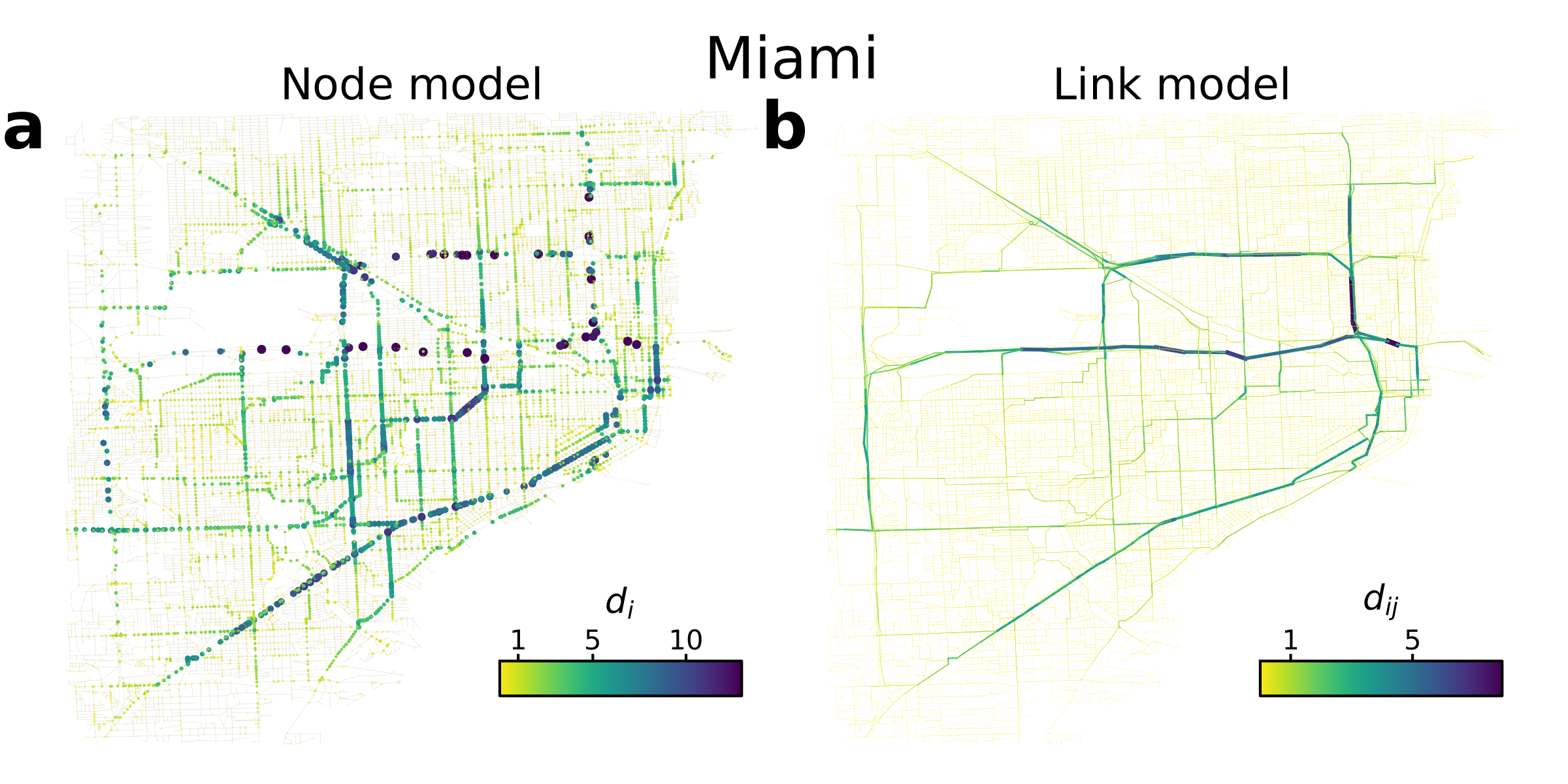}
  \end{center}
  \caption[\textbf{Analysis of congestion hotspots in Miami.}]{\textbf{Analysis of congestion hotspots in Miami.} Congestion hotspots observed in Miami for \textbf{a} the node model and \textbf{b} the link model with destinations distributed according to community venues. Both maps where generated with $\eta_{\rm data}=0.31$.} \label{miami}
\end{figure*}

\begin{figure*}[!htbp]
  \begin{center}
  \includegraphics[width=6in]{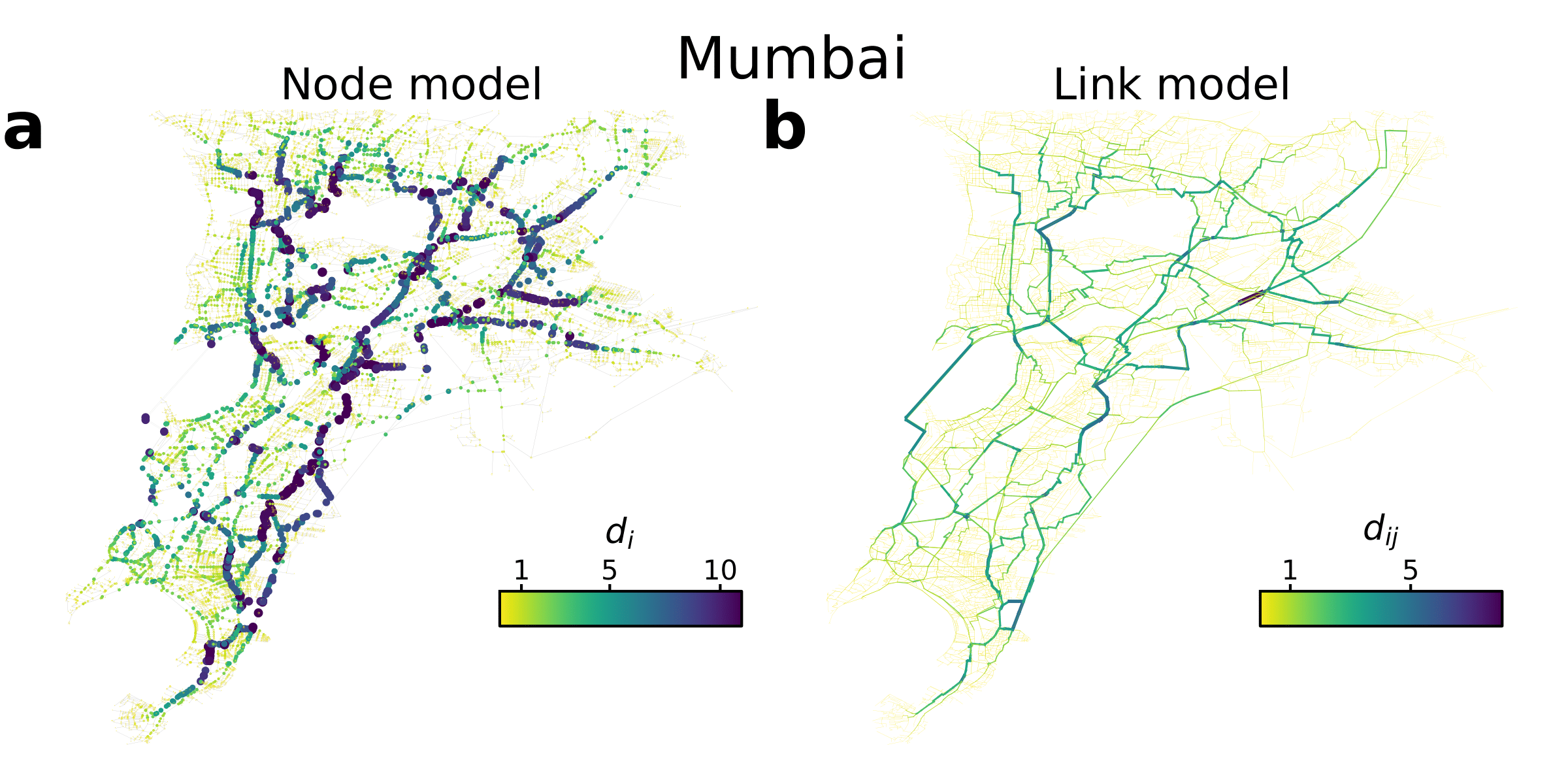}
  \end{center}
  \caption[\textbf{Analysis of congestion hotspots in Mumbai.}]{\textbf{Analysis of congestion hotspots in Mumbai.} Congestion hotspots observed in Mumbai for \textbf{a} the node model and \textbf{b} the link model with destinations distributed according to community venues. Both maps where generated with $\eta_{\rm data}=0.65$.} \label{mumbai}
\end{figure*}

\begin{figure*}[!htbp]
  \begin{center}
  \includegraphics[width=6in]{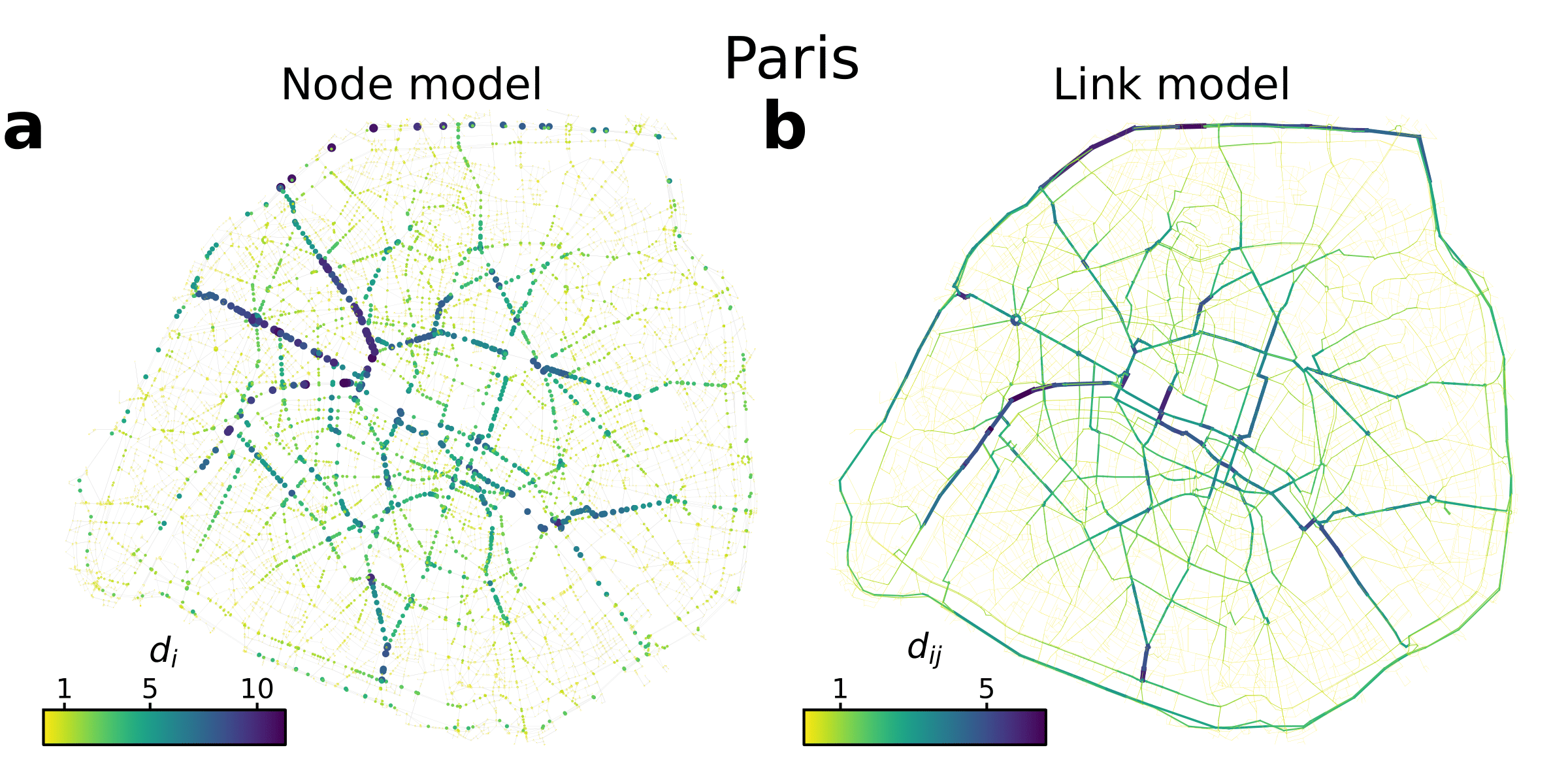}
  \end{center}
  \caption[\textbf{Analysis of congestion hotspots in Paris.}]{\textbf{Analysis of congestion hotspots in Paris.} Congestion hotspots observed in Paris for \textbf{a} the node model and \textbf{b} the link model with destinations distributed according to community venues. Both maps where generated with $\eta_{\rm data}=0.39$.} \label{paris}
\end{figure*}

\begin{figure*}[!htbp]
  \begin{center}
  \includegraphics[width=6in]{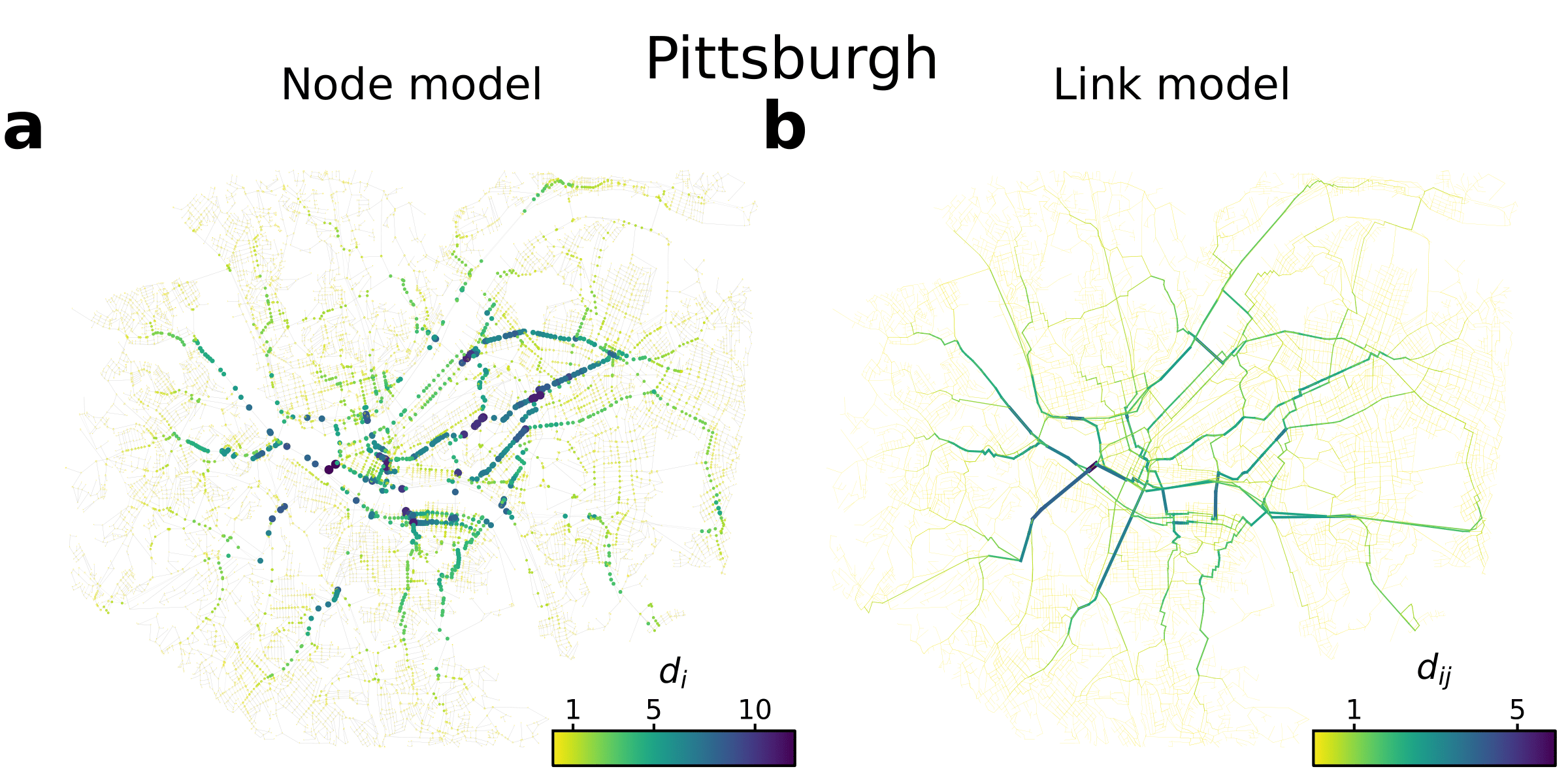}
  \end{center}
  \caption[\textbf{Analysis of congestion hotspots in Pittsburgh.}]{\textbf{Analysis of congestion hotspots in Pittsburgh.} Congestion hotspots observed in Pittsburgh for \textbf{a} the node model and \textbf{b} the link model with destinations distributed according to community venues. Both maps where generated with $\eta_{\rm data}=0.21$.} \label{pittsburgh}
\end{figure*}

\begin{figure*}[!htbp]
  \begin{center}
  \includegraphics[width=6in]{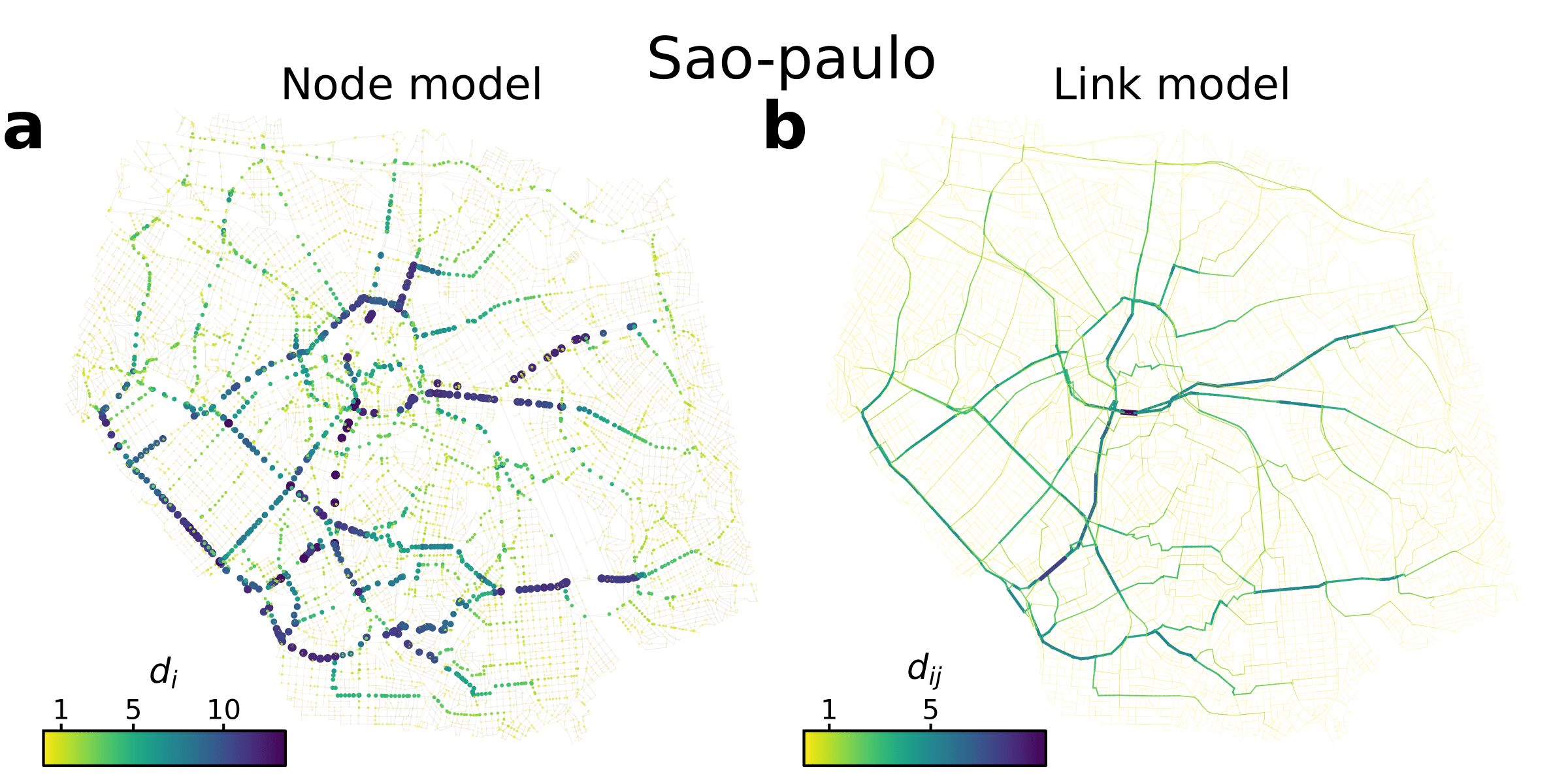}
  \end{center}
  \caption[\textbf{Analysis of congestion hotspots in Sao Paulo.}]{\textbf{Analysis of congestion hotspots in Sao Paulo.} Congestion hotspots observed in Sao Paulo for \textbf{a} the node model and \textbf{b} the link model with destinations distributed according to community venues. Both maps where generated with $\eta_{\rm data}=0.45$.} \label{sao-paulo}
\end{figure*}

\begin{figure*}[!htbp]
  \begin{center}
  \includegraphics[width=6in]{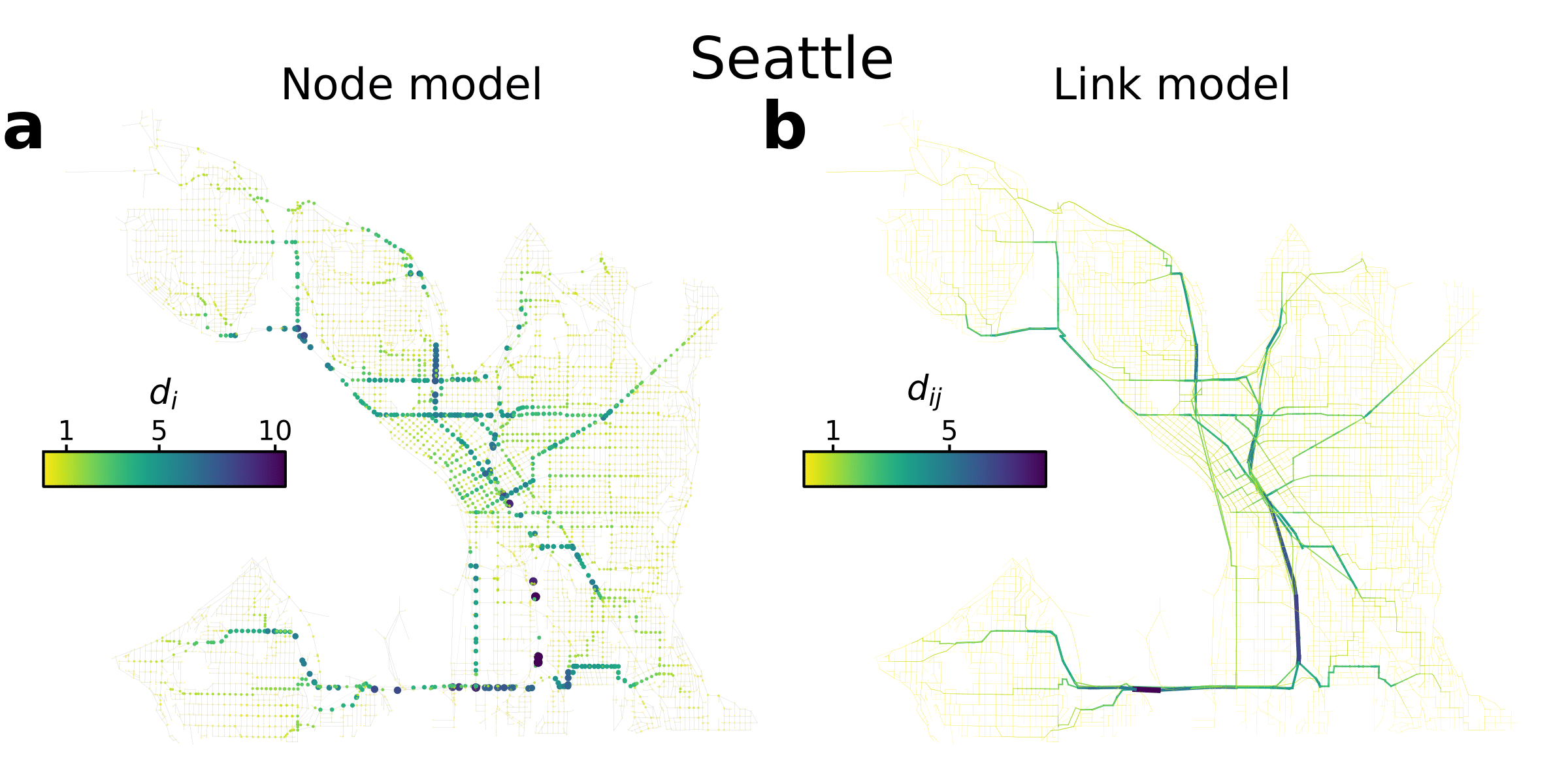}
  \end{center}
  \caption[\textbf{Analysis of congestion hotspots in Seattle.}]{\textbf{Analysis of congestion hotspots in Seattle.} Congestion hotspots observed in Seattle for \textbf{a} the node model and \textbf{b} the link model with destinations distributed according to community venues. Both maps where generated with $\eta_{\rm data}=0.31$.} \label{seattle}
\end{figure*}

\begin{figure*}[!htbp]
  \begin{center}
  \includegraphics[width=6in]{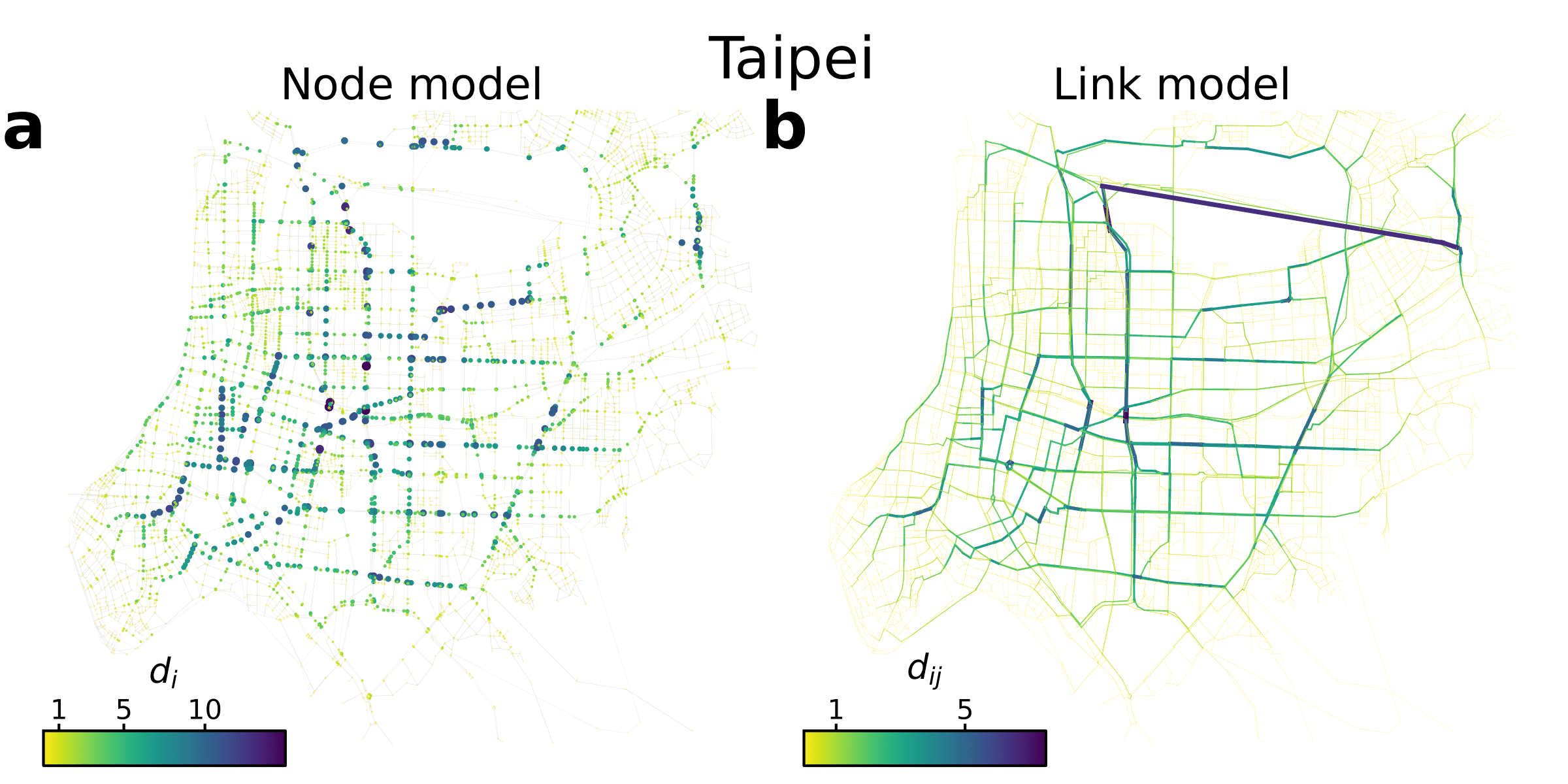}
  \end{center}
  \caption[\textbf{Analysis of congestion hotspots in Taipei.}]{\textbf{Analysis of congestion hotspots in Taipei.} Congestion hotspots observed in Taipei for \textbf{a} the node model and \textbf{b} the link model with destinations distributed according to community venues. Both maps where generated with $\eta_{\rm data}=0.35$.} \label{taipei}
\end{figure*}

\begin{figure*}[!htbp]
  \begin{center}
  \includegraphics[width=6in]{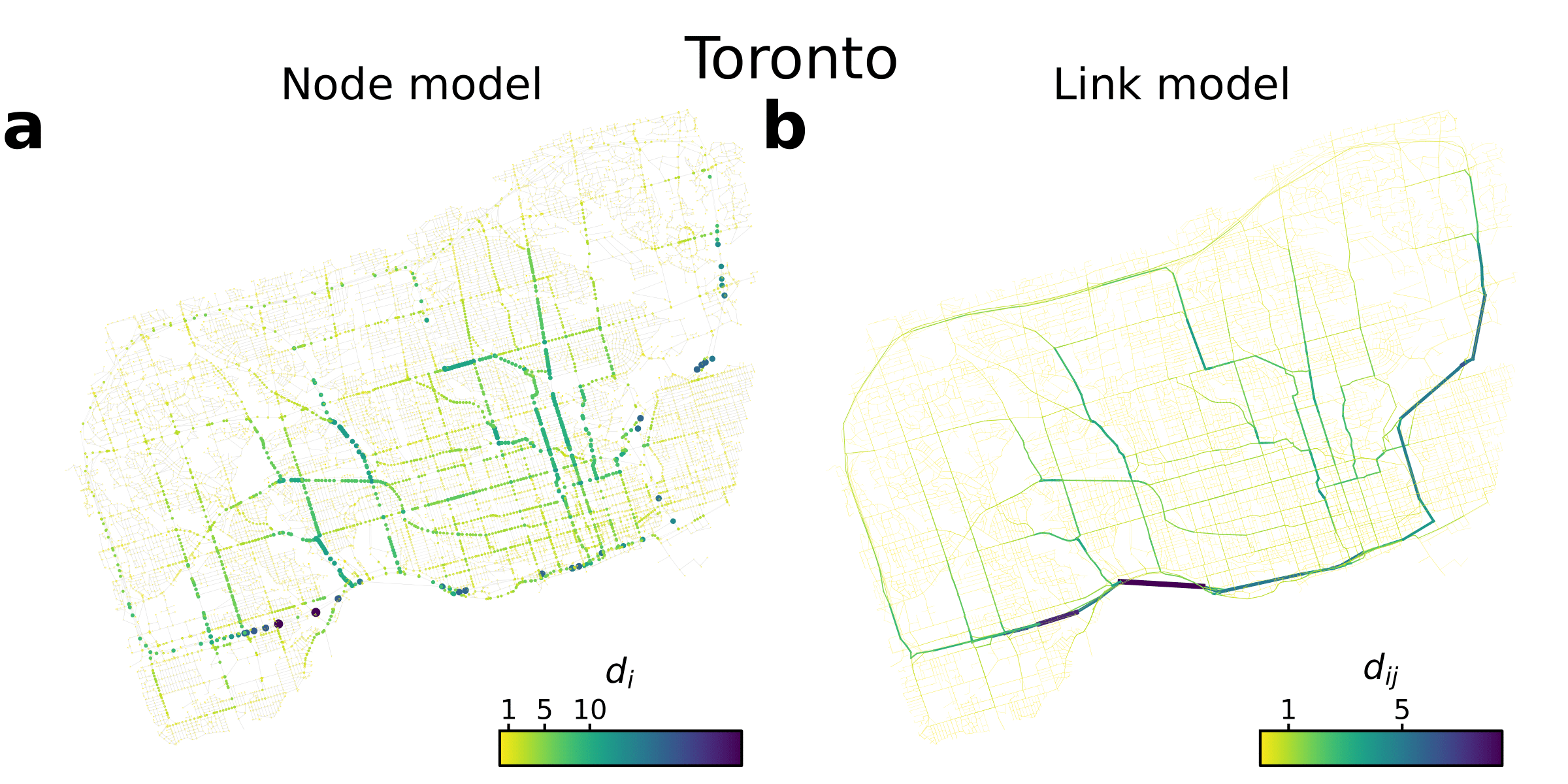}
  \end{center}
  \caption[\textbf{Analysis of congestion hotspots in Toronto.}]{\textbf{Analysis of congestion hotspots in Toronto.} Congestion hotspots observed in Toronto for \textbf{a} the node model and \textbf{b} the link model with destinations distributed according to community venues. Both maps where generated with $\eta_{\rm data}=0.33$.} \label{toronto}
\end{figure*}

\begin{figure*}[!htbp]
  \begin{center}
  \includegraphics[width=6in]{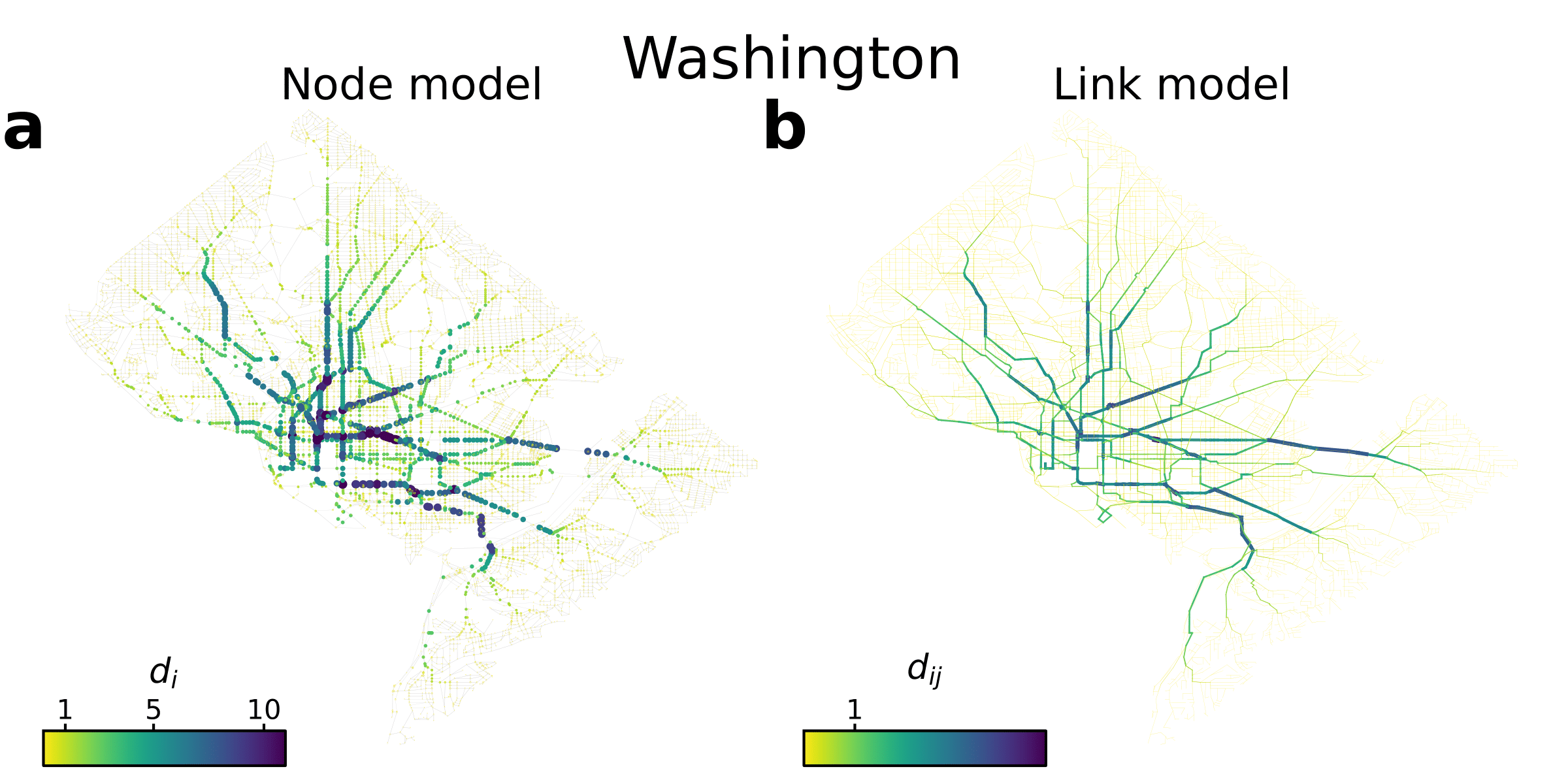}
  \end{center}
  \caption[\textbf{Analysis of congestion hotspots in Washington.}]{\textbf{Analysis of congestion hotspots in Washington.} Congestion hotspots observed in Washington for \textbf{a} the node model and \textbf{b} the link model with destinations distributed according to community venues. Both maps where generated with $\eta_{\rm data}=0.29$. } \label{washington}
\end{figure*}

\clearpage

\section{Correlation with traffic counts and observed delays}

\begin{figure*}[!htbp]
  \begin{center}
  \includegraphics[width=0.5\textwidth]{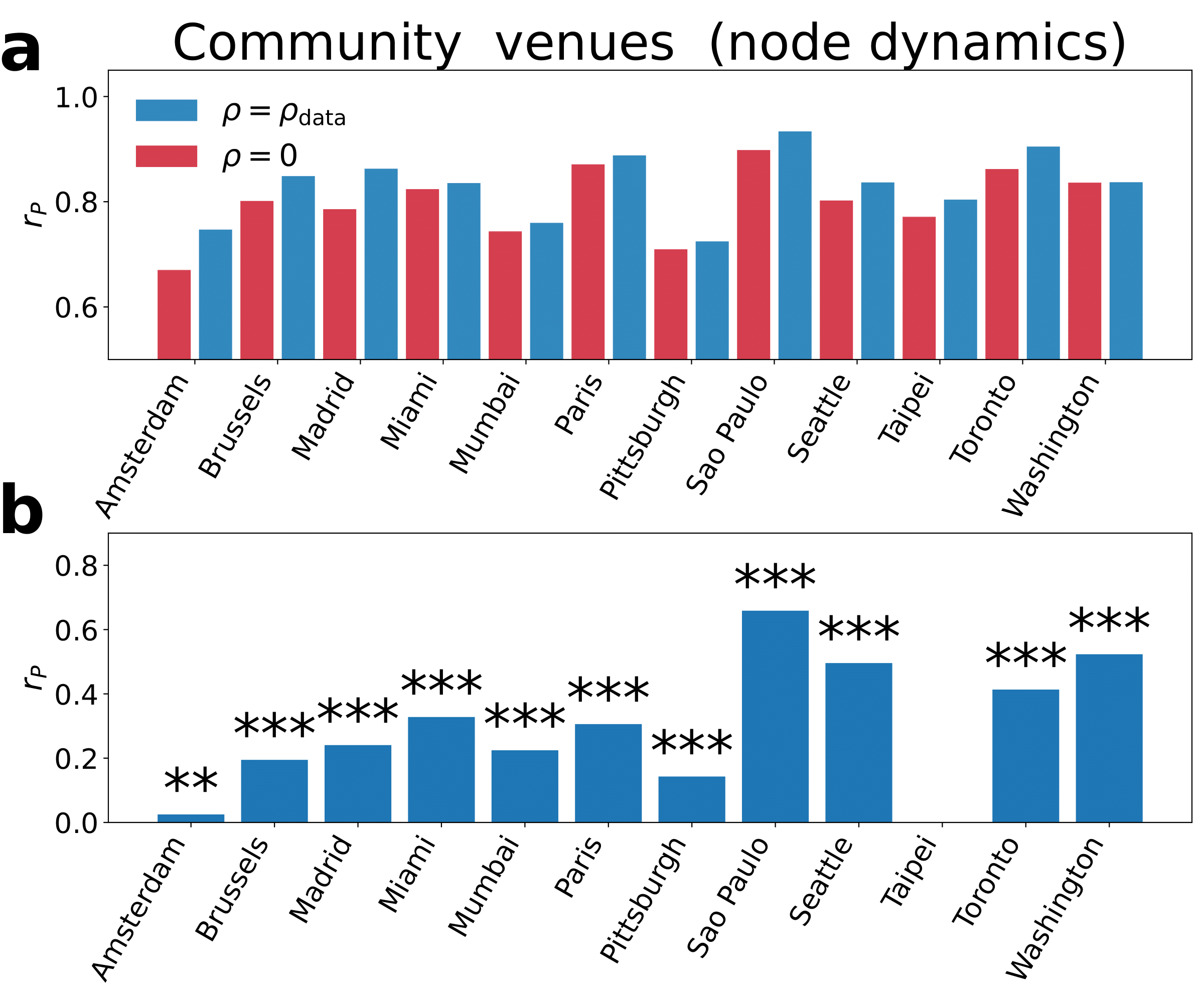}
  \end{center}
  \caption[\textbf{Correlation between the real and modelled delays in the node dynamics when destinations are distributed according to the community POIs.}]{\textbf{Correlation between the real and modelled delays in the node dynamics when destinations are distributed according to the community POIs.} (\textbf{a}) Comparison between the Pearson correlation coefficient obtained between the travel times from Uber Data \cite{uber} during the mornning peak ($8-10$am) in a set of cities and the travel times obtained for $\rho=0$ (red) and $\rho=\rho_{\rm data}$ (blue). (\textbf{b}) Pearson correlation coefficient between the delay observed in the data and in the model. Asterisks indicate the level of significance ($*$ p-value$<0.05$, $**$ p-value$<0.01$, $***$ p-value$<0.001$. The injection rate for each city $\rho_{\rm data}$ is set to match $\eta$ with the percentage of delay observed in the Tom Tom traffic index data \cite{tomtom}. } \label{934node}
\end{figure*}

\begin{figure*}[!htbp]
  \begin{center}
  \includegraphics[width=0.5\textwidth]{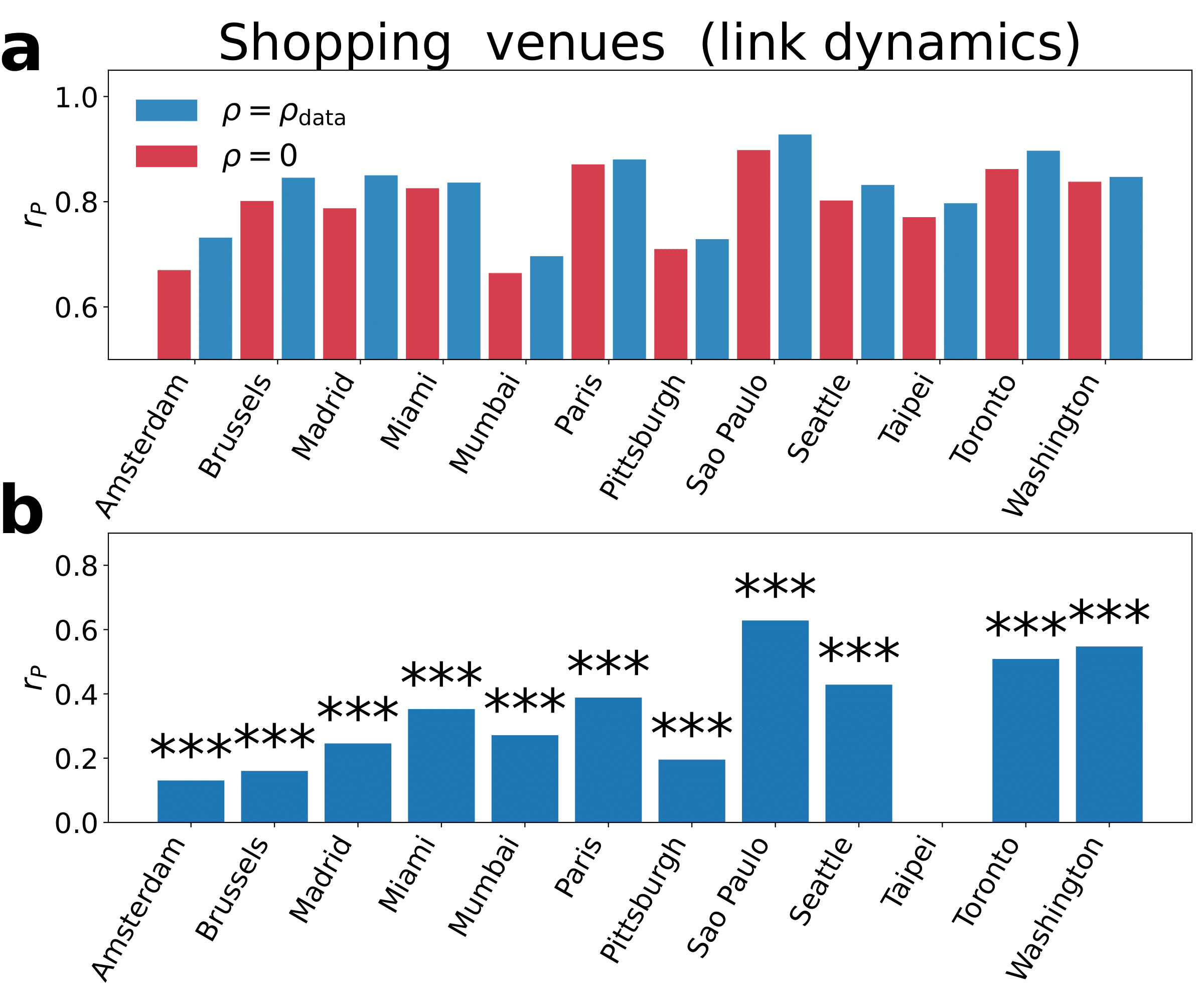}
  \end{center}
    \caption[\textbf{Correlation between the real and modelled delays in the link dynamics when destinations are distributed according to the shopping POIs.}]{\textbf{Correlation between the real and modelled delays in the link dynamics when destinations are distributed according to the shopping POIs.} (\textbf{a}) Comparison between the Pearson correlation coefficient obtained between the travel times from Uber Data \cite{uber} during the mornning peak ($8-10$am) in a set of cities and the travel times obtained for $\rho=0$ (red) and $\rho=\rho_{\rm data}$ (blue). (\textbf{b}) Pearson correlation coefficient between the delay observed in the data and in the model. Asterisks indicate the level of significance ($*$ p-value$<0.05$, $**$ p-value$<0.01$, $***$ p-value$<0.001$). The injection rate for each city $\rho_{\rm data}$ is set to match $\eta$ with the percentage of delay observed in the Tom Tom traffic index data \cite{tomtom}.  } \label{6link}
\end{figure*}

\begin{figure*}[!htbp]
  \begin{center}
  \includegraphics[width=0.5\textwidth]{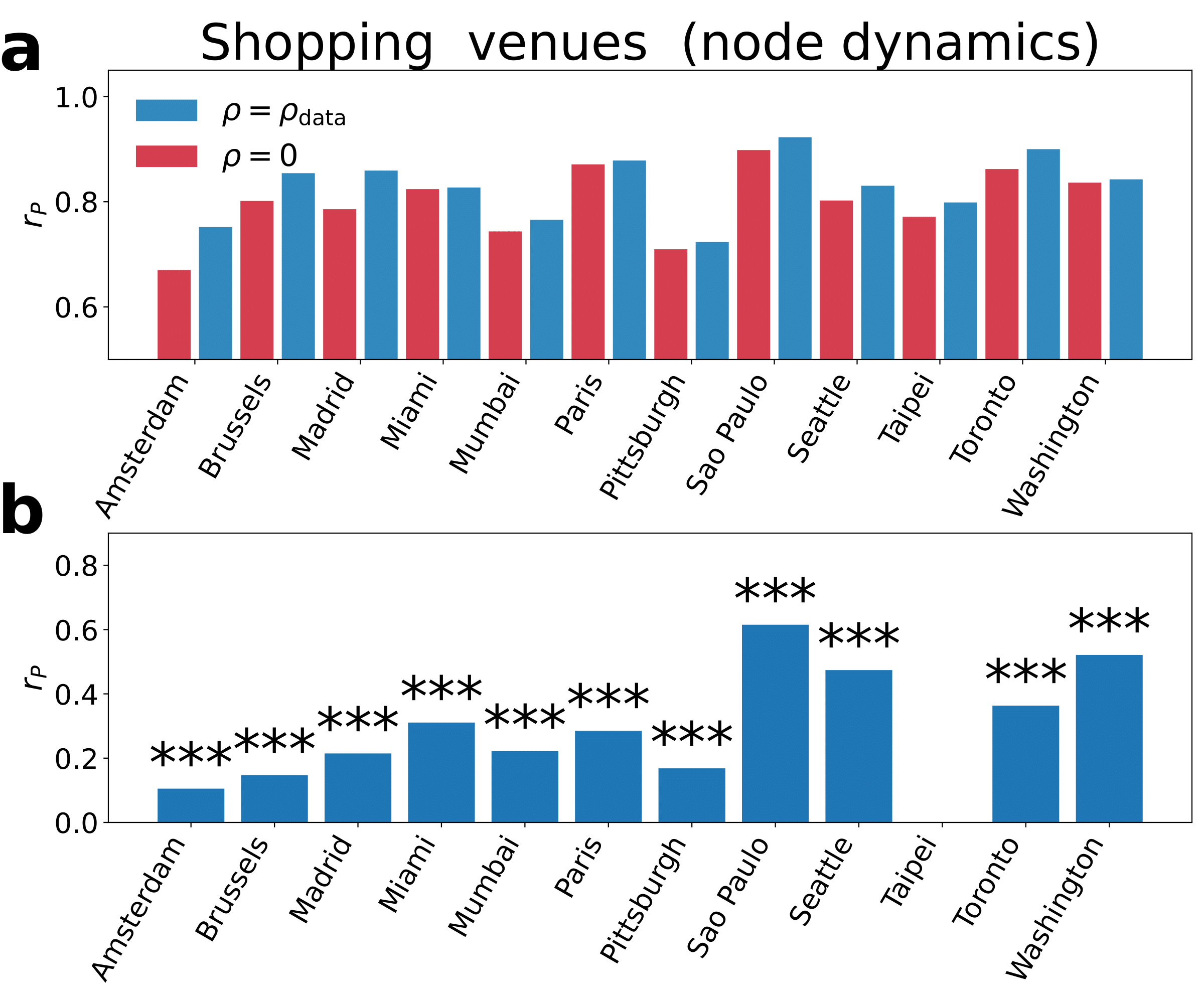}
  \end{center}
    \caption[ \textbf{Correlation between the real and modelled delays in the node dynamics when destinations are distributed according to the shopping POIs.}]{\textbf{Correlation between the real and modelled delays in the node dynamics when destinations are distributed according to the shopping POIs.} (\textbf{a}) Comparison between the Pearson correlation coefficient obtained between the travel times from Uber Data \cite{uber} during the mornning peak ($8-10$am) in a set of cities and the travel times obtained for $\rho=0$ (red) and $\rho=\rho_{\rm data}$ (blue). (\textbf{b}) Pearson correlation coefficient between the delay observed in the data and in the model. Asterisks indicate the level of significance ($*$ p-value$<0.05$, $**$ p-value$<0.01$, $***$ p-value$<0.001$. The injection rate for each city $\rho_{\rm data}$ is set to match $\eta$ with the percentage of delay observed in the Tom Tom traffic index data \cite{tomtom}. } 
    \label{6node}
\end{figure*}

\begin{figure*}[!htbp]
  \begin{center}
  \includegraphics[width=0.5\textwidth]{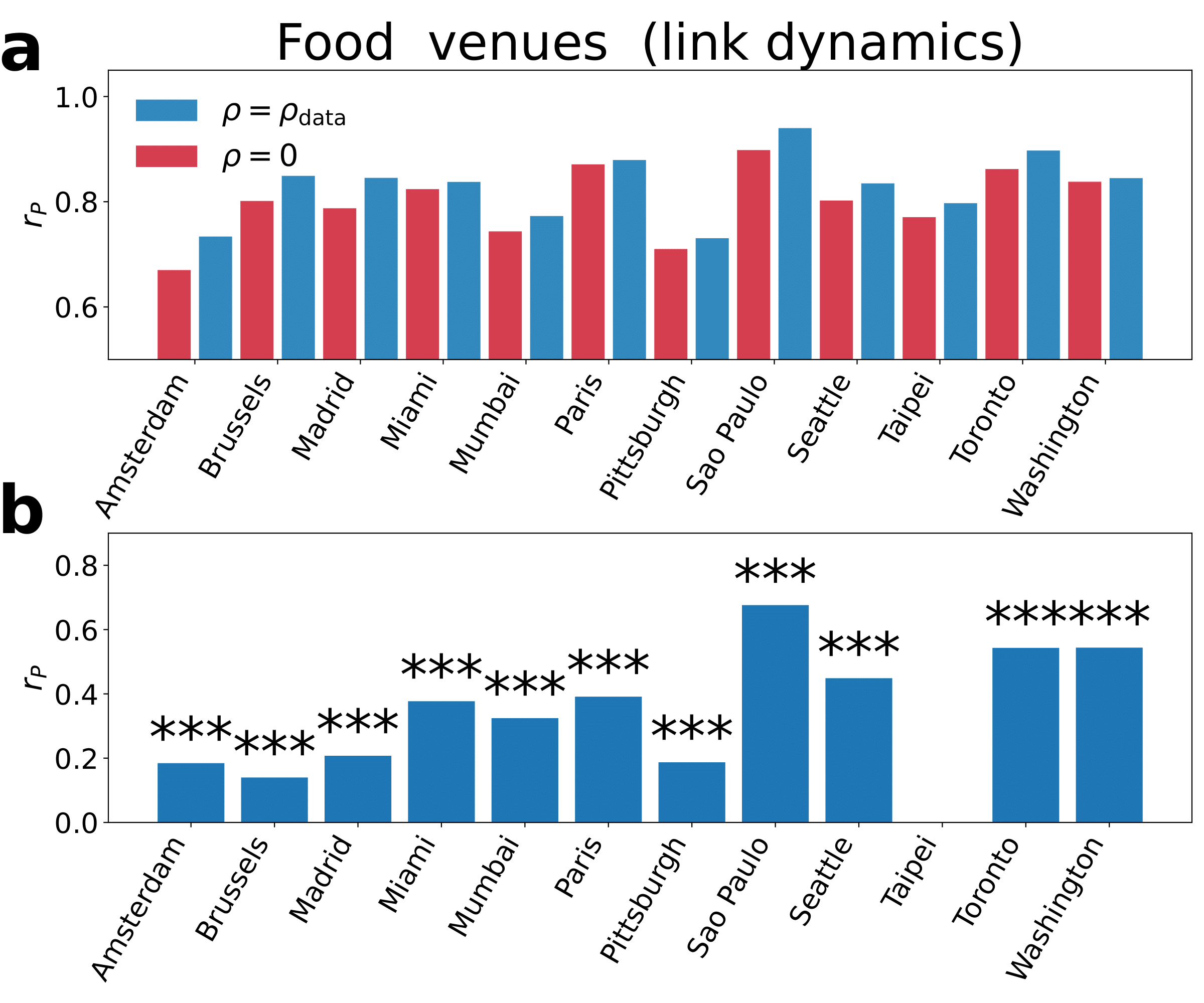}
  \end{center}
    \caption[\textbf{Correlation between the real and modelled delays in the link dynamics when destinations are distributed according to the food POIs.}]{\textbf{Correlation between the real and modelled delays in the link dynamics when destinations are distributed according to the food POIs.} (\textbf{a}) Comparison between the Pearson correlation coefficient obtained between the travel times from Uber Data \cite{uber} during the mornning peak ($8-10$am) in a set of cities and the travel times obtained for $\rho=0$ (red) and $\rho=\rho_{\rm data}$ (blue). (\textbf{b}) Pearson correlation coefficient between the delay observed in the data and in the model. Asterisks indicate the level of significance ($*$ p-value$<0.05$, $**$ p-value$<0.01$, $***$ p-value$<0.001$). The injection rate for each city $\rho_{\rm data}$ is set to match $\eta$ with the percentage of delay observed in the Tom Tom traffic index data \cite{tomtom}.  } \label{7link}
\end{figure*}

\begin{figure*}[!htbp]
  \begin{center}
  \includegraphics[width=0.5\textwidth]{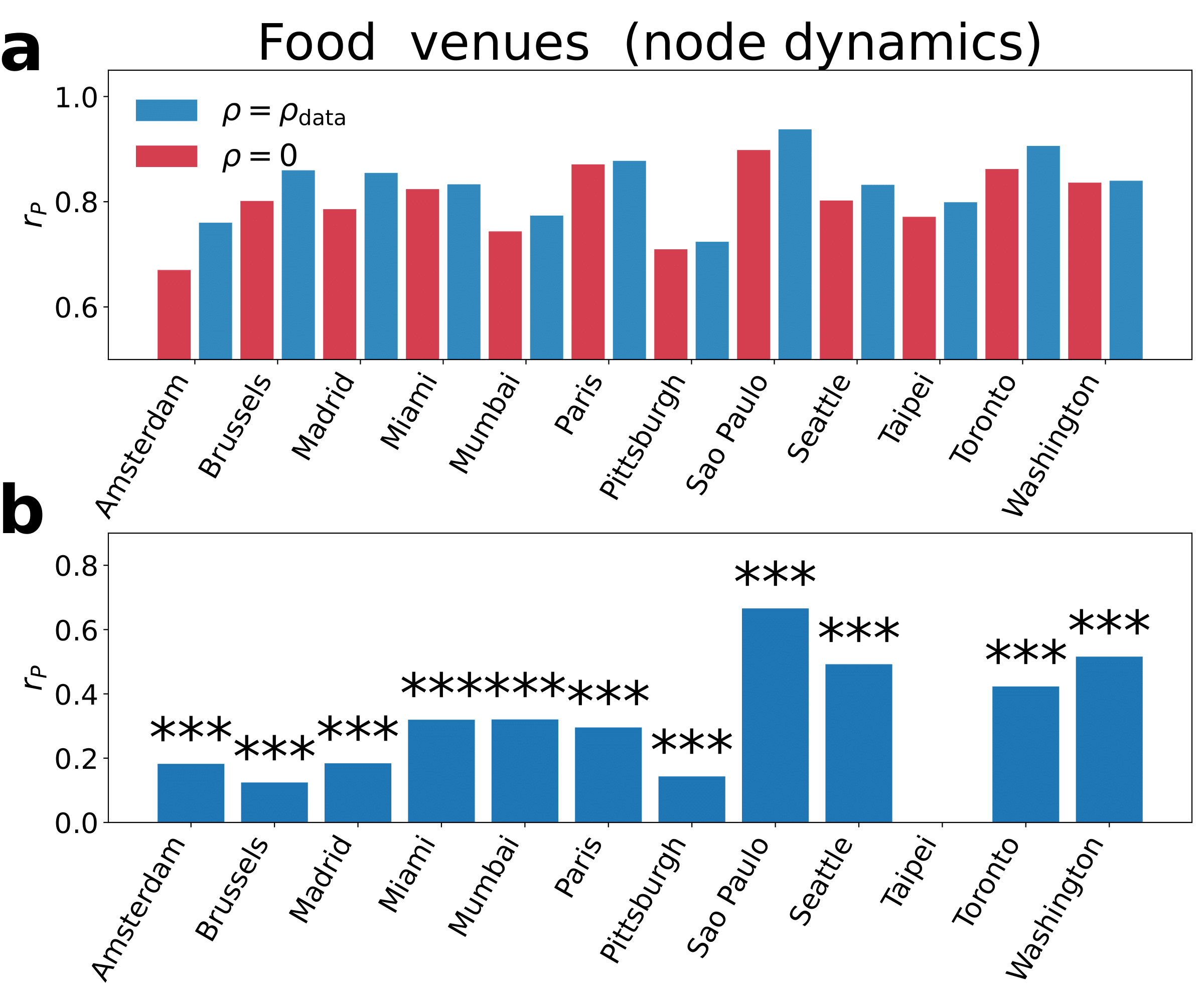}
  \end{center}
  \caption[\textbf{Correlation between the real and modelled delays in the node dynamics when destinations are distributed according to the food POIs.}]{\textbf{Correlation between the real and modelled delays in the node dynamics when destinations are distributed according to the food POIs.} (\textbf{a}) Comparison between the Pearson correlation coefficient obtained between the travel times from Uber Data \cite{uber} during the mornning peak ($8-10$am) in a set of cities and the travel times obtained for $\rho=0$ (red) and $\rho=\rho_{\rm data}$ (blue) as detailed in Eq. \ref{traveltimes}. (\textbf{b}) Pearson correlation coefficient between the delay observed in the data and in the model. Asterisks indicate the level of significance ($*$ p-value$<0.05$, $**$ p-value$<0.01$, $***$ p-value$<0.001$. The injection rate for each city $\rho_{\rm data}$ is set to match $\eta$ with the percentage of delay observed in the Tom Tom traffic index data \cite{tomtom}. } \label{7node}
\end{figure*}

\begin{figure*}[!htbp]
  \begin{center}
  \includegraphics[width=0.5\textwidth]{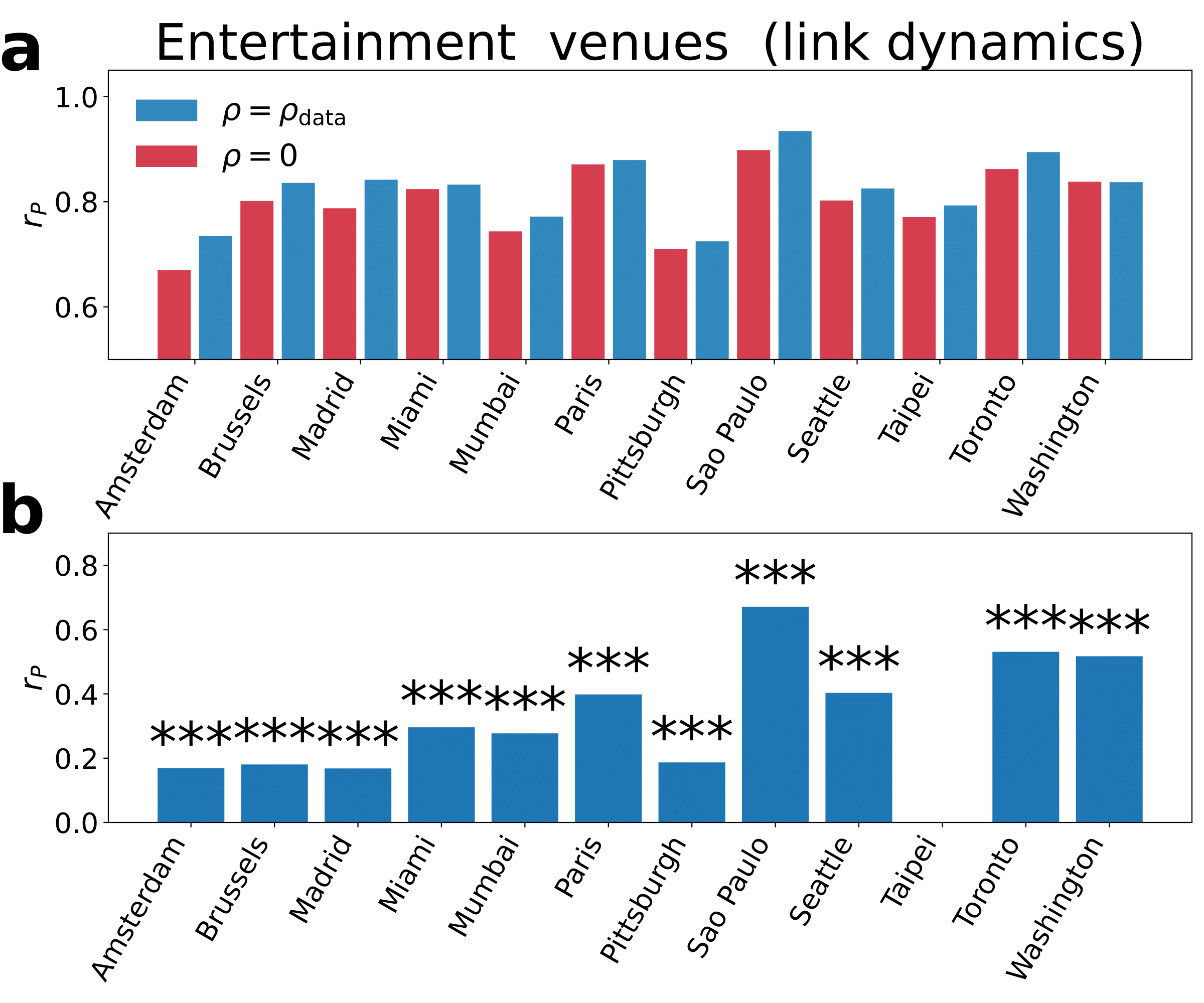}
  \end{center}
    \caption[\textbf{Correlation between the real and modelled delays in the link dynamics when destinations are distributed according to the entertainment POIs.}]{\textbf{Correlation between the real and modelled delays in the link dynamics when destinations are distributed according to the entertainment POIs.} (\textbf{a}) Comparison between the Pearson correlation coefficient obtained between the travel times from Uber Data \cite{uber} during the mornning peak ($8-10$am) in a set of cities and the travel times obtained for $\rho=0$ (red) and $\rho=\rho_{\rm data}$ (blue). (\textbf{b}) Pearson correlation coefficient between the delay observed in the data and in the model. Asterisks indicate the level of significance ($*$ p-value$<0.05$, $**$ p-value$<0.01$, $***$ p-value$<0.001$). The injection rate for each city $\rho_{\rm data}$ is set to match $\eta$ with the percentage of delay observed in the Tom Tom traffic index data \cite{tomtom}.  } \label{12link}
\end{figure*}

\begin{figure*}[!htbp]
  \begin{center}
  \includegraphics[width=0.5\textwidth]{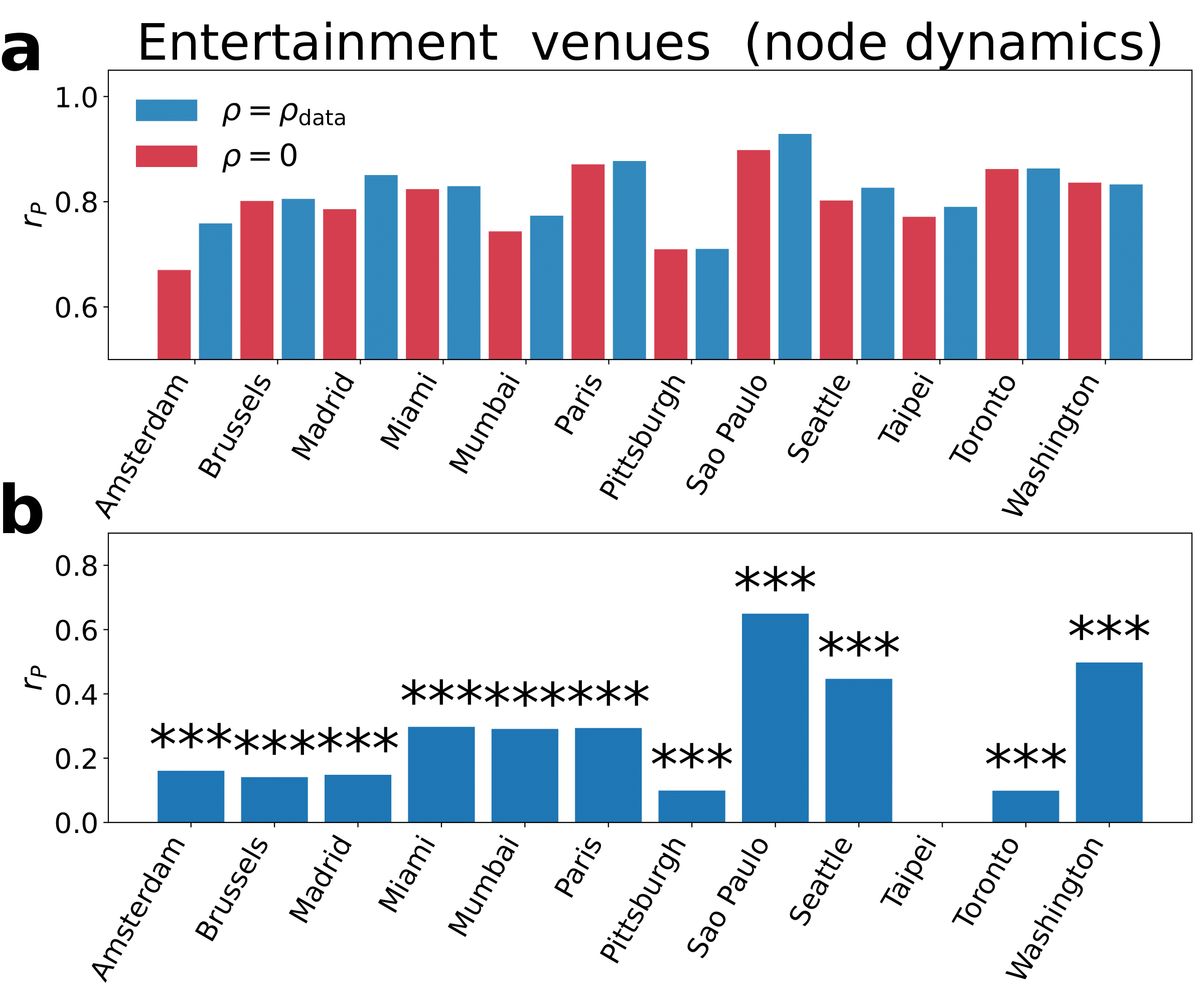}
  \end{center}
  \caption[\textbf{Correlation between the real and modelled delays in the node dynamics when destinations are distributed according to the entertainment POIs.}]{\textbf{Correlation between the real and modelled delays in the node dynamics when destinations are distributed according to the entertainment POIs.} (\textbf{a}) Comparison between the Pearson correlation coefficient obtained between the travel times from Uber Data \cite{uber} during the mornning peak ($8-10$am) in a set of cities and the travel times obtained for $\rho=0$ (red) and $\rho=\rho_{\rm data}$ (blue). (\textbf{b}) Pearson correlation coefficient between the delay observed in the data and in the model. Asterisks indicate the level of significance ($*$ p-value$<0.05$, $**$ p-value$<0.01$, $***$ p-value$<0.001$. The injection rate for each city $\rho_{\rm data}$ is set to match $\eta$ with the percentage of delay observed in the Tom Tom traffic index data \cite{tomtom}. } \label{12node}
\end{figure*}

\begin{figure*}[!htbp]
  \begin{center}
  \includegraphics[width=0.5\textwidth]{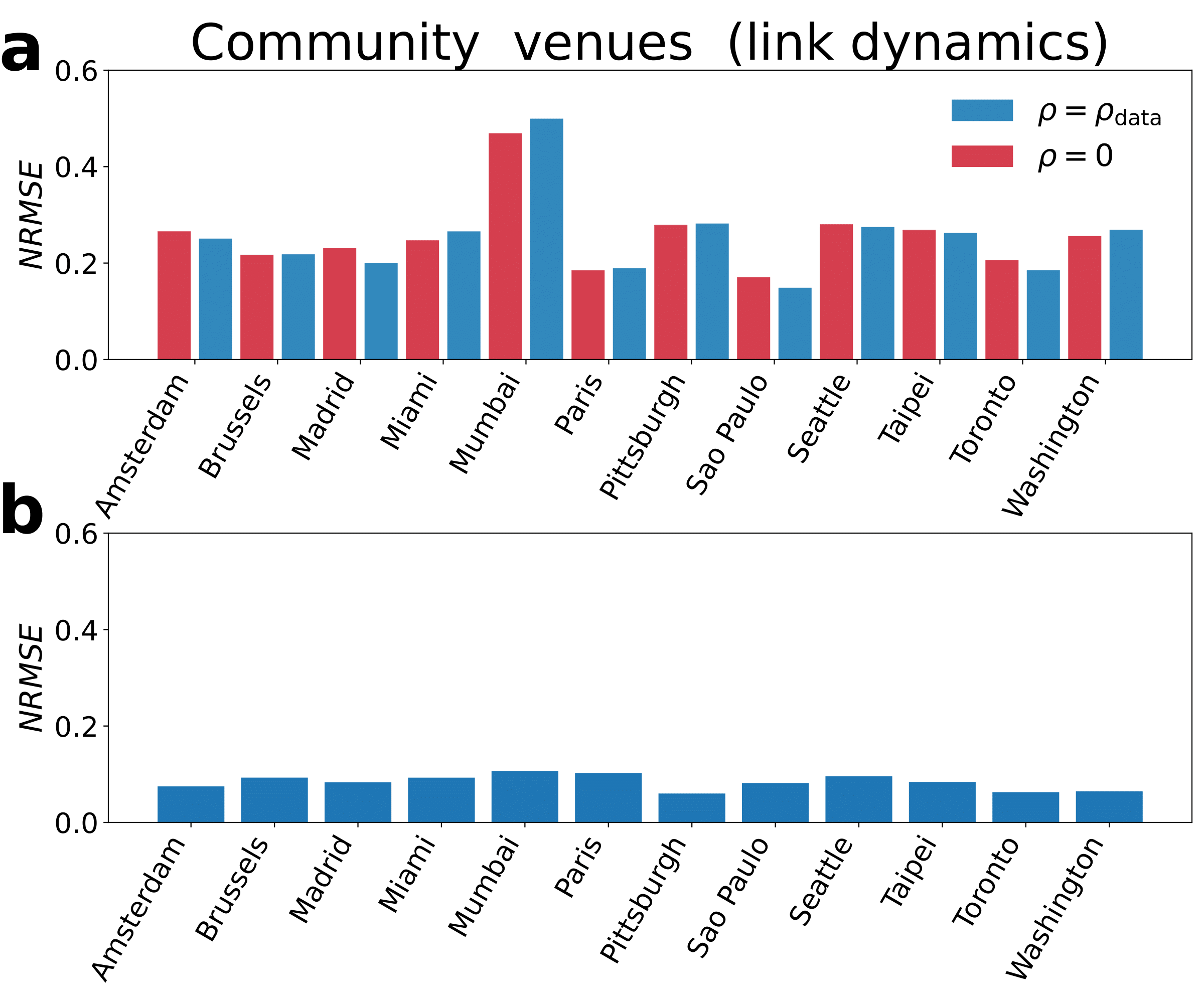}
  \end{center}
  \caption[\textbf{Analysis of the normalized mean squared error for the Correlation the delays in the node dynamics when destinations are distributed according to the community POIs.}]{\textbf{Correlation between the real and modelled delays in the node dynamics when destinations are distributed according to the community POIs.} (\textbf{a}) Normalized root mean squared error (NRMSE) obtained by dividing the  standard deviation of the residuals by the sample mean for the regression between the travel times from Uber Data \cite{uber} during the mornning peak ($8-10$am) in a set of cities and the travel times obtained for $\rho=0$ (red) and $\rho=\rho_{\rm data}$ (blue). (\textbf{b}) Normalized root mean squared error (NRMSE) obtained by dividing the  standard deviation of the residuals by the sample mean for the regression  between the delay observed in the data and in the model. Asterisks indicate the level of significance ($*$ p-value$<0.05$, $**$ p-value$<0.01$, $***$ p-value$<0.001$. The injection rate for each city $\rho_{\rm data}$ is set to match $\eta$ with the percentage of delay observed in the Tom Tom traffic index data \cite{tomtom}. } \label{934linknrmse}
\end{figure*}

\begin{figure*}[!htbp]
  \begin{center}
  \includegraphics[width=0.5\textwidth]{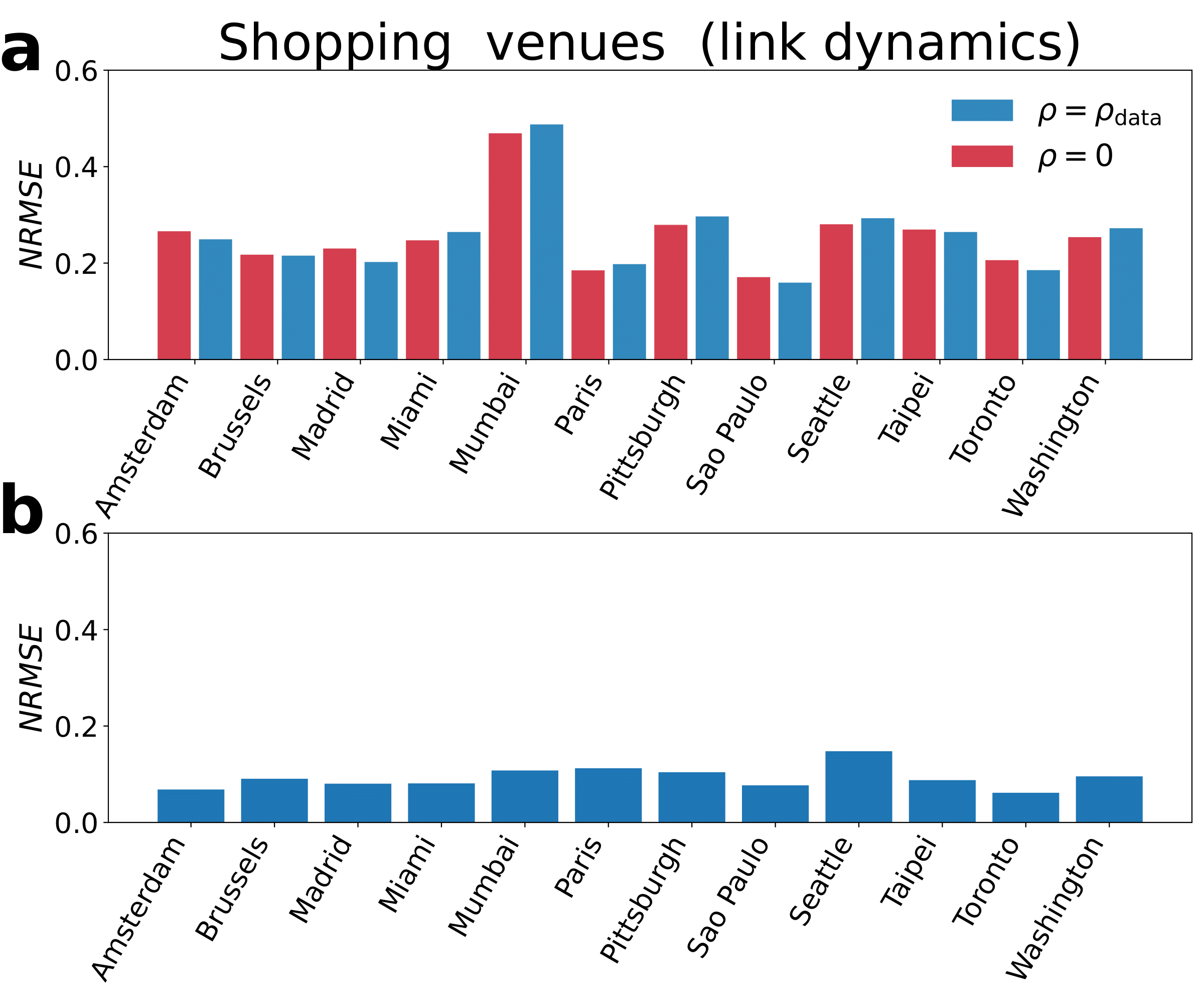}
  \end{center}
  \caption[\textbf{Analysis of the normalized mean squared error for the Correlation the delays in the node dynamics when destinations are distributed according to the community POIs.}]{\textbf{Correlation between the real and modelled delays in the node dynamics when destinations are distributed according to the community POIs.} (\textbf{a}) Normalized root mean squared error (NRMSE) obtained by dividing the  standard deviation of the residuals by the sample mean for the regression between the travel times from Uber Data \cite{uber} during the mornning peak ($8-10$am) in a set of cities and the travel times obtained for $\rho=0$ (red) and $\rho=\rho_{\rm data}$ (blue). (\textbf{b}) Normalized root mean squared error (NRMSE) obtained by dividing the  standard deviation of the residuals by the sample mean for the regression  between the delay observed in the data and in the model. Asterisks indicate the level of significance ($*$ p-value$<0.05$, $**$ p-value$<0.01$, $***$ p-value$<0.001$. The injection rate for each city $\rho_{\rm data}$ is set to match $\eta$ with the percentage of delay observed in the Tom Tom traffic index data \cite{tomtom}. } \label{934nodenrmse}
\end{figure*}

\begin{figure*}[!htbp]
  \begin{center}
  \includegraphics[width=0.5\textwidth]{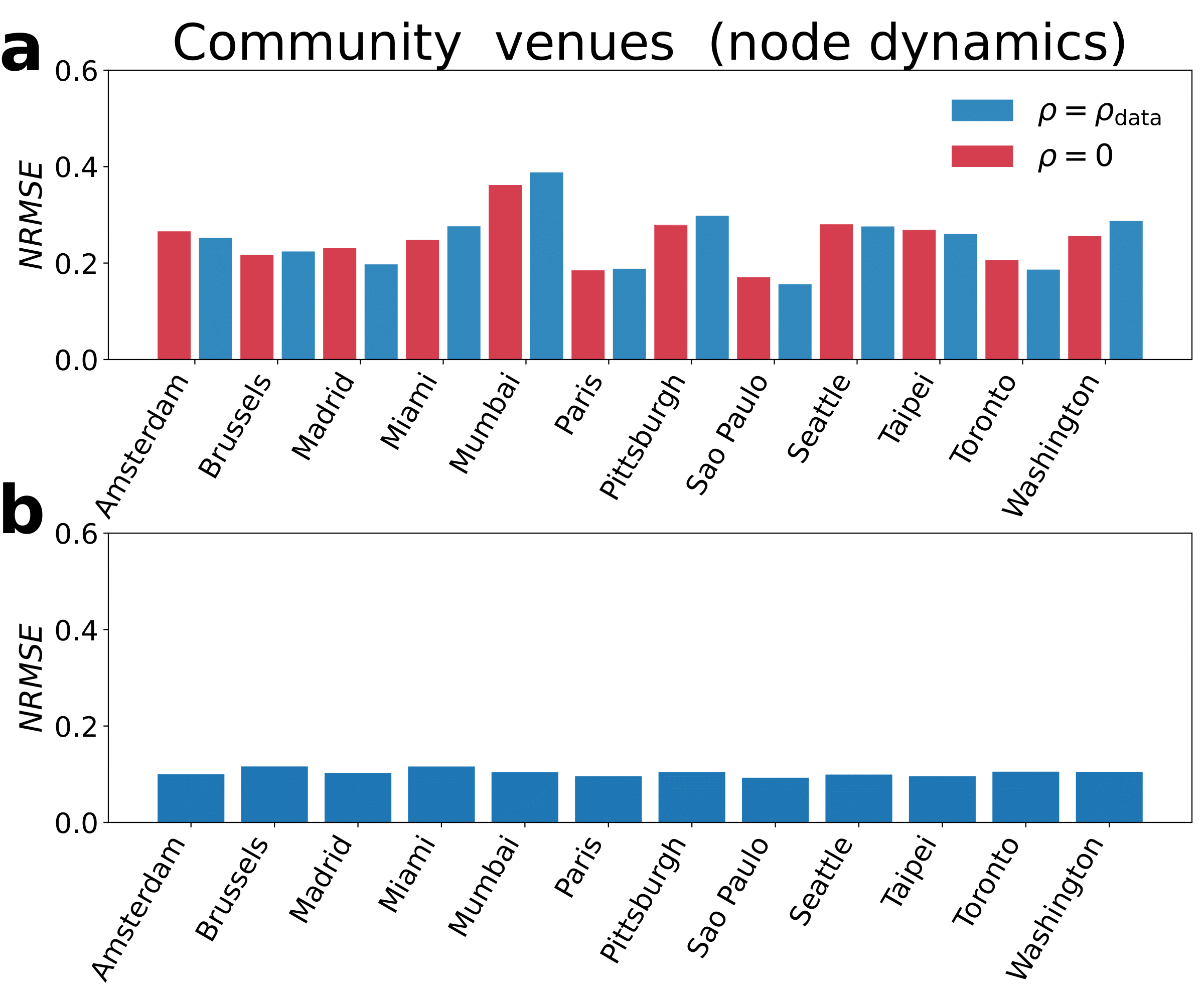}
  \end{center}
  \caption[\textbf{Analysis of the normalized mean squared error for the Correlation the delays in the link dynamics when destinations are distributed according to the shopping POIs.}]{\textbf{Correlation between the real and modelled delays in the link dynamics when destinations are distributed according to the shopping POIs.} (\textbf{a}) Normalized root mean squared error (NRMSE) obtained by dividing the  standard deviation of the residuals by the sample mean for the regression between the travel times from Uber Data \cite{uber} during the mornning peak ($8-10$am) in a set of cities and the travel times obtained for $\rho=0$ (red) and $\rho=\rho_{\rm data}$ (blue) as detailed in Eq. \ref{traveltimes}. (\textbf{b}) Normalized root mean squared error (NRMSE) obtained by dividing the  standard deviation of the residuals by the sample mean for the regression  between the delay observed in the data and in the model. Asterisks indicate the level of significance ($*$ p-value$<0.05$, $**$ p-value$<0.01$, $***$ p-value$<0.001$. The injection rate for each city $\rho_{\rm data}$ is set to match $\eta$ with the percentage of delay observed in the Tom Tom traffic index data \cite{tomtom}. } \label{6linknrmse}
\end{figure*}

\begin{figure*}[!htbp]
  \begin{center}
  \includegraphics[width=0.5\textwidth]{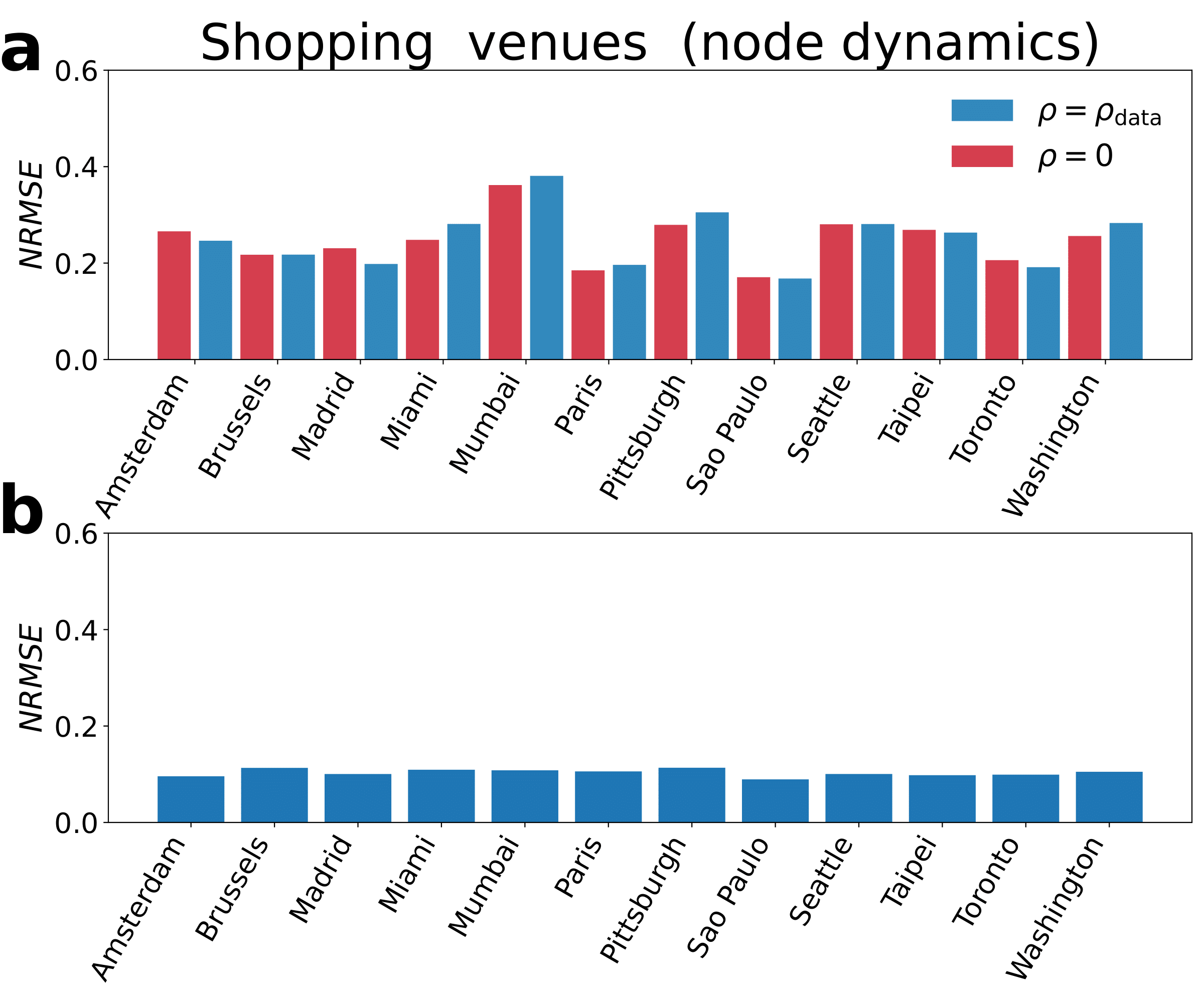}
  \end{center}
  \caption[\textbf{Analysis of the normalized mean squared error for the Correlation the delays in the node dynamics when destinations are distributed according to the shopping POIs.}]{\textbf{Correlation between the real and modelled delays in the node dynamics when destinations are distributed according to the shopping POIs.} (\textbf{a}) Normalized root mean squared error (NRMSE) obtained by dividing the  standard deviation of the residuals by the sample mean for the regression between the travel times from Uber Data \cite{uber} during the mornning peak ($8-10$am) in a set of cities and the travel times obtained for $\rho=0$ (red) and $\rho=\rho_{\rm data}$ (blue). (\textbf{b}) Normalized root mean squared error (NRMSE) obtained by dividing the  standard deviation of the residuals by the sample mean for the regression  between the delay observed in the data and in the model. Asterisks indicate the level of significance ($*$ p-value$<0.05$, $**$ p-value$<0.01$, $***$ p-value$<0.001$. The injection rate for each city $\rho_{\rm data}$ is set to match $\eta$ with the percentage of delay observed in the Tom Tom traffic index data \cite{tomtom}. } \label{6nodenrmse}
\end{figure*}

\begin{figure*}[!htbp]
  \begin{center}
  \includegraphics[width=0.5\textwidth]{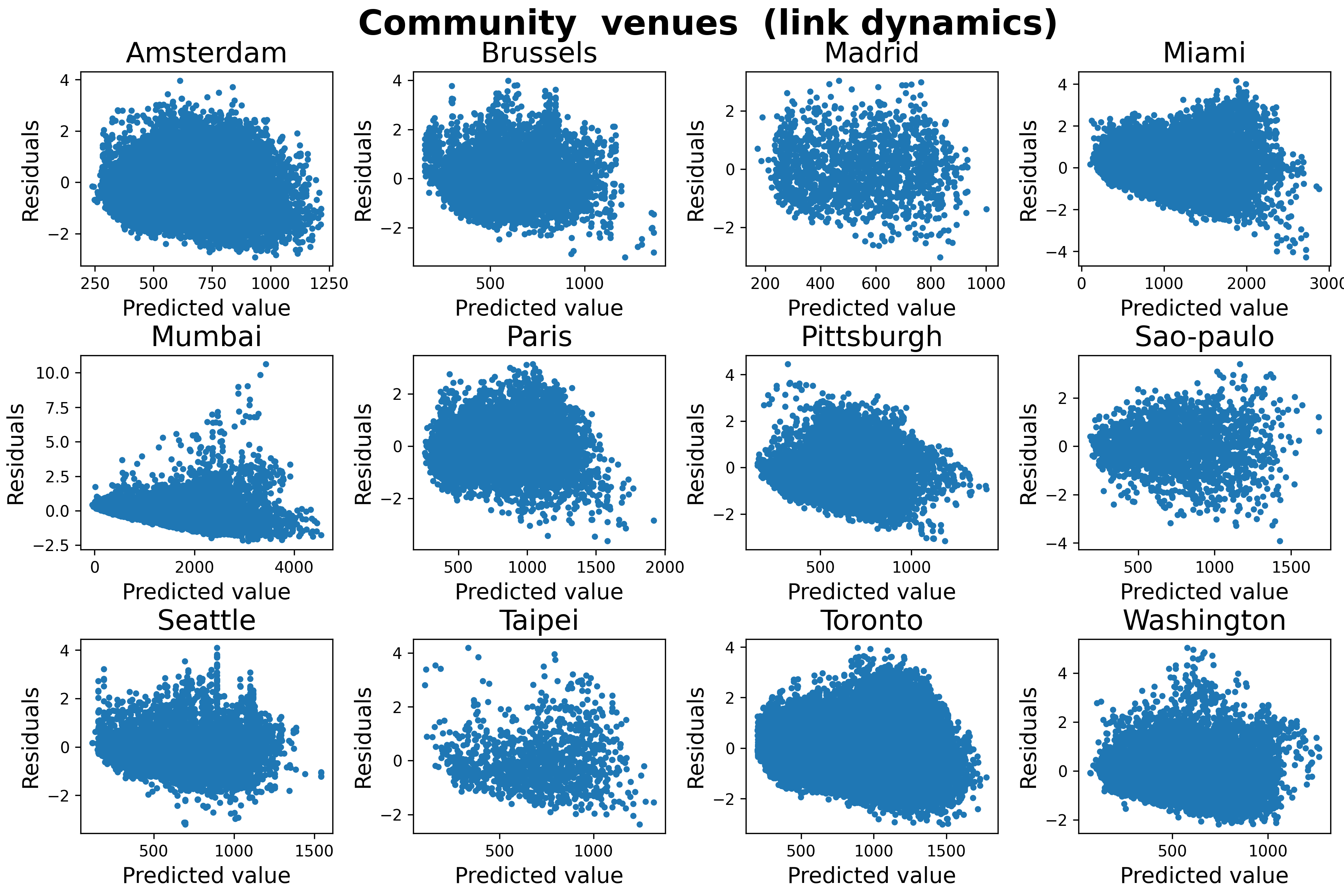}
  \end{center}
  \caption[\textbf{Residual analysis for the regression between the travel times in the link dynamics when destinations are distributed according to the community POIs.}]{\textbf{Residual analysis for the regression between the travel times in the link dynamics when destinations are distributed according to the community POIs.} (\textbf{a}) Residual analisis for the regression between the travel times from Uber Data \cite{uber} during the morning peak ($8-10$am) in a set of cities and the travel times obtained for $\rho=\rho_{\rm data}$. } \label{934linkresidualsdeln}
\end{figure*}

\begin{figure*}[!htbp]
  \begin{center}
  \includegraphics[width=0.5\textwidth]{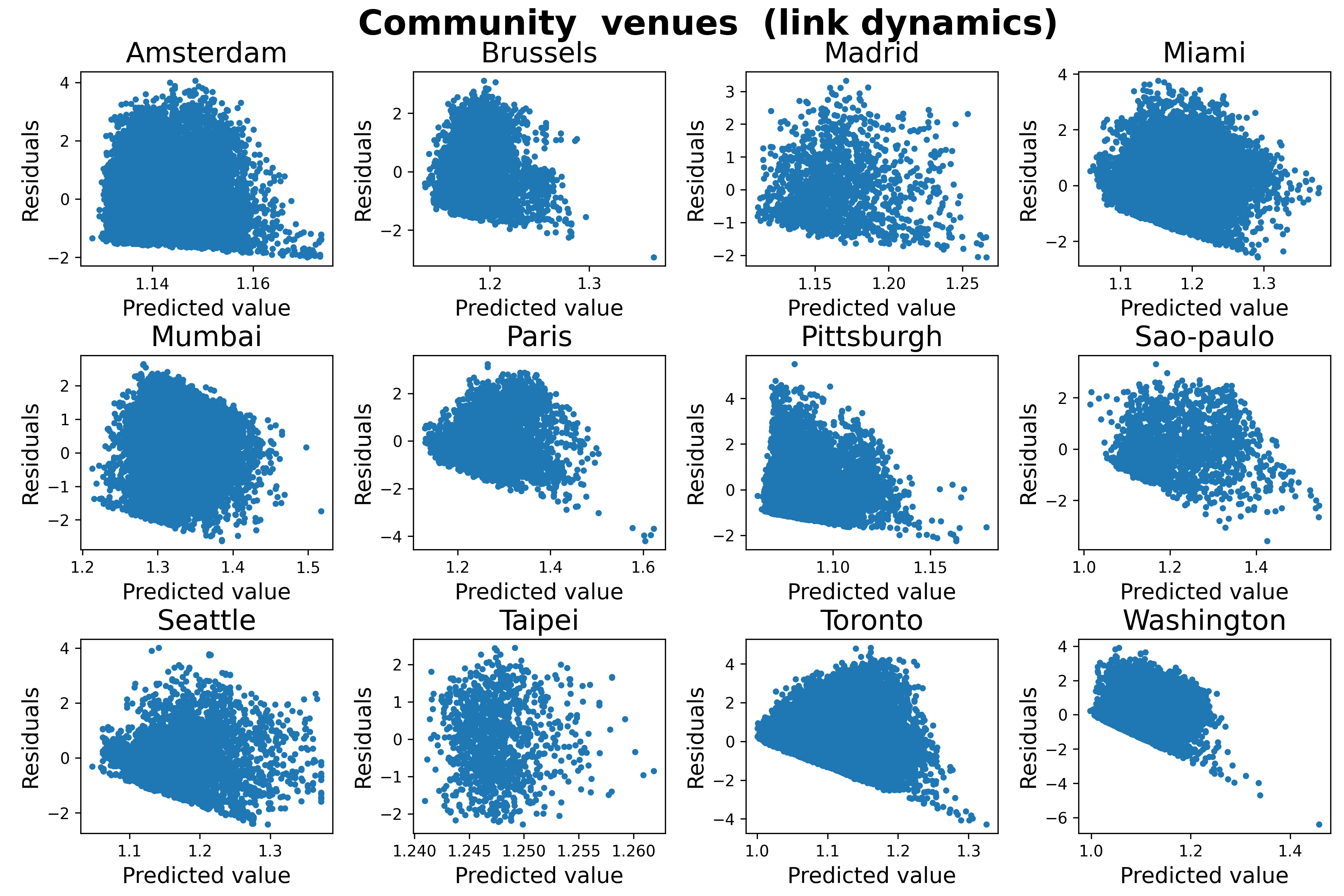}
  \end{center}
  \caption[\textbf{Residual analysis for the regression between the delays in the link dynamics when destinations are distributed according to the community POIs.}]{\textbf{Residual analysis for the regression between the delays in the link dynamics when destinations are distributed according to the community POIs.} (\textbf{a}) Residual analysis for the regression between the travel times from Uber Data \cite{uber} during the morning peak ($8-10$am) in a set of cities and the travel times obtained for $\rho=\rho_{\rm data}$. } \label{934linkresidualsdel}
\end{figure*}

\begin{figure*}[!htbp]
  \begin{center}
  \includegraphics[width=0.5\textwidth]{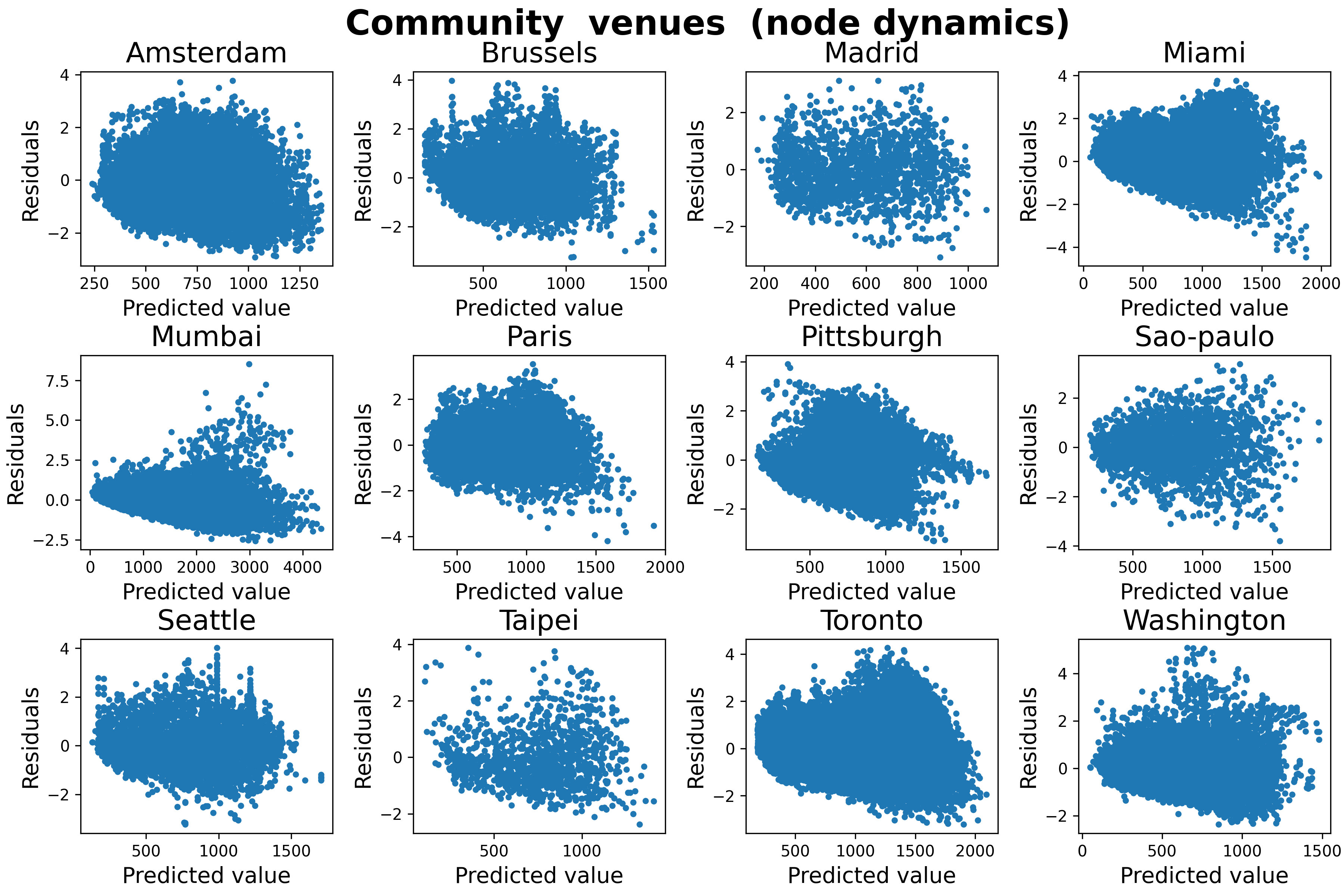}
  \end{center}
  \caption[\textbf{Residual analysis for the regression between the travel times in the node dynamics when destinations are distributed according to the community POIs.}]{\textbf{Residual analysis for the regression between the travel times in the node dynamics when destinations are distributed according to the community POIs.} (\textbf{a}) Residual analysis for the regression between the travel times from Uber Data \cite{uber} during the morning peak ($8-10$am) in a set of cities and the travel times obtained for $\rho=\rho_{\rm data}$. } \label{934noderesidualsdeln}
\end{figure*}

\begin{figure*}[!htbp]
  \begin{center}
  \includegraphics[width=0.5\textwidth]{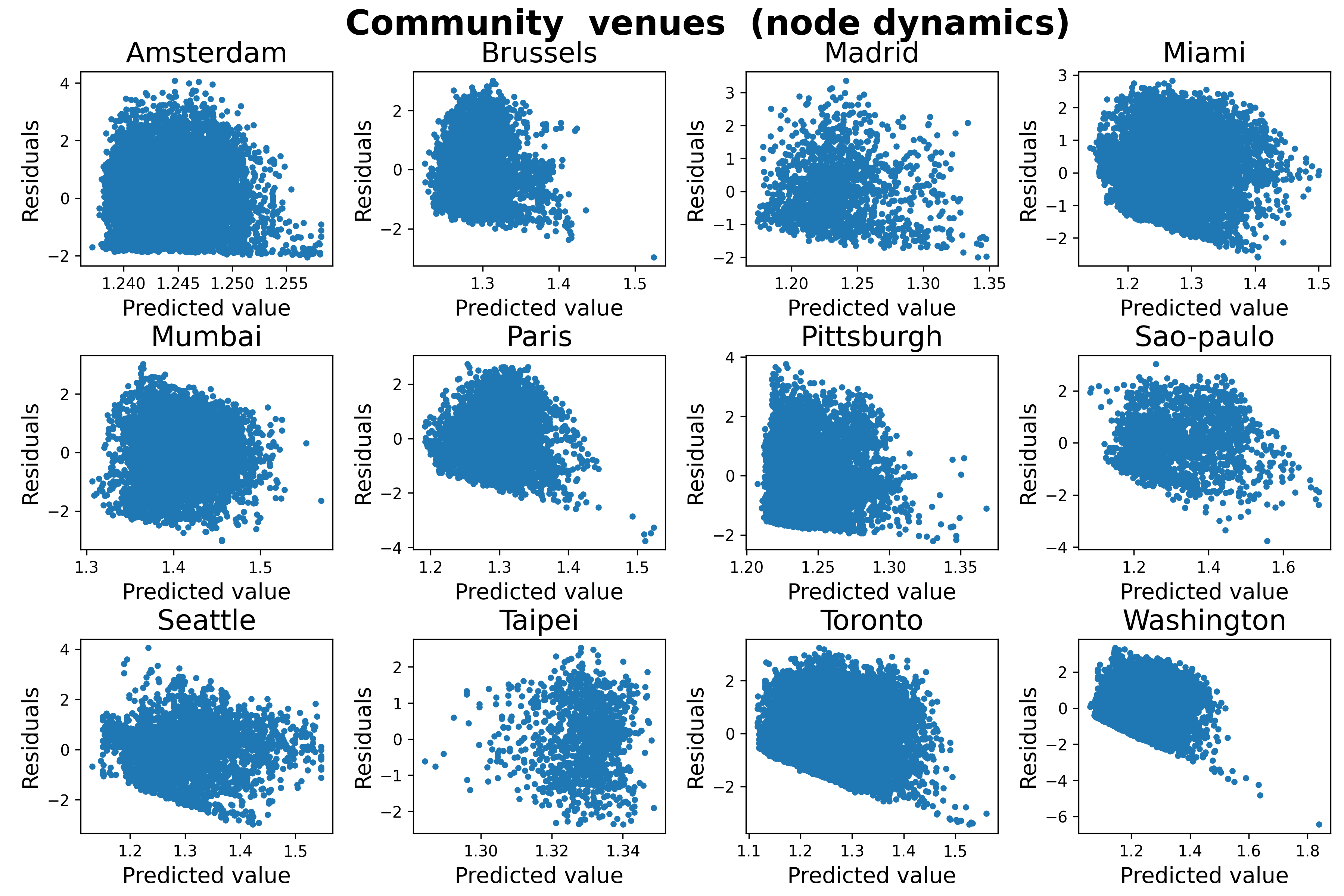}
  \end{center}
  \caption[\textbf{Residual analysis for the regression between the delays in the node dynamics when destinations are distributed according to the community POIs.}]{\textbf{Residual analysis for the regression between the delays in the node dynamics when destinations are distributed according to the community POIs.} (\textbf{a}) Residual analysis for the regression between the delay observed in the Uber Data \cite{uber} during the morning peak ($8-10$am) and in the model for $\rho=\rho_{\rm data}$. } \label{934noderesidualsdel}
\end{figure*}

\begin{figure*}[!htbp]
  \begin{center}
  \includegraphics[width=0.5\textwidth]{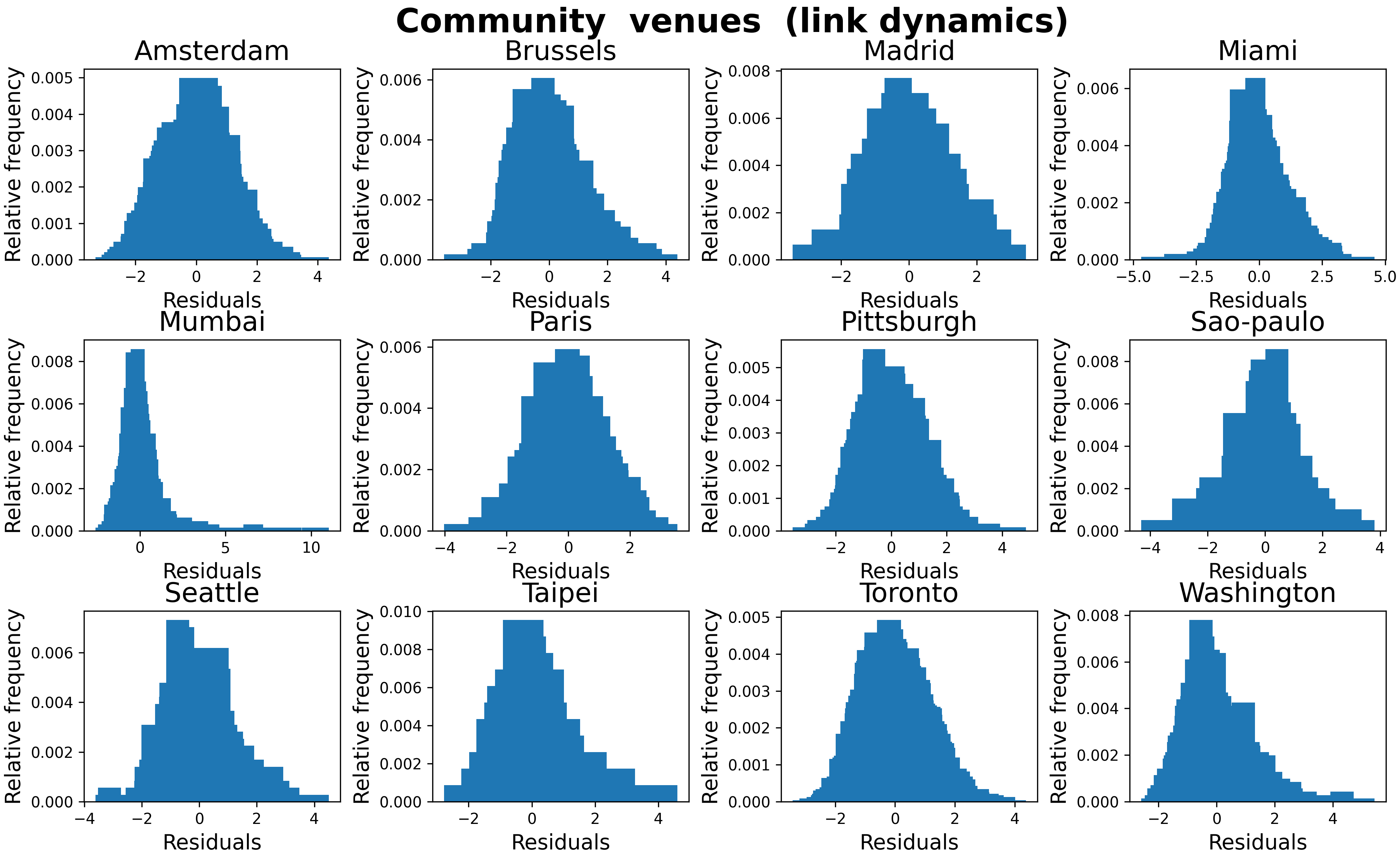}
  \end{center}
  \caption[\textbf{Distribution of residuals for the regression between the travel times in the link dynamics when destinations are distributed according to the community POIs.}]{\textbf{Distribution of residuals for the regression between the travel times in the link dynamics when destinations are distributed according to the community POIs.} (\textbf{a}) Distribution of residuals for the regression between the travel times from Uber Data \cite{uber} during the morning peak ($8-10$am) in a set of cities and the travel times obtained for $\rho=\rho_{\rm data}$. } \label{934linkdistdeln}
\end{figure*}

\begin{figure*}[!htbp]
  \begin{center}
  \includegraphics[width=0.5\textwidth]{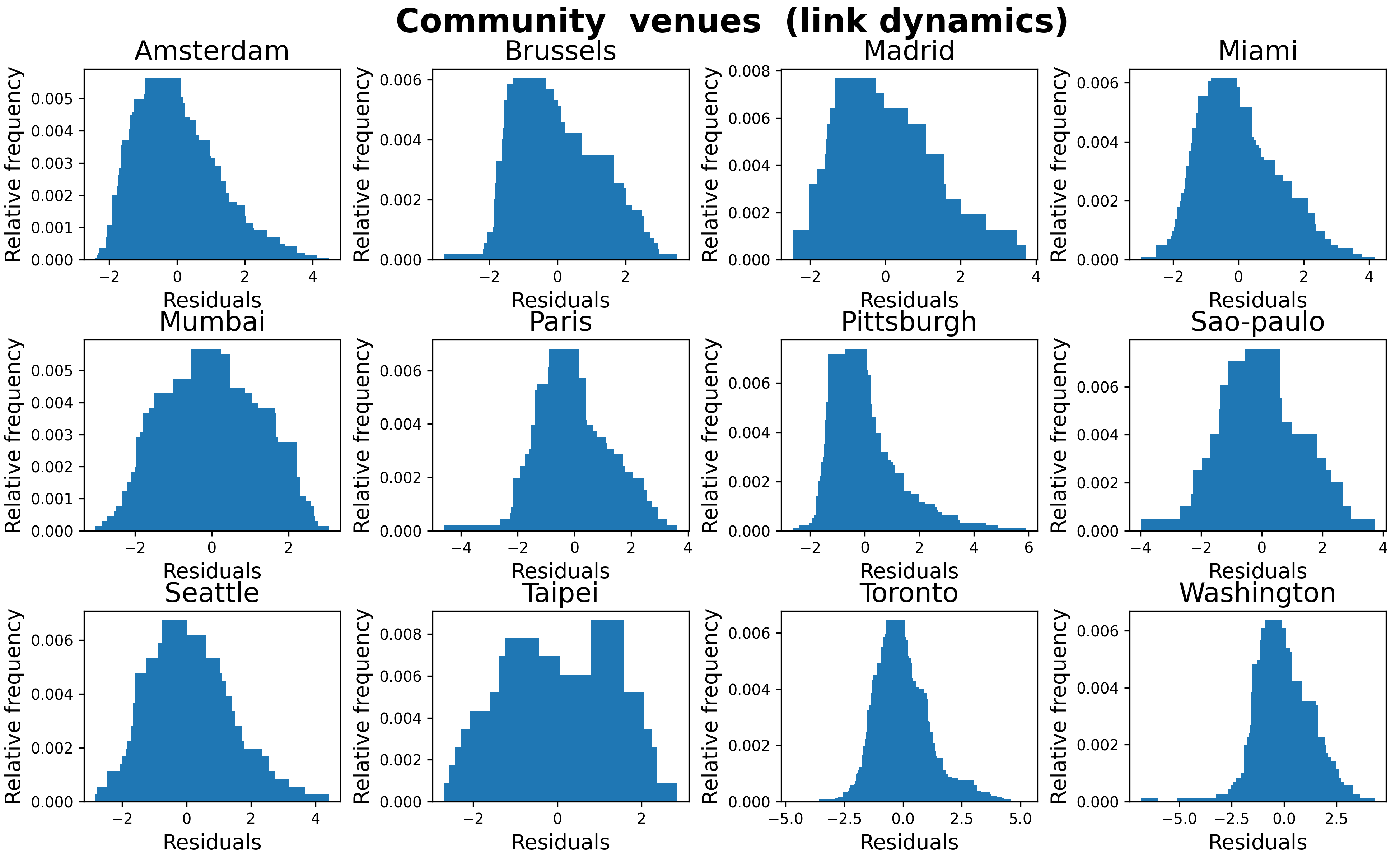}
  \end{center}
  \caption[\textbf{Distribution of residuals for the regression between the delays in the link dynamics when destinations are distributed according to the community POIs.}]{\textbf{Distribution of residuals for the regression between the delays in the link dynamics when destinations are distributed according to the community POIs.} ((\textbf{a}) Distribution of residuals for the regression between  the delay observed in the Uber Data \cite{uber} during the morning peak ($8-10$am) and in the model for $\rho=\rho_{\rm data}$. } \label{934linkdistdel}
\end{figure*}

\begin{figure*}[!htbp]
  \begin{center}
  \includegraphics[width=0.5\textwidth]{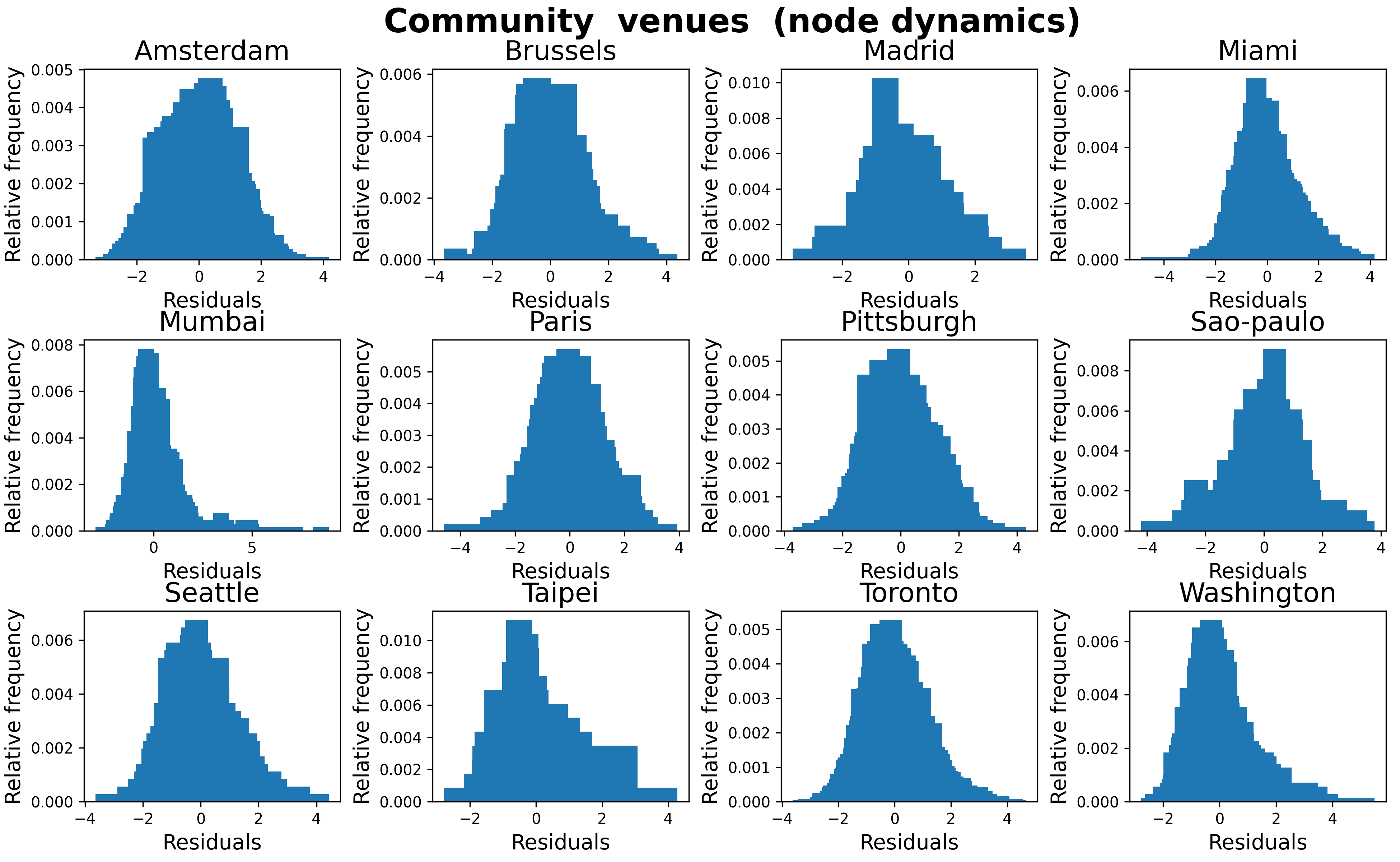}
  \end{center}
  \caption[\textbf{Distribution of residuals for the regression between the travel times in the node dynamics when destinations are distributed according to the community POIs.}]{\textbf{Distribution of residuals for the regression between the travel times in the node dynamics when destinations are distributed according to the community POIs.} (\textbf{a}) Distribution of residuals for the regression between the travel times from Uber Data \cite{uber} during the morning peak ($8-10$am) in a set of cities and the travel times obtained for $\rho=\rho_{\rm data}$.} \label{934nodedistdeln}
\end{figure*}

\begin{figure*}[!htbp]
  \begin{center}
  \includegraphics[width=0.5\textwidth]{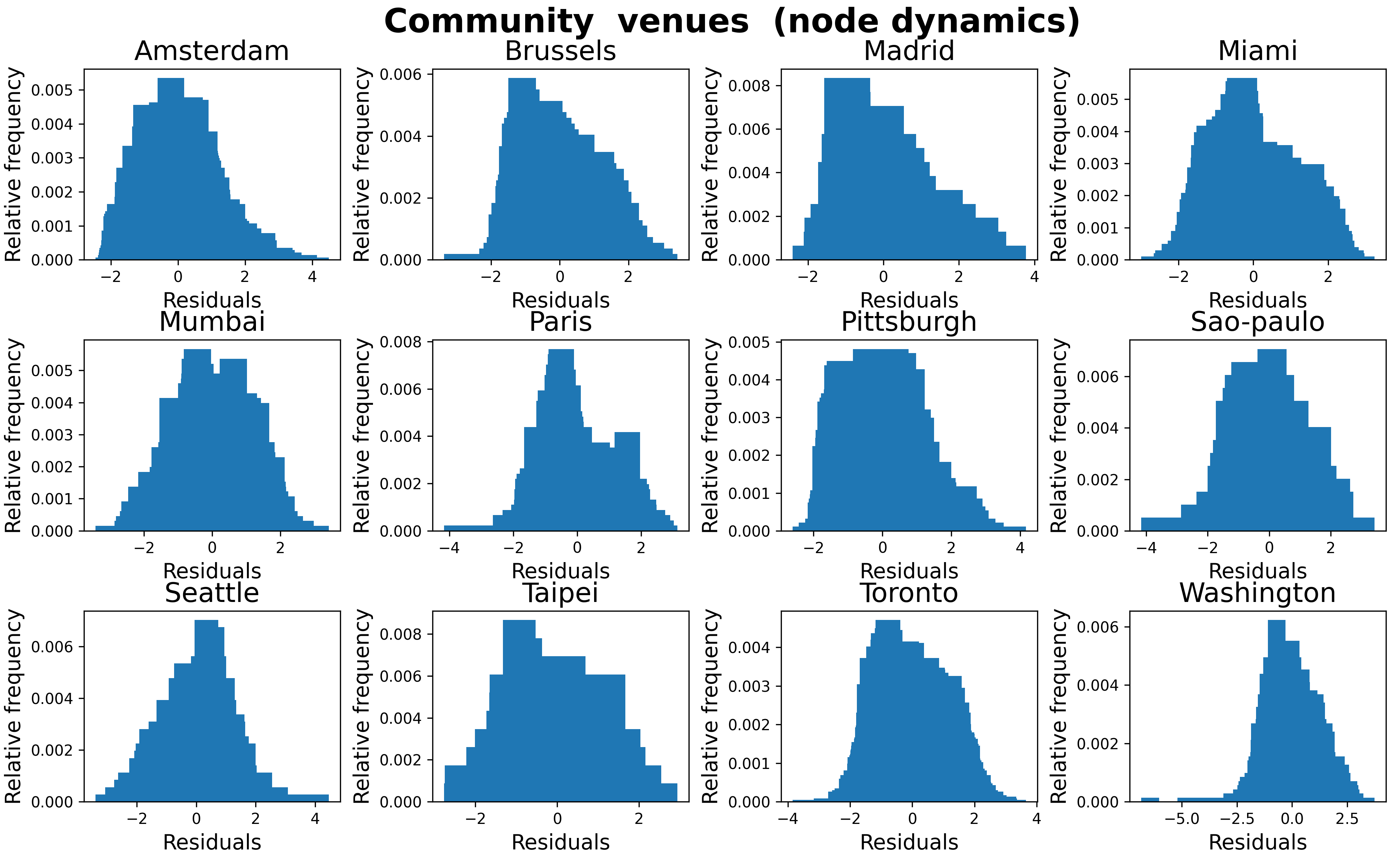}
  \end{center}
  \caption[\textbf{Distribution of residuals for the regression between the delays in the node dynamics when destinations are distributed according to the community POIs.}]{\textbf{Distribution of residuals for the regression between the delays in the node dynamics when destinations are distributed according to the community POIs.}  (\textbf{a}) Distribution of residuals for the regression between  the delay observed in the Uber Data \cite{uber} during the morning peak ($8-10$am) and in the model for $\rho=\rho_{\rm data}$. } \label{934nodedistdel}
\end{figure*}

We analyze here the correlations obtained for the expected delay with a different distribution of destinations and other models. In Fig. \ref{6link} we provide the results for the link model when destinations are distributed according to the shopping venues where we observe that there is also an increase on the prediction power from its non-delayed counterpart. By focusing only the delay, there is also a significant correlation similar to the results in Fig. 7 of the main paper.
For comparison we display in Figs. \ref{934node} and \ref{6node} the same analysis for the node model. As it is shown there, although the correlations are still present, they are lower than for the link model, specially if we focus only on the delay. We have also analyzed the case of food venues (Figs. \ref{7link} and \ref{7node}) and entertainment venues (Figs. \ref{12link} and \ref{12node}).
Additional results regarding the normalized mean squared error is shown in Figs. \ref{934linknrmse}, \ref{934nodenrmse}, \ref{6linknrmse} and \ref{6nodenrmse}
We also provide in Figs. \ref{934linkdistdel}, \ref{934linkdistdeln}, \ref{934linkresidualsdel},\ref{934linkresidualsdeln} the analysis of residuals for the travel times and delay regression in the case of community venues for the link model and in Figs. \ref{934nodedistdel},\ref{934nodedistdeln}, \ref{934noderesidualsdel}, \ref{934noderesidualsdeln} for the node model.

\begin{table}
 \caption{Pearson correlation between the traffic counts in Madrid and the flow of vehicles according to our model. The analysis has been performed including the 10$\%$ of links with highest flow. }
\begin{center}
\begin{tabular}{ |c|c| } 
 \hline
 Venue & $r_P$  \\ 
  \hline
Community & 0.49$^{***}$ \\ 
  \hline
Outdoors & 0.30$^{***}$ \\
 \hline
 Nightlife & 0.39$^{***}$ \\
  \hline
  Shopping & 0.41$^{***}$ \\
   \hline
  Food & 0.40$^{***}$ \\
    \hline
Travel & 0.44 $^{***}$ \\
    \hline
Entertainment 0.43  $^{***}$ \\
\hline
\end{tabular}
\label{countsmadrid}
\end{center}
\end{table}

\bibliography{references}

\end{document}